\documentclass[final]{siamltex}
\usepackage{amssymb}
\usepackage{amsfonts}
\usepackage{latexsym,bm}
\usepackage{mathrsfs}
\usepackage{amsmath}
\usepackage{graphicx}
\usepackage{placeins}
\usepackage[ruled,vlined]{algorithm2e}
\usepackage{amsmath}
\usepackage{listings}
\usepackage{xcolor}
\usepackage{makecell}
\usepackage{bbding}
\usepackage{hyperref}
\usepackage{graphicx} 
\usepackage{float}
\usepackage{tabularx} 
\usepackage{booktabs} 
\usepackage{subfig} 
\usepackage{algorithmic}
\newtheorem{Theorem}{Theorem}
\allowdisplaybreaks

\def\min{\mbox{min}}
\usepackage[top=2.3cm,bottom=1.2cm,left=2.8cm,right=2.8cm,includehead,includefoot
         ]{geometry}

\title{A Mixed-Form PINNs (MF-PINNs) for Solving the Coupled Stokes-Darcy Equations \thanks{L. Shan is supported by Guangdong Basic and Applied Basic Research Foundation (2024A1515010294) and STU Scientifc Research Initiation Grant (NTF25007T,NTF21006)}}
\author{
Li Shan\thanks{%
Corresponding author. Department of Mathematics, Shantou University, Shantou, China.  ({\tt lishan@stu.edu.cn})}
\and
Xi Shen\thanks{%
Department of Mathematics, Shantou University, Shantou, China.  ({\tt 24xshen@stu.edu.cn})}}

\date{}
\begin{document}
\maketitle
\begin{abstract}
Parallel physical information neural networks (P-PINNs) have been widely used to solve systems with multiple coupled physical fields, such as the coupled Stokes-Darcy equations with Beavers-Joseph-Saffman (BJS) interface conditions. However, excessively high or low physical constants in partial differential equations (PDE) often lead to ill-conditioned loss functions and can even cause the failure of training numerical solutions for PINNs. To solve this problem, we develop a new kind of enhanced parallel PINNs, MF-PINNs, in this article. Our MF-PINNs combines the velocity-pressure form (VP) with the stream-vorticity form (SV) and add them with adjusted weights to the total loss functions. The results of numerical experiments show our MF-PINNs have successfully improved the accuracy of the streamline fields and the pressure fields when kinematic viscosity and permeability tensor range from $ 10^{-4} $ to $ 10^{4} $. Thus, our MF-PINNs hold promise for more chaotic PDE systems involving turbulent flows. Additionally, we also explore the best combination of the activation functions and their periodicity. And we also try to set the initial learning rate and design its decay strategies. The code and data associated with this paper are available at \href{https://github.com/shxshx48716/MF-PINNs.git}{https://github.com/shxshx48716/MF-PINNs.git}. 
\end{abstract}

\pagestyle{myheadings} \thispagestyle{plain} \markboth{L.~Shan and X.~Shen}{ A Mixed Form PINNs (MF-PINNs) for Solving the Coupled Stokes-Darcy Equations}

\section{Introduction}
Stokes-Darcy coupling models arise in several applications, such as interaction between surface and groundwater flows, oil reservoirs in vuggy porous media, and industrial filtrations. In mathematical modeling, the Stokes and Darcy equations are employed to describe free fluid flows and porous media seepage, respectively. Additional equations are introduced to comply with physical laws, such as mass conservation, normal stress balance, and the BJS conditions \cite{ref-journal12}. 

\setlength{\parindent}{2em}
The rapid advancement of artificial intelligence has increased the applications for deep neural networks, such as PINNs \cite{ref-journal1}, as a new approach for solving PDE. Moreover, parallel PINNs and region decomposition strategies \cite{ref-journal4,ref-journal6,ref-journal16,ref-journal17,ref-journal18,ref-journal19,ref-journal13} use multiple GPUs to train multiple neural networks in parallel. Above, all of these methods are designed to handle coupled models with multiple physical fields and media, including the coupled Stokes-Darcy system. Compared with traditional numerical methods, finite difference method, finite element method, finite volume method, spectral method, etc., PINNs offer several advantages for coupled systems: $ \mathit{(i)} $ no need for mesh generation; $ \mathit{(ii)} $ handling boundary conditions more flexibly; $ \mathit{(iii)} $ multi-scale systems of overdetermined equations; $ \mathit{(iv)} $ enriched interpolation (activation) functional spaces. However, how to mitigate the gradient competition between multi-objective loss functions and accurately capture the frequency of PDEs remains an open research question.


\setlength{\parindent}{2em}
The current research for balancing gradient competition between boundary errors and PDE errors is as follows: second-order optimization perspective, a new quasi-Newton method \cite{ref-journal31} ; dual cone gradient descent \cite{ref-journal29} ; neural tangent kernel theory \cite{ref-journal28,ref-journal30} ; multi-magnitude PINNs \cite{ref-journal32} ; conflict-free inverse gradients \cite{ref-journal33} , etc. Several studies have discretized equation systems to solve the coupling among the multiple physical fields: semi implicit method for pressure linked equations (SIMPLE) \cite{ref-book1} ; component-consistent 
pressure correction \cite{ref-journal36}, etc. Few experiments have studied the gradient competition between coupled equations \cite{ref-journal34}, etc. In brief, these methods have explored various approaches to correct the ill-conditioned numerical formats and have achieved favorable improvements. Thus, we try to develop a new type of PINNs, MF-PINNs, which decouples the equations and rebalances the loss functions to mitigate the gradient competition among different physical quantities.

\setlength{\parindent}{2em}
Recently, a wide variety of operator mappings have been widely applied to PINNs. For instance, adaptive activation functions strategies \cite{ref-journal38}, Fourier feature PINNs (FFPINNs) \cite{ref-journal37}, DNN for approximating nonlinear operators \cite{ref-journal39,ref-journal40}, etc. Besides, several studies have revealed the basic logic to make improvements, such as decomposition based DNN \cite{ref-journal41} based on the Frequency Principle \cite{ref-journal42} , etc. In all, these theories and approaches enhance the fitting and generalization capabilities of PINNs by developing the operators mappings. Hence, we aim to identify the proper periods of all multiple physical fields and construct a proper neural operator basis adaptive to the problem in order to improve accuracy and reduce extra computational costs.

\setlength{\parindent}{2em}
Plenty of research has been devoted to developing optimal strategies for learning rate scheduling to improve PINNs. For example, physics-constrained neural networks with the minimax architecture \cite{ref-journal43}, residual adaptive networks \cite{ref-journal44}, preprocessing for weights and bias \cite{ref-journal45}, etc. Consequently, we aim to develop stable and universal learning rate decay strategies for improving. 

\setlength{\parindent}{2em}
In order to solve these problems above, we first explain why the traditional PINNs sometimes fail to converge to the analytical solutions under extreme physical constants. Then, we innovate a new kind of enhanced PINNs, MF-PINNs. We decouple the multiple physics fields of the Stokes-Darcy equations and add mixed-form equations into the loss functions. These improvements create well-conditioned loss functions for PINNs and mitigate gradient competition between muiltiple physical fields. Besides, we research the impact of the periodicity of activation functions and apply a fast and universal learning rate decay strategy for training PINNs.

\setlength{\parindent}{2em}
The organization of this paper is as follows: In Section~\ref{sec:Physical modeling}, we introduce the coupled Stokes-Darcy model and decouple the velocity and pressure fields. In Section~\ref{sec:Algorithm framework}, we present the traditional parallel PINNs and apply the multi-scale operator-decoupled equations to develop MF-PINNs. In Section~\ref{sec:Numerical test}, we conduct numerical experiments to validate the effectiveness of our MF-PINNs and provide a detailed analysis based on the results. In Section~\ref{sec:Conclusions and prospects}, we summarize several key suggestions for training PINNs.

\section{Physical modeling}\label{sec:Physical modeling}
\setlength{\parindent}{2em} To begin with, we define a symbolic declaration in Section~\ref{sec:Symbol declaration}. Next, the coupled Stokes-Darcy system is established by physics laws in Section~\ref{sec:The velocity-pressure form of the Stokes-Darcy system}. Furthermore, we decouple the velocity and pressure in Section~\ref{sec:The stream-vorticity form of the Stokes-Darcy system}.

\subsection{Symbol declaration}\label{sec:Symbol declaration}

\begin{enumerate}
	\item The subscript $ s $ means the Stokes system and the subscript $ d $ means the Darcy system. And the subscript $ NN $ represents a numerical solution of PINNs. 
	
	\item The $ \Omega $ represents a given domain with boundary $ \partial \Omega $, and $ \Gamma $ represents a certain subset of  $ \partial \Omega $. The $ \mu (\Omega) $ represents the measurement of the region $ \Omega $. The $ \mathbf{n}_{s} $ and $ \mathbf{n}_{d} $ respectively represent the outward normal vectors of the domain. The $ \boldsymbol{\tau} $ represents the tangential vector. Their relationships are as follows: 
	\begin{align*}
		\Omega = \Omega_s \cup \Omega_d,
		\Gamma = \partial \Omega_s \cap \partial \Omega_d, 
		\Gamma_s = \partial \Omega_s \setminus \Gamma,
		\Gamma_d = \partial \Omega_d \setminus \Gamma.
	\end{align*}
	
	\item The bolded vector  $ \mathbf{u} = \left[ u, v \right]^{T} $ in 2D or $ \mathbf{u} = \left[ u_{1}, u_{2}, u_{3} \right]^{T} $ in 3D stands for the velocity field and the components paired with the Cartesian coordinates, $ x $, $ y $, and $ z $. Similarly, the $ \Psi $ in 2D or the $ \mathbf{\Psi} = \left[ \Psi_{1}, \Psi_{2}, \Psi_{3} \right]^{T} $ in 3D stands for Streamline field. The $ \boldsymbol{\omega} = \left[ \omega_{1}, \omega_{2}, \omega_{3} \right]^{T} $ in 3D represents the vorticity field. And the unbold scalar $ p $ represents the pressure field.	
	
	\item In a 3D steady flow the streamline field is the family of curves $\mathbf{\Psi}$ and its rotation is the velocity field, $ \nabla \times \mathbf{\Psi} = \mathbf{u} $. The vorticity $\boldsymbol{\omega}$ is the rotation of the velocity field $\mathbf{u}$, $ \boldsymbol{\omega} = \nabla \times \mathbf{u} $. It measures the local rotation of fluid.
	
	\item The $ \mathbb{K} $ represents an n-dimensional matrix, and it is a constant. The $ \mathbb{I} $ represents the identity matrix. The $ \mathbf{1} = \left[ 1, \cdots ,1 \right]  $  is a row vector of shape $ 1 \times n $ in nD. Its every element is $ 1 $.
	\item The symbol $ \circ $ denotes the composite functions (mappings), and it means composite functions (mappings) are performed from right to left. The order of differential operators is the same.
	\item The symbol $ \left\| \cdot \right\|_{2}  $ represents the Euclidean norm of a matrix or vector. The $ err\mathit{L}_{2} $ means the relative Euclidean norm. 
	\item The x $ \gg $ y  or  x $ \ll $ y respectively means that x is much higher than or much lower than y.
	\item $ \theta $  is the adaptive parameter of the activation functions. Furthermore, $ a $ and $ b $ are respectively adaptive parameters of the parallel PINNs in the Stokes and Darcy domains.
	
\end{enumerate}

\clearpage

\subsection {The velocity-pressure form of the Stokes-Darcy system}\label{sec:The velocity-pressure form of the Stokes-Darcy system} 
\begin{figure}[H]
	\centering
	\includegraphics[width=10cm]{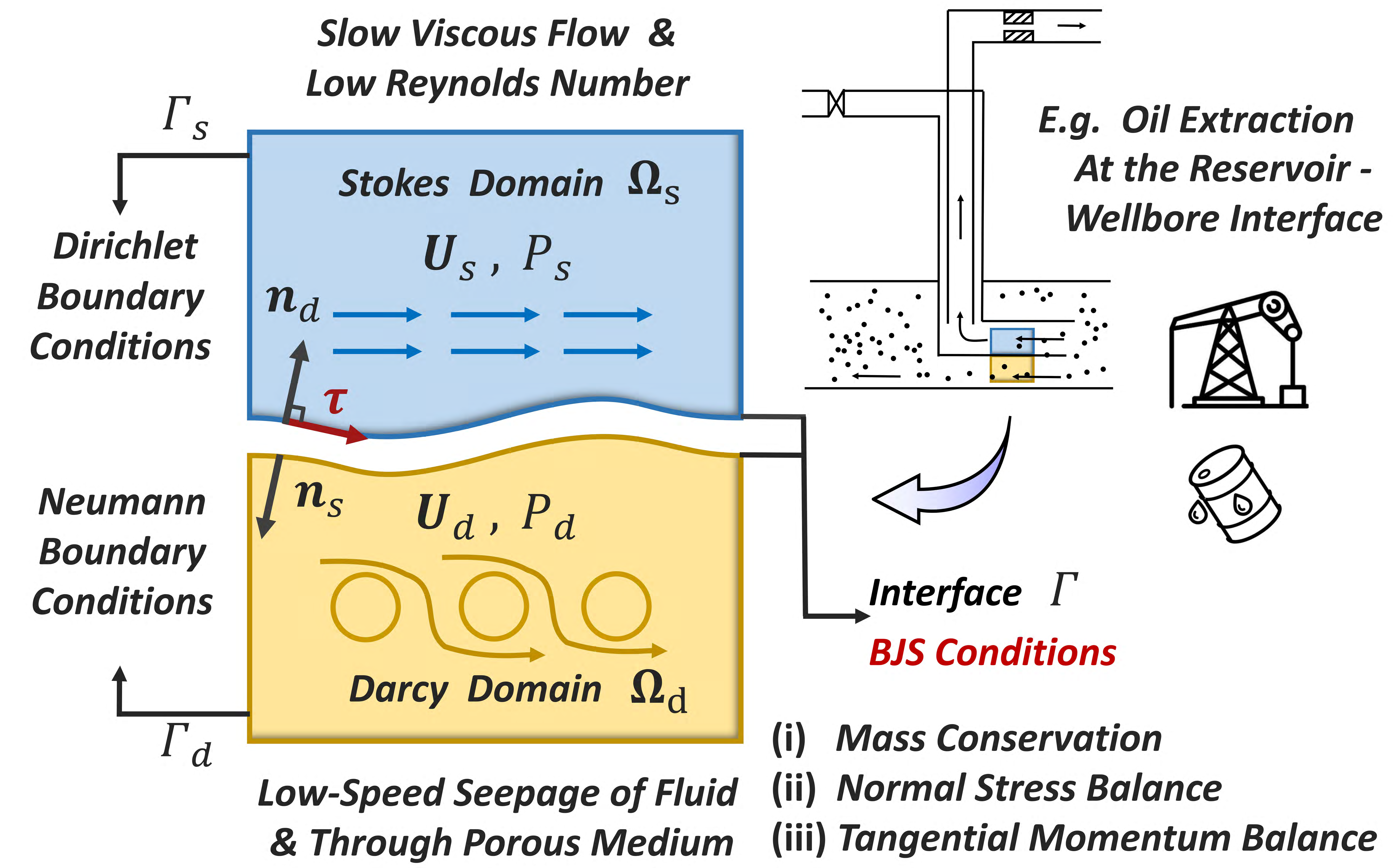}
	\caption{This cartoon overviews the coupled Stokes-Darcy model with the BJS interface conditions.\label{SD.pdf}}
\end{figure}

Here, we introduce the equations of the coupled Stokes-Darcy equations in the VP form \cite{ref-journal62} in the following four parts and Fig.~\ref{SD.pdf}.
\\
\textbf{Stokes' law} 

\setlength{\parindent}{2em} The steady incompressible Stokes equations describe the motion of viscous fluids when the inertial forces are negligible compared to the viscous forces: 
\begin{subequations}\label{eq:Stokes}
	\begin{alignat}{3} 	
		-\nabla \cdot \mathcal{T}(\mathbf{u}_s, p_s) &= \mathbf{f}_s,  & \qquad & \mathbf{x} \in \Omega_s,	\label{eq:Stokes2} \\
		\nabla \cdot \mathbf{u}_{s} &= 0,  & \qquad &\mathbf{x} \in \Omega_s, \label{eq:Stokes1} \\	
		\mathbf{u}_s &= \mathbf{g}_{\Gamma_s},  & \qquad &\mathbf{x} \in \Gamma_{s}, \label{eq:BC_Stokes} 	
	\end{alignat}
\end{subequations}

\setlength{\parindent}{0em} where $ \mathbf{u}_s $ represents the fluid velocity, \(p_s\) represents the kinematic pressure, $ \mathbf{f}_s $ represents the external force (homogeneous or inhomogeneous term), \(\nu > 0\) represents the kinematic viscosity of the fluid,  $ \mathcal{T}(\mathbf{u}_s, p_s) = 2\nu \mathcal{D}(\mathbf{u}_s) - p_s \mathbb{I} $ represents the stress tensor, and $ \mathcal{D}(\mathbf{u}_s) =  (\nabla \mathbf{u}_s + (\nabla \mathbf{u}_s)^T)/2 $ represents the deformation tensor. The (\ref{eq:Stokes2}) could be simplified as  $ \nabla p_{s} - \nu \Delta \mathbf{u}_{s} = \mathbf{f}_{s} $ when the Stokes fluid is incompressible embodied in (\ref{eq:Stokes1}).

\setlength{\parindent}{0em} \textbf{Darcy's Law} 

\setlength{\parindent}{2em} The steady incompressible Darcy equations describe fluid flow within the porous medium: 
\begin{subequations}\label{eq:Darcy}
	\begin{alignat}{3}
		\nu \mathbb{K}^{-1} \mathbf{u}_d + \nabla p_d &= \mathbf{f}_d, & \qquad & \mathbf{x} \in \Omega_d, \label{eq:Darcy2}\\
		\nabla \cdot \mathbf{u}_{d} & = 0 , & \qquad & \mathbf{x} \in \Omega_d, \label{eq:Darcy1} \\
		\mathbf{u}_d \cdot \mathbf{n}_d & = \mathbf{g}_{\Gamma_d}, & \qquad & \mathbf{x} \in \Gamma_{d}, \label{eq:BC_Darcy}
	\end{alignat}
\end{subequations}

\setlength{\parindent}{0em}  where $ \mathbf{u}_d $ represents the fluid velocity, $ p_d $ represents the dynamic pressure, $ \mathbf{f}_d $ represents the external force source term, and permeability tensor $ \mathbb{K} $ represents a positive symmetric tensor. Although tensor $ \mathbb{K} $ may vary, it usually keeps  $ \mathbb{K} = \kappa \mathbb{I} $.

\setlength{\parindent}{0em} \textbf{Interface conditions} 

\setlength{\parindent}{2em} The well-known Beavers-Joseph-Saffman boundary conditions describe the flow characteristics at the interface between the free-flow region of Stokes equation and the porous medium region of Darcy equation:
\begin{subequations}\label{eq:Interface}
	\begin{alignat}{3}
		\mathbf{u}_s \cdot \mathbf{n}_s + \mathbf{u}_d \cdot \mathbf{n}_d &= 0, & \qquad &\mathbf{x} \in \Gamma,  \label{eq:Interface1} \\
		2\nu \mathbf{n}_s \cdot \mathcal{D}(\mathbf{u}_s) \cdot \mathbf{n}_s  - p_s + p_d &= g_{\Gamma_{1}} , & \qquad &\mathbf{x} \in \Gamma, \label{eq:Interface2} \\
		2\mathbf{n}_s \cdot \mathcal{D}(\mathbf{u}_s) \cdot \boldsymbol{\tau}  +\alpha \mathbb {K}^{-1/2} \mathbf{u}_s \cdot \boldsymbol{\tau} & = g_{\Gamma_{2}}, & \qquad & \mathbf{x} \in \Gamma, \label{eq:Interface3}
	\end{alignat}
\end{subequations}

\setlength{\parindent}{0em} where the parameter $ \alpha $ is a constant affected by the friction.

\setlength{\parindent}{2em} The first equation (\ref{eq:Interface1}) stands for the continuity of normal velocity to keep the mass conservation, the second equation (\ref{eq:Interface2}) stands for the continuity of normal stress to keep the equilibrium condition, and the last equation (\ref{eq:Interface3}) stands for frictional effects at the interface in order to keep the tangential velocity slip condition \cite{ref-journal12}, as is shown in Fig.~\ref{SD.pdf}. 

\setlength{\parindent}{0em} \textbf{Pressure conditions}

\setlength{\parindent}{2em} The pressure is non-unique due to an additional constant hidden in this system, because $ p_{s} - p_{d} $ and $ (p_{s} + C) - (p_{d} + C) $ are equivalent in (\ref{eq:Interface2}). Thus, we could add (\ref{eq:Pressure}) to fix the reference frame of the pressure field: 

\begin{align}		
	\left( \int_{\Omega_{s}} p_s \, d{\Omega_{s}} + \int_{\Omega_{d}} p_d \, d{\Omega_{d}} \right)  \Big/ \left( \mu(\Omega_{s}) + \mu(\Omega_{d}) \right) = C_{p} , \quad \mathbf{x} \in \Omega_{s} \,\,\, or \,\,\, \mathbf{x} \in \Omega_{d}. \label{eq:Pressure}
\end{align}


\clearpage

\subsection {The stream-vorticity form of the Stokes-Darcy system}\label{sec:The stream-vorticity form of the Stokes-Darcy system}
 Below we would infer the decoupled form of SV form of Stokes-Darcy equations \cite{ref-journal61}. 
\begin{Theorem}
	The streamline field $ \mathbf{\Psi}_s $ and pressure field  $ p_s $ of steady Stokes equations could be decoupled as the form $ \mathcal{L}_{1}(\mathbf{\Psi}_s,\mathbf{f}_s) = 0 $ and $ \mathcal{L}_{2}(p_s,\mathbf{f}_s) = 0 $. The $  \mathcal{L}_{1}(\mathbf{\Psi}_s,\mathbf{f}_s) = 0 $ is a fourth-order equation without $ p_s $ and $ \mathcal{L}_{2}(p_s,\mathbf{f}_s) = 0 $ is an elliptic equation without $ \mathbf{\Psi}_s $.
\end{Theorem}


\begin{proof}
	We note the rotation of streamline field $ \nabla \times \mathbf{\Psi}_s $ and gradient of pressure field $ \nabla p_s $ of Stokes equation are coupled by the kinematic viscosity $ \nu $ in (\ref{eq:Stokes}).
	We could apply differential operators $ \mathbf{1} \cdot \nabla \times $ and $ \nabla \cdot $ to both sides of (\ref{eq:Stokes}). Hence, we could get (\ref{eq:Stokes1122}):  
	\begin{subequations}
		\begin{align}
			\mathcal{L}_{1}(\mathbf{\Psi}_s,\mathbf{f}_s) &= \!\!\!\!\!\! \overbrace{\mathbf{1} \cdot \nabla \times \nabla p_s }^{\text{\textbf{Gradient} has no \textbf{Rotation}.}} \!\!\!\!\!\!  - \,\,\, \mathbf{1} \cdot \nabla \times ( \mathbf{f}_s + \nu \Delta  (\nabla \times \mathbf{\Psi}_s) ) \,\,\, =0,\label{eq:Stokes111}\\
			\mathcal{L}_{2}(p_s,\mathbf{f}_s) &= \,\,\, \underbrace{\nabla  \cdot ( - \nu \Delta (\nabla \times \mathbf{\Psi}_s) )}_{\text{\textbf{Rotation} has no \textbf{Divergence}.}} \,\,\, + \,\,\, \underbrace{\nabla \cdot \nabla p_s }_{\Delta \mathit{p_s}} \,\,\, - \,\,\, \nabla \cdot \mathbf{f}_s \,\,\, = 0. \label{eq:Stokes222}  
		\end{align}
		\label{eq:Stokes1122}
	\end{subequations}
\setlength{\parindent}{2em}
For example, if we apply differential operators $ \mathbf{1} \cdot \nabla \times $  to both sides of (\ref{eq:Stokes}) in 3D, we will notice  (\ref{eq:ps0}):
\begin{align}
	\nabla \times \nabla p_{s} = \begin{bmatrix} 0 & -\frac{\partial}{\partial z} & \frac{\partial}{\partial y} \\ \frac{\partial}{\partial z} & 0 & -\frac{\partial}{\partial x} \\ -\frac{\partial}{\partial y} & \frac{\partial}{\partial x} & 0 \end{bmatrix} \begin{bmatrix} \frac{\partial}{\partial x} \\ \frac{\partial}{\partial y} \\ \frac{\partial}{\partial z} \end{bmatrix} p_{s} = \mathbf{0}. \label{eq:ps0}
\end{align}
\setlength{\parindent}{2em} And finally we could get (\ref{eq:l1}):
\begin{align}
	\mathcal{L}_{1}(\mathbf{\Psi}_s,\mathbf{f}_s) = -\begin{bmatrix} 0 & -\frac{\partial}{\partial z} & \frac{\partial}{\partial y} \\ \frac{\partial}{\partial z} & 0 & -\frac{\partial}{\partial x} \\ -\frac{\partial}{\partial y} & \frac{\partial}{\partial x} & 0 \end{bmatrix}
	\left( 
	\begin{bmatrix}
		f_{s1} \\
		f_{s2} \\
		f_{s3}
	\end{bmatrix}
	+ \nu
	\left(
	\frac{\partial^{2}}{\partial x^{2}} +
    \frac{\partial^{2}}{\partial y^{2}} +
    \frac{\partial^{2}}{\partial z^{2}}
    \right)	 
	\begin{bmatrix} 0 & -\frac{\partial}{\partial z} & \frac{\partial}{\partial y} \\ \frac{\partial}{\partial z} & 0 & -\frac{\partial}{\partial x} \\ -\frac{\partial}{\partial y} & \frac{\partial}{\partial x} & 0 \end{bmatrix}
	\begin{bmatrix}
		\mathbf{\Psi}_{s1} \\
		\mathbf{\Psi}_{s2} \\
		\mathbf{\Psi}_{s3}
	\end{bmatrix}
    \right)
	= \mathbf{0}. \label{eq:l1}
\end{align}

\setlength{\parindent}{0em}
If we apply differential operators $ \nabla \cdot $ to both sides of (\ref{eq:Stokes}) in 3D, we will notice (\ref{eq:sphi0}):
\begin{align}
	- \nu \nabla \cdot \nabla \times \Delta \mathbf{\Psi}_s 
	= 
	- \nu   
	\left(
	\frac{\partial^{2}}{\partial x^{2}} +
	\frac{\partial^{2}}{\partial y^{2}} +
	\frac{\partial^{2}}{\partial z^{2}}
	\right)
	\begin{bmatrix}
		\frac{\partial}{\partial x} \\
		\frac{\partial}{\partial y} \\
		\frac{\partial}{\partial z}
	\end{bmatrix}^T
	\begin{bmatrix} 0 & -\frac{\partial}{\partial z} & \frac{\partial}{\partial y} \\ \frac{\partial}{\partial z} & 0 & -\frac{\partial}{\partial x} \\ -\frac{\partial}{\partial y} & \frac{\partial}{\partial x} & 0 \end{bmatrix}
	\begin{bmatrix}
		\mathbf{\Psi}_{s1} \\
		\mathbf{\Psi}_{s2} \\
		\mathbf{\Psi}_{s3}
	\end{bmatrix}
	= 0. \label{eq:sphi0}
\end{align}
\setlength{\parindent}{2em}
And finally we could get (\ref{eq:l2}):
\begin{align}
	\mathcal{L}_{2}(p_s,\mathbf{f}_s) = \begin{bmatrix}
		\frac{\partial}{\partial x} \\
		\frac{\partial}{\partial y} \\
		\frac{\partial}{\partial z}
	\end{bmatrix}^T
	\begin{bmatrix}
		\frac{\partial}{\partial x} \\
		\frac{\partial}{\partial y} \\
		\frac{\partial}{\partial z}
	\end{bmatrix} p_{s} 
	-
	\begin{bmatrix}
		\frac{\partial}{\partial x} \\
		\frac{\partial}{\partial y} \\
		\frac{\partial}{\partial z}
	\end{bmatrix} ^T
	\begin{bmatrix}
		f_1 \\
		f_2 \\
		f_3
	\end{bmatrix}
	= \left( \frac{\partial^2 }{\partial x^2} + \frac{\partial^2 }{\partial y^2} + \frac{\partial^2 }{\partial z^2}\right)  p_{s}
	-
	\begin{bmatrix}
		\frac{\partial}{\partial x} \\
		\frac{\partial}{\partial y} \\
		\frac{\partial}{\partial z}
	\end{bmatrix}^T 
	\begin{bmatrix}
		f_{s1} \\
		f_{s2} \\
		f_{s3}
	\end{bmatrix}
	= 0. \label{eq:l2}
\end{align}

\setlength{\parindent}{2em} We notice using Cramer's rule to solve for the partial derivative of (\ref{eq:Stokes}) could result in (\ref{eq:Stokes1122}), but the indefinite integral only determines the original function with an additional constant. So (\ref{eq:Stokes}) are sufficient but unnecessary conditions for (\ref{eq:Stokes1122}). A 2D case can be derived from a 3D one.
\end{proof}

\begin{Theorem}
	The streamline field $ \mathbf{\Psi}_d $ and pressure field  $ p_d $ of steady Darcy equations could be decoupled as the form $ \mathcal{L}_{3}(\mathbf{\Psi}_d,\mathbf{f}_d) = 0 $ and $ \mathcal{L}_{4}(p_d,\mathbf{f}_d) = 0 $. The $  \mathcal{L}_{3}(\mathbf{\Psi}_d,\mathbf{f}_d) = 0 $ is a fourth-order equation without $ p_d $ and $ \mathcal{L}_{4}(p_d,\mathbf{f}_d) = 0 $ is a equation without $ \mathbf{\Psi}_d $.
\end{Theorem}


\begin{proof}
	We note that the rotation of streamline field $ \nabla \times \mathbf{\Psi}_d $ and gradient of pressure field $ \nabla p_d $ of the Darcy equation are coupled by the permeability and Reynolds number ratio $ \nu \mathbb{K}^{-1} $ in (\ref{eq:Darcy}).
	We could apply differential operators $ \mathbf{1} \cdot \nabla \times $ and $ \nabla \cdot \mathbb{K} $ to both sides of (\ref{eq:Darcy}). Hence, we could get (\ref{eq:Darcy1122}):
	\begin{subequations}\label{eq:Darcy1122}
		\begin{align}
			\mathcal{L}_{3}(\mathbf{\Psi}_d,\mathbf{f}_d) &= \,\,\,  \mathbf{1} \cdot \nabla \times ( \nu \mathbb{K}^{-1}\nabla \times \mathbf{\Psi}_d - \mathbf{f}_d ) \,\,\, + \!\!\!\!\!\! \overbrace{\mathbf{1} \cdot \nabla \times \nabla p_d }^{\text{\textbf{Gradient} has no \textbf{Rotation}.}} \!\!\!\!\!\! = 0,  \label{eq:Darcy111} \\ 
			\mathcal{L}_{4}(p_{d},\mathbf{f}_d) &= \underbrace{\nu \nabla \cdot \mathbb{K} \mathbb{K}^{-1} \nabla \times \mathbf{\Psi}_{d}}_{\text{\textbf{Rotation} has no \textbf{Divergence}.}} + \, \, \, \nabla \cdot \mathbb{K} ( \nabla p_d - \mathbf{f}_d ) \,\,\, = 0. \label{eq:Darcy222}				
		\end{align}
	\end{subequations}
	\setlength{\parindent}{2em}
	For example, if we apply differential operators $ \mathbf{1} \cdot \nabla \times $  to both sides of (\ref{eq:Darcy}) in 3D, we will notice  (\ref{eq:pd0}):
	\begin{align}
		\nabla \times \nabla p_{d} = 
		\begin{bmatrix} 0 & -\frac{\partial}{\partial z} & \frac{\partial}{\partial y} \\ \frac{\partial}{\partial z} & 0 & -\frac{\partial}{\partial x} \\ -\frac{\partial}{\partial y} & \frac{\partial}{\partial x} & 0 
		\end{bmatrix} 
		\begin{bmatrix} \frac{\partial}{\partial x} \\ \frac{\partial}{\partial y} \\ \frac{\partial}{\partial z} 
		\end{bmatrix}
		p_{d}
		= \mathbf{0}. \label{eq:pd0}
	\end{align}
	\setlength{\parindent}{2em}
	And finally we could get (\ref{eq:l3}):
	\begin{align}
		\mathcal{L}_{3}(\mathbf{\Psi}_d,\mathbf{f}_d) = \nu \begin{bmatrix} 0 & -\frac{\partial}{\partial z} & \frac{\partial}{\partial y} \\ \frac{\partial}{\partial z} & 0 & -\frac{\partial}{\partial x} \\ -\frac{\partial}{\partial y} & \frac{\partial}{\partial x} & 0 \end{bmatrix}
		\left( 
		\begin{bmatrix} 
			k_{11} & k_{12} & k_{13} \\ k_{21} & k_{22} & k_{23} \\ k_{31} & k_{32} & k_{33} 
		\end{bmatrix}^{-1}
		\begin{bmatrix}
			\mathbf{\Psi}_{d1} \\
			\mathbf{\Psi}_{d2} \\
			\mathbf{\Psi}_{d3}
		\end{bmatrix}
		-
		\begin{bmatrix}
			f_{d1} \\
			f_{d2} \\
			f_{d3}
		\end{bmatrix}
		\right)
		= 0. \label{eq:l3}
	\end{align}
	\setlength{\parindent}{2em}
	In a similar way, if we apply differential operators $ \nabla \cdot $ to both sides of (\ref{eq:Darcy}) in 3D, we will notice (\ref{eq:dphi0}):
	
	\begin{align}
		\nu \nabla \cdot \mathbb{K} \mathbb{K}^{-1} \nabla \times \mathbf{\Psi}_{d}
		=
		\nu
		\begin{bmatrix}
			\frac{\partial}{\partial x} \\
			\frac{\partial}{\partial y} \\
			\frac{\partial}{\partial z}
		\end{bmatrix}^T
		\begin{bmatrix} 
			k_{11} & k_{12} & k_{13} \\ k_{21} & k_{22} & k_{23} \\ k_{31} & k_{32} & k_{33} 
		\end{bmatrix}
		\begin{bmatrix} 
			k_{11} & k_{12} & k_{13} \\ k_{21} & k_{22} & k_{23} \\ k_{31} & k_{32} & k_{33} 
		\end{bmatrix}^{-1}
		\begin{bmatrix} 0 & -\frac{\partial}{\partial z} & \frac{\partial}{\partial y} \\ \frac{\partial}{\partial z} & 0 & -\frac{\partial}{\partial x} \\ -\frac{\partial}{\partial y} & \frac{\partial}{\partial x} & 0 \end{bmatrix}
		\begin{bmatrix}
			\mathbf{\Psi}_{d1} \\
			\mathbf{\Psi}_{d2} \\
			\mathbf{\Psi}_{d3}
		\end{bmatrix}
		=0.	
		\label{eq:dphi0}
	\end{align}
	\setlength{\parindent}{2em}
	And finally we could get (\ref{eq:l4}):
	\begin{align}
		\mathcal{L}_{4}(p_d,\mathbf{f}_d) =
		\nu 
		\begin{bmatrix}
			\frac{\partial}{\partial x} \\
			\frac{\partial}{\partial y} \\
			\frac{\partial}{\partial z}
		\end{bmatrix}^T
		\begin{bmatrix} 
			k_{11} & k_{12} & k_{13} \\ k_{21} & k_{22} & k_{23} \\ k_{31} & k_{32} & k_{33} 
		\end{bmatrix}
		\left(  
		\begin{bmatrix} \frac{\partial}{\partial x} \\ \frac{\partial}{\partial y} \\ \frac{\partial}{\partial z} 
		\end{bmatrix}
		p_{d}
		-
		\begin{bmatrix}
			f_{d1} \\
			f_{d2} \\
			f_{d3}
		\end{bmatrix}
		\right) 
		= \mathbf{0}. \label{eq:l4}
	\end{align}
	\setlength{\parindent}{2em}
	We notice using Cramer's rule to solve for the partial derivative of (\ref{eq:Darcy}) could result in (\ref{eq:Darcy1122}), but the indefinite integral only determines the original function with an additional constant. So (\ref{eq:Darcy}) are sufficient but unnecessary conditions for (\ref{eq:Darcy1122}). A 2D case can be derived from a 3D one.
\end{proof}

\clearpage

\section{Algorithm framework}\label{sec:Algorithm framework}

\setlength{\parindent}{2em} First of all, we list the numerical solution forms of DNN and PINNS in Section~\ref{sec:Numerical solutions of DNN}. Secondly, we explain how PINNs solves the coupled Stokes-Darcy equations in Section~\ref{sec:Physical information drives optimization}. Thirdly, we construct improved loss functions by using VP form equations and SV form equations from Section~\ref{sec:The stream-vorticity form of the Stokes-Darcy system} to alleviate gradient competition in Section~\ref{sec:Gradient competition and MF-PINNs}. Lastly, we adopt several training strategies to accelerate converging and improve stability in Section~\ref{sec:Activation functions with high-frequency features} and Section~\ref{sec:Optimizer and learning rate decay}.

\subsection{Numerical solutions of DNN}\label{sec:Numerical solutions of DNN}
\setlength{\parindent}{2em} DNN could be formed by the composition of multiple nonlinear and linear mappings belonging to undetermined  weights and bias. In other words, PINNs could be regarded as a kind of numerical solutions (NN solutions) composed of various functions (\ref{eq:NNolutions}): 

\begin{align}
	\hspace{-1cm} u_{NN}(\mathbf{x},\mathbf{w},\mathbf{b} ) = \sum_{n=1}^{N_{m}} w_{m,n} \mathcal{F}_{m-1,n}\left( \cdots \sum_{n=1}^{N_{2}} w_{2,n} \mathcal{F}_{1,n}( \sum_{n=1}^{N_{1}} w_{1,n} x_{n} + b_{1,n}) + b_{2,n} \cdots \right) + b_{m,n},
	\label{eq:NNolutions}
\end{align}

\setlength{\parindent}{0em} where \( \Theta = \{w_{m,n}, b_{m,n}\}_{m=1,2,\cdots,M} \) are undetermined parameters groups of the PINNs solutions, \( w_{m,n} \) is the \(n^{th}\) weight of the \(m^{th}\) linear layer, \( b_{m,n} \) is the  \(n^{th}\) bia of the \(m^{th}\) linear layer and \( \mathcal{F}_{m,n} \) is the \(n^{th}\) activation function paired with the \(m^{th}\) linear layer. 

\setlength{\parindent}{2em}
This feature endows PINNs with several advantages — extensive fitting capabilities \cite{ref-journal20}, rapid computing \cite{ref-journal63}, various well-proposed function spaces, and superior generalization and transfer learning performance for extrapolation \cite{ref-journal21}.  In the Stokes-Darcy problems, we had better select sufficient smooth activation functions like \textit{tanh, sigmoid, sin} $ \in C^{\infty} $ or $ P^{n}[x](n>1) $, but we could not select \textit{ReLU} $\in C^{0}$ and its family \cite{ref-journal15}.

\subsection{Physical information drives optimization}\label{sec:Physical information drives optimization}

\setlength{\parindent}{2em} A significant work for PINNs is that we need to design the total loss function $ \mathcal{J}( \mathbf{x}, \Theta )  $ based on the  boundary conditions and equations in order to induce PINNs solutions (\ref{eq:NNolutions}) to converge to analytical solutions:

\begin{align}
	\hspace{0cm}
	\mathcal{J}( \mathbf{x}_{f_s}, \Theta )  = \frac{1}{N_{f_s}} \sum_{n=1}^{N_{f_s}}  \left\| -\nu   \nabla \times \Delta \mathbf{\Psi}_{sNN} ( \mathbf{x}_n^{f_s})     + \nabla p_{sNN}( \mathbf{x}_n^{f_s})  - \mathbf{f}_s( \mathbf{x}_n^{f_s})  \right\|_{2}^{2},
	\label{eq:lossfs} 
\end{align}
\begin{align}
	\hspace{0cm}
	\mathcal{J}( \mathbf{x}_{f_d}, \Theta)  =  \frac{1}{N_{f_d}} \sum_{n=1}^{N_{f_d}} \left\| \nu \mathbb{K}^{-1}  \nabla \times \mathbf{\Psi}_{dNN}( \mathbf{x}_n^{f_d})   + \nabla p_{dNN}( \mathbf{x}_n^{f_d})  - \mathbf{f}_d( \mathbf{x}_n^{f_d})  \right\|_{2}^{2},
	\label{eq:lossfd}  
\end{align}
\begin{align}
	\begin{aligned}	
		\mathcal{J}( \mathbf{x}_{\Gamma}, \Theta)  = \frac{1}{N_{\Gamma}} \sum_{n=1}^{N_{u\Gamma}}  
		& \left[  \left\| \nabla \times \mathbf{\Psi}_{sNN}( \mathbf{x}_n^{\Gamma})  \cdot \mathbf{n}_s + \nabla \times \mathbf{\Psi}_{dNN}( \mathbf{x}_n^{\Gamma})  \cdot \mathbf{n}_d \right\|_{2}^{2}   \right.  
		\\
		&  +  \left\| 2\nu \mathbf{n}_s \cdot \mathcal{D}( \nabla \times \mathbf{\Psi}_{sNN}( \mathbf{x}_n^{\Gamma}) )  \cdot \mathbf{n}_s - p_{sNN}( \mathbf{x}_n^{\Gamma})  + p_{dNN}( \mathbf{x}_n^{\Gamma})  - g_{\Gamma_{1}}( \mathbf{x}_n^{\Gamma})  \right\|_{2}^{2} 
		\\
		& \left. +  \left\| 2 \mathbf{n}_s \cdot \mathcal{D}( \nabla \times \mathbf{\Psi}_{sNN}( \mathbf{x}_n^{\Gamma}) )  \cdot \boldsymbol{\tau}
		  + \alpha \mathbb{K}^{-1/2} \nabla \times \mathbf{\Psi}_{sNN}( \mathbf{x}_n^{\Gamma})  \cdot \tau - g_{\Gamma_{2}}( \mathbf{x}_n^{\Gamma})   \right\|_{2}^{2}  \right],  \label{eq:lossgamma} 
	\end{aligned}	
\end{align}	
\begin{align}
	\hspace{0cm} 
	\mathcal{J}( \mathbf{x}_{u_s}, \Theta)  =  \frac{1}{N_{u_s}} \sum_{n=1}^{N_{u_s}} \left\| \nabla \times \mathbf{\Psi}_{sNN}( \mathbf{x}_n^{u_s})  - \mathbf{g}_{\Gamma_s}( \mathbf{x}_n^{u_s})  \right\|_{2}^{2},
	\label{eq:lossus} 
\end{align}
\begin{align}
	\hspace{0cm} 
	\mathcal{J}( \mathbf{x}_{u_d}, \Theta)  =  \frac{1}{N_{u_d}} \sum_{n=1}^{N_{u_d}} \left\|  \nabla \times \mathbf{\Psi}_{dNN}( \mathbf{x}_n^{u_d})   \cdot \mathbf{n}_d  - \mathbf{g}_{\Gamma_d}( \mathbf{x}_n^{u_d}) \right\|_{2}^{2},
	\label{eq:lossud} 
\end{align}	 
\begin{align}
	\mathcal{J}( \mathbf{x}, \Theta)  =  \lambda_{f_s}\mathcal{J}( \mathbf{x}_{f_s}, \Theta)  + \lambda_{f_d}  \mathcal{J}( \mathbf{x}_{f_d}, \Theta)  + 
	\lambda_{\Gamma} \mathcal{J}( \mathbf{x}_{\Gamma}, \Theta)  +
	\lambda_{u_s} \mathcal{J}( \mathbf{x}_{u_s}, \Theta)  +
	\lambda_{u_d} \mathcal{J}( \mathbf{x}_{u_d}, \Theta)   
	\label{eq:lossall},
\end{align}

	\setlength{\parindent}{0em} where (\ref{eq:lossfs}) stands for the loss of the Stokes equation, (\ref{eq:lossfd}) stands for the loss of the Darcy equation, (\ref{eq:lossgamma}) stands for the loss on the coupled interface,  (\ref{eq:lossus}) and (\ref{eq:lossud}) respectively stand for the loss on the boundary conditions of the Stokes equation and the Darcy equation, and different coefficients $ \lambda $ stand for the difference in importance of the five loss functions in different PINNs. 
	 
	\setlength{\parindent}{2em} Finally, we add the five loss functions with different coefficients $ \lambda $ to get the total loss function (\ref{eq:lossall}). Therefore, based on \cite{ref-journal20} and \cite{ref-journal8}, we could conclude:

	\begin{Theorem}
	For any \( \varepsilon > 0 \), there exists a wide and deep enough neural network \( \mathcal{F}_{NN}(\mathbf{x}, \Theta) = (\mathbf{\Psi}_{NN}, \nabla p_{NN}) \) with sufficiently large degrees of freedom \( DOF_{\Theta} \) to make:
	\[
	\underset{\Theta}{\min} \mathcal{J}(\mathbf{x}, \Theta) = \mathcal{J}(\mathbf{x}, \hat{\Theta}) < \varepsilon.
	\]
	Moreover, the following error bounds hold:
	\[
	\left\| \nabla \times \mathbf{\Psi}_{NN}(\mathbf{x}, \hat{\Theta}) - \mathbf{u}(\mathbf{x}) \right\|_{2} < C_{1} \varepsilon^{n_{1}},
	\]
	\[
	\left\| \nabla p_{NN}(\mathbf{x}, \hat{\Theta}) - \nabla p(\mathbf{x}) \right\|_{2} < C_{2} \varepsilon^{n_{2}},
	\]
	where \( C_{1} \) and \( C_{2} \) are known constants, and \( n_{1} \) and \( n_{2} \) are positive integers.   
	\end{Theorem}
	
	\setlength{\parindent}{2em}  Remarking on Section~\ref{sec:The velocity-pressure form of the Stokes-Darcy system}, we consider that the pressure field of the PINNs numerical solutions, $ p_{sNN} $ and $ p_{dNN}$, differ from the analytical solutions, $ p_{s}$ and $ p_{d}$ by a constant $ C_{p} $. Thus, we set \texttt{bias = 'False'} for the last linear layer, and we developed (\ref{eq:renewP1}) and (\ref{eq:renewP2}) to correct $ p_{NNs} $ and $ p_{NNd}$ based on (\ref{eq:Pressure}), after updating the weights and bias per epoch.
	  
\begin{align}		
    \left( \sum_{m=1}^{N_{s}} \left(p_{sNN}(\mathbf{x}_{m},\hat{\Theta}) - p_{s}(\mathbf{x}_{m})\right)  +  \sum_{n=1}^{N_{d}} \left(p_{dNN}(\mathbf{x}_{n},\hat{\Theta}) - p_{d}(\mathbf{x}_{n})\right)  \right) \Big/ (N_{s}+N_{d}) \approx C_{pNN} - C_{p}, \, \mathbf{x} \in \Omega_{s} \cup \Omega_{d}. \label{eq:renewP1}
\end{align}
\begin{align}
	\widetilde{p}_{NN}(\mathbf{x},\hat{\Theta}) = p_{NN}(\mathbf{x},\hat{\Theta}) - C_{pNN} + C_{p}, \label{eq:renewP2}
\end{align}

\begin{figure}[H]
	\centering
	\includegraphics[width=8.8cm]{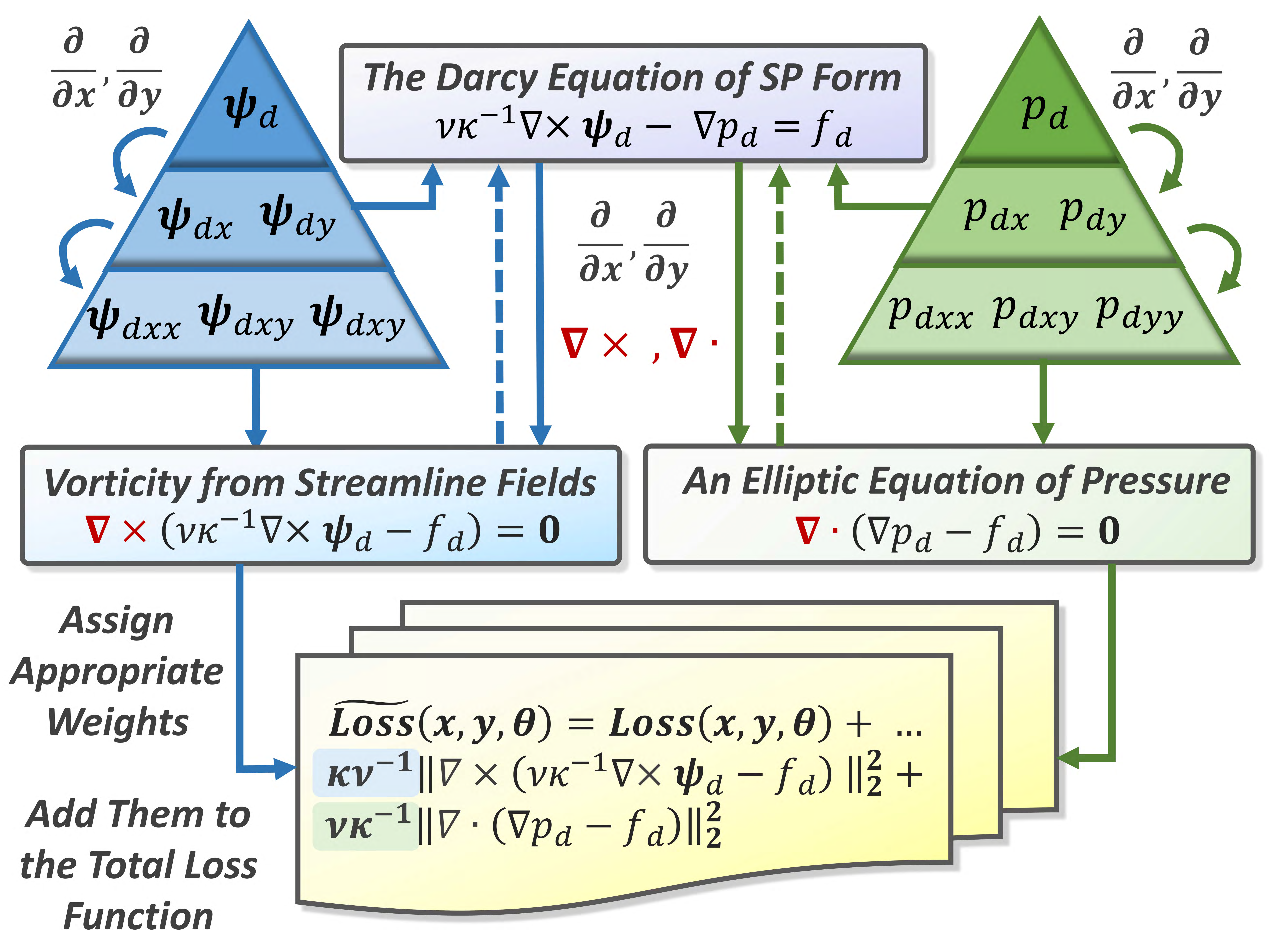}
	\\
	\caption{This cartoon shows how we combine the VP form and the SV form and design the total new loss function with appropriate weights in the Darcy domain. \label{gradcomp.pdf}}
\end{figure}

\subsubsection{Gradient competition and MF-PINNs}\label{sec:Gradient competition and MF-PINNs} 
Gradient competition may be a potential risk to the training of PINNs. If we directly construct the loss function in Section~\ref{sec:Physical information drives optimization} for the Stokes-Darcy system, it may lead to the failure of PINNs training, possibily. A major reason why PINNs might fail is that extreme high or low constants ($ \kappa$, $\nu $, etc.) in the multi-objective loss functions (\ref{eq:lossall}) may create competition in the gradients. What's worse, the PINNs with such ill-conditioned loss functions may lead to the fact that some physical quantities have sufficient accuracy, but the optimization direction of other physical quantities is opposite to the optimal point, as is shown in Fig.~\ref{gradcomp.pdf}.

\setlength{\parindent}{2em} The following are the potential risks for PINNs in Stokes-Darcy system:

\begin{enumerate}
	\item $ \nu \gg 1 $ or $ \nu \ll 1 $ may cause the gradient competition between $ \mathbf{u}_{s} $ and $ p_{s} $. 
	\item $ \nu / \kappa \gg 1 $ or $ \nu / \kappa  \ll 1  $ may cause the gradient competition between $ \mathbf{u}_{d} $ and $ p_{d} $.
	\item $ \nu  \gg \nu / \kappa  $ or $ \nu \ll \nu / \kappa  $ may cause the gradient competition between the Stokes system and the Darcy system.
	\item Extreme gradient may cause the gradient competition in total loss among the boundary, the interface, and the inner points.
	\item $ \nu \gg 1 $ may cause the gradient exploration for the loss of the inner points during backward.  
\end{enumerate} 

\setlength{\parindent}{2em} For example, if we chose  $ \nu = 1 $ and $ \mathbb{K} = 10^{-4}\mathbb{I} $, the ratio of the updating gradients of $ \nabla \times \mathbf{\Psi}_{dNN} $ and $ \nabla p_{dNN} $ would be approximately $ 10^{4}:1 $. These choices
may result in $ \nabla p_{dNN} $ being very insignificant compared to $ \nabla \times \mathbf{\Psi}_{dNN} $, and the loss function (\ref{eq:lossfd}) may improperly become an ill-conditioned form (\ref{eq:lossfdfdfd}):

\begin{align}
	\begin{aligned}
		\hspace{0cm}
		\mathcal{J}( \mathbf{x}_{f_d}, \Theta)  &=  \frac{1}{N_{f_d}} \sum_{n=1}^{N_{f_d}} \left\| \mathbf{10^{4}}  \nabla \times \mathbf{\Psi}_{dNN}( \mathbf{x}_n^{f_d})   + \nabla p_{dNN}( \mathbf{x}_n^{f_d})  - \mathbf{f}_d( \mathbf{x}_n^{f_d})  \right\|_{2}^{2}
		\\
		& \approx \frac{1}{N_{f_d}} \sum_{n=1}^{N_{f_d}} \left\| \mathbf{10^{4}}  \nabla \times \mathbf{\Psi}_{dNN}( \mathbf{x}_n^{f_d})   - \mathbf{f}_d( \mathbf{x}_n^{f_d})  \right\|_{2}^{2}.
		\label{eq:lossfdfdfd}  
	\end{aligned}
\end{align}

\setlength{\parindent}{2em} As a terrible result, $ \nabla p_{dNN} $ neither is trained nor converges to $ \nabla p_{d} $ by mistake in Fig.~\ref{gradcomp.pdf}. We could infer that the error function of $ \nabla \times \mathbf{\Psi}_{dNN} $ approaches its minimum point, but the error function of $ \nabla p_{dNN} $ may be very far from that one in value. Finally, we could validate this inference in the following numerical experiments in Section~\ref{sec:Numerical examples }.


\setlength{\parindent}{2em} To deal with these problems, we innovate the \textbf{Mixed-Form PINNs} (\textbf{MF-PINNs}). It is a kind of enhanced PINNs to deal with ill-conditioned loss functions under extreme physical constants. As shown in Fig.~\ref{gradcomp.pdf}, our MF-PINNs redesign the coefficients of SV form and combine it with VP form (\ref{eq:lossall}) to get the new total loss functions (\ref{eq:lossalldc}): 

\begin{align}
	\begin{aligned}
		\hspace{0cm}
		& \widetilde{\mathcal{J}}( \mathbf{x}_{f_s}, \Theta ) = 		\frac{\mathcal{J}( \mathbf{x}_{f_s}, \Theta )}{max(\nu,1)}  + \frac{1}{\nu } \left\| \mathcal{L}_{1}(\mathbf{\Psi}_{sNN},\mathbf{f}_s) \right\|_{2}^{2} + \nu \left\| \mathcal{L}_{2}(p_{sNN},\mathbf{f}_s) \right\|_{2}^{2}
		\\
		& = \frac{\mathcal{J}( \mathbf{x}_{f_s}, \Theta )}{max(\nu,1)}  + \frac{1}{\nu N_{f_s}} \sum_{n=1}^{N_{f_s}}  \left\|  \nu \nabla \times  \nabla \times \Delta \mathbf{\Psi}_{sNN}( \mathbf{x}_{n}^{f_s}) + \nabla \times  \mathbf{f}_s( \mathbf{x}_{n}^{f_s}) \right\|_{2}^{2} + \frac{\nu}{N_{f_s}} \sum_{n=1}^{N_{f_s}}  \left\| {\Delta \mathit{p_{sNN}( \mathbf{x}_{n}^{f_s})}} - \nabla \cdot \mathbf{f}_s( \mathbf{x}_{n}^{f_s}) \right\|_{2}^{2} ,  
		\label{eq:lossfsdc}  
	\end{aligned}
\end{align}
\begin{align}
	\begin{aligned}
		\hspace{0cm}
		& \widetilde{\mathcal{J}}( \mathbf{x}_{f_d}, \Theta ) = 
		\frac{\mathcal{J}( \mathbf{x}_{f_d}, \Theta )}{max(\nu,1)}  + \frac{\kappa}{\nu} \left\| \mathcal{L}_{3}(\mathbf{\Psi}_{dNN},\mathbf{f}_d) \right\|_{2}^{2} + \frac{\nu}{\kappa} \left\| \mathcal{L}_{4}(p_{dNN},\mathbf{f}_d) \right\|_{2}^{2}
		\\ 
		& = \frac{\mathcal{J}( \mathbf{x}_{f_d}, \Theta )}{max(\nu,1)} + \frac{\kappa}{\nu N_{f_d}} \sum_{n=1}^{N_{f_d}} \left\| \frac{\nu}{\kappa} \nabla \times \nabla \times \mathbf{\Psi}_{dNN}( \mathbf{x}_{n}^{f_d}) - \nabla \times \mathbf{f}_d( \mathbf{x}_{n}^{f_d}) \right\|_{2}^{2} +  \frac{\nu}{\kappa N_{f_d}} \sum_{n=1}^{N_{f_d}} \left\| \Delta p_{dNN}( \mathbf{x}_{n}^{f_d}) - \nabla \cdot \mathbf{f}_d( \mathbf{x}_{n}^{f_d})  \right\|_{2}^{2}, 
		\label{eq:lossfddc}  
	\end{aligned}
\end{align}
\begin{align}
	\widetilde{\mathcal{J}}( \mathbf{x}, \Theta)  =  \lambda_{f_s} \widetilde{\mathcal{J}}( \mathbf{x}_{f_s}, \Theta)  + \lambda_{f_d}  \widetilde{\mathcal{J}}( \mathbf{x}_{f_d}, \Theta)  + \lambda_{\Gamma} \mathcal{J}( \mathbf{x}_{\Gamma}, \Theta)  +
	\lambda_{u_s} \mathcal{J}( \mathbf{x}_{u_s}, \Theta)  +
	\lambda_{u_d} \mathcal{J}( \mathbf{x}_{u_d}, \Theta)   
	\label{eq:lossalldc},
\end{align}

\setlength{\parindent}{0em} where the coefficients $ 1 / \nu $ and $ \nu $ assigned to $ \mathcal{L}_{1}(\mathbf{\Psi}_{sNN},\mathbf{f}_s) $ and $\mathcal{L}_{2}(p_{sNN},\mathbf{f}_s) $ alleviate the gradient competition between $ \mathbf{\Psi}_{sNN} $ and $ p_{sNN} $ for (\ref{eq:lossfsdc}), the coefficients $ \kappa / \nu $ and $ \nu / \kappa $ assigned to $ \mathcal{L}_{3}(\mathbf{\Psi}_{dNN}, \mathbf{f}_d) $ and $\mathcal{L}_{4}(p_{dNN}, \mathbf{f}_{d}) $ alleviate the gradient competition between $\mathbf{\Psi}_{dNN} $ and $ p_{dNN} $ for (\ref{eq:lossfddc}), and additionally, we multiply  (\ref{eq:lossall}) by the coefficient $ 1 / max(\nu,1) $ to prevent the gradient explosion during backward because $ \nu $ may be much greater than $ 1 $. 

\setlength{\parindent}{2em} As an ideal result, these skills of our MF-PINNs could not only alleviate the gradient competition among different physical quantities, but also accelerate the convergence of PINNs numerical solutions during per epoch. Therefore, our MF-PINNs makes it possible to precisely solve each physical field under extreme physical constants. Next, We would compare the differences between our MF-PINNs and several other PINNs models in Section~\ref{sec:Algorithm design}, and verify the effectiveness of our MF-PINNs under extreme physical constants in Section~\ref{sec:Numerical test}.

\setlength{\parindent}{2em} In addition, we could only apply the differential operator to the unmodified loss functions by the automatic differentiation technique. So we have no need to deduce the detailed form \ref{eq:lossfddc} of the decoupled equations in programming. 

\begin{figure}[H]
	\centering 
	\includegraphics[width=16cm]{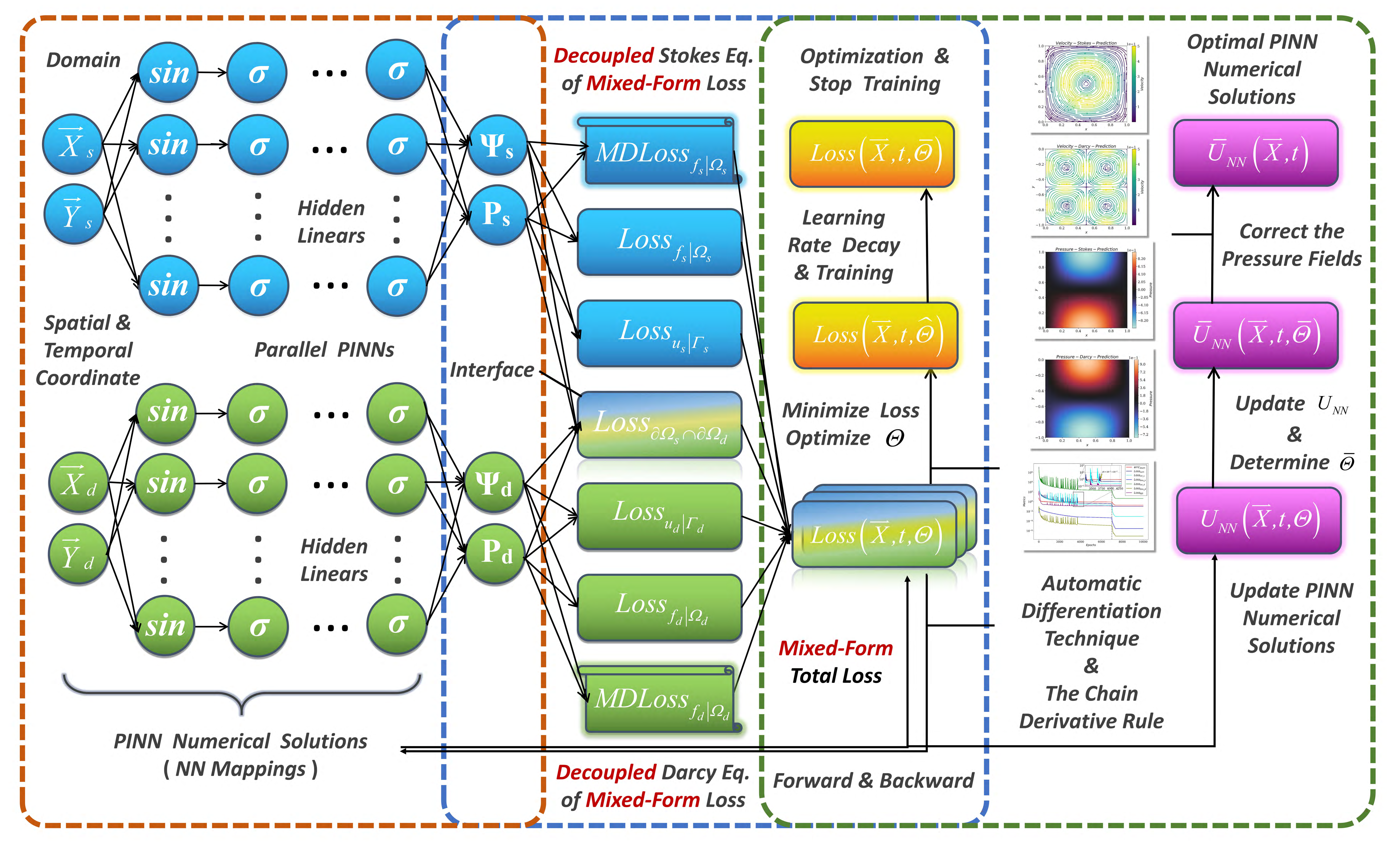}
	\caption{This picture illustrates the framework of our MF-PINNs for solving coupled Stokes-Darcy Problems.\label{NNframe.pdf}}
\end{figure}

\subsection{Activation functions with high-frequency features}\label{sec:Activation functions with high-frequency features}
PINNs with Fourier features is a way to solve multi-frequency PDE problems. Thus, we improve the activation functions of the first nonlinear layer of NNs for embedding high-frequency features. In detail, we replace $  \mathcal{F}_{1,n}\left( \mathbf{w}\mathbf{x}+b\right)  = \mathit{tanh}\left( \theta \left( \mathbf{w}\mathbf{x}+b \right) \right)  $ as $ \mathcal{\widetilde{F}}_{1,n}\left( \mathbf{w}\mathbf{x}+b\right)   = \mathit{tanh} \left(  \mathit{sin}\left( 2 \pi \theta \left( \mathbf{w}\mathbf{x}+b \right) / \mathit{T}  \right) \right)  $ in Fig.~\ref{NNframe.pdf}, where the physical periodicity $ T_{i} $ can be obtained from the non-homogeneous term $ \mathbf{f} $ or the boundary conditions $ \mathbf{g}_{\Gamma} $. Then we choose the common multiple of the period $ T $ as the period of the first activation function.

\setlength{\parindent}{2em} 
In Section~\ref{sec:Numerical test}, we would verify that our improvement not only enhances the fitting ability and extension capability of PINNs but also keeps them easy to code and train without consuming additional computing resources.

\subsection{Optimizer and learning rate decay}\label{sec:Optimizer and learning rate decay} In this section, we introduced the combination of optimizers we used, Adam \& L-BFGS, and the learning rate decay strategy we design. In the following Section~\ref{sec:Numerical examples }, we would verify the effect of the learning rate decay strategy we designed.

\setlength{\parindent}{2em} 
The Adam optimizer is widely used in deep learning tasks, especially for initial processing of large datasets and complex models. However, results from Adam may not always be sufficiently precise. Hence, we would use Adam optimizer in the early stage of training PINNs (from $ 1^{st} $ to $ 7000^{th} $ epochs), and we also design the adaptive interval learning rate decay strategy for Adam. We adopt \texttt{ReduceLROnPlateau} and set \texttt{Initial\_LR\_Adam = 10\textsuperscript{-3}}, \texttt{threshold = 10\textsuperscript{-4}}, \texttt{factor = 10\textsuperscript{-1}}, \texttt{patience = 10\textsuperscript{2}}, \texttt{cooldown = 10\textsuperscript{2}}, while the rest are default. 


\setlength{\parindent}{2em}
L-BFGS is a highly efficient quasi-Newton optimization algorithm, and it does well in handling large-scale datasets and high-dimensional parameter spaces. L-BFGS achieves a higher order of convergence, but it requires that the parameter groups be sufficiently close to the optimal points. Therefore, we would use the L-BFGS optimizer in the later stage of training PINNs (from $ 7001^{st} $ to $ 10000^{th} $ epochs), and we also design the adaptive interval learning rate decay strategy  for L-BFGS. We adopt \texttt{ReduceLROnPlateau} and set \texttt{Initial\_LR\_L-BFGS = 10\textsuperscript{-1}}, \texttt{threshold = 10\textsuperscript{-3}}, \texttt{factor = 10\textsuperscript{-1}}, \texttt{patience = 10}, \texttt{cooldown = 10\textsuperscript{2}}, while the rest are default. 

\subsection{Algorithm design}\label{sec:Algorithm design}

\setlength{\parindent}{2em} In this section, we list several optimization algorithms. Their performance would be compared in the following numerical examples of Section~\ref{sec:Numerical test}:
\begin{itemize}
	\item
	\textbf{AS-DNN} : We use the \textbf{Deep Neural Networks} to fit the \textbf{Analytical Solutions} (\ref{eq:ana_solu}) directly. Therefore, the AS-DNN could display the maximal fitting capability of PINNs in theory. Next, we will use AS-DNN to compare with several PINNs with unsupervised learning in the fixed size of NNs and common input data. 
	\item	\textbf{PINNs} : We design the loss functions $ \mathcal{J}(\mathbf{x}, \Theta) $ directly, without adding any weight. That is $ \lambda_{f_s} =\lambda_{f_d} = \lambda_{u_d} =\lambda_{u_d} = \lambda_{\Gamma} = 1  $ for (\ref{eq:lossall}).
	\item	\textbf{AT-PINNs} : We take examples from the \textbf{Sharp-PINNs} \cite{ref-journal20} to alternately train parallel PINNs. In detail, we alternately train different loss functions paired with different region-decomposed NNs, respectively. That is, $ \lambda_{f_s} = \lambda_{u_s} =\lambda_{\Gamma} = 1, \lambda_{f_d} = \lambda_{u_d} = 0 $ for updating argument $ \Theta_s \subsetneqq \Theta  $ of the Stokes system and $ \lambda_{f_d} = \lambda_{u_d} = \lambda_{\Gamma} = 1, \lambda_{f_s} = \lambda_{u_s} = 0 $ for updating argument $ \Theta_d \subsetneqq \Theta $ of the Darcy system in (\ref{eq:lossall}). What's more, we change the loss functions (region-decomposed NNs) every $ 100 $ epochs during the Adam training stage. But there is no change during the L-BFGS training stage. 
	\item	\textbf{MW-PINNs} \cite{ref-journal5} : We design the $ \mathcal{J}(\mathbf{x}, \Theta) $ based on their different importance, which could be quantitatively described as appropriate ratios. An appropriate group of \textbf{Multiple Weights} is $ \lambda_{f_s} = \lambda_{u_s} = 1 / v, \lambda_{f_d} = \lambda_{u_d} = \kappa / v, \lambda_{\Gamma} = 1 $ for (\ref{eq:lossall}).
	\item	\textbf{MF-PINNs (Ours)} : We have derived the VP form and SV form of both Stokes and Darcy equations by using the automatic differential operators. Next, we apply multiple weights for the new total loss sfunction $ \widetilde{\mathcal{J}}(\mathbf{x}, \Theta) $ with \textbf{Mixed Forms}. The multiple weights are $ \lambda_{u_s} = \lambda_{u_d} = 10^{2}, \lambda_{f_d} = \kappa, \lambda_{f_s} = \lambda_{\Gamma} = 1 $ for (\ref{eq:lossalldc}).
\end{itemize}
\clearpage
\section{Numerical test}\label{sec:Numerical test}

\subsection{Model parameter}\label{sec:Model parameter} In this section, we list the parameters and size of the NNs for our experiments. Here are some notifications:

\begin{enumerate}
	\item We divide the $ 127 \times 127 $ square grids with the same size in the Stokes and Darcy domains respectively, and then we input the coordinates of the cell grid nodes as labeled data into the PINNs. We notice that there are $ 128 $ points shared on the interface shared  by the Stokes domain and the Darcy domain.
	\item We use parallel PINNs to solve the coupled Stokes-Darcy equations. Both parallel PINNs have $ 4 $ hidden layers $ \times \, 70 $ neurons. All kinds of PINNs in our article use the same neural networks with the same size. And the strategies for the activation functions are shown in Section~\ref{sec:Activation functions with high-frequency features}, unless we have special statements in the following ablation experiments.
	\item In Table~\ref{tabpara}, we apply the optimizer paired with the adaptive learning rate strategies in Section~\ref{sec:Optimizer and learning rate decay} to the specified number of epochs.
	\item All the experiments in this article are under the same configuration -- \texttt{CPU:16 vCPU AMD EPYC 9K84 96-Core Processor}, \texttt{GPU: H20-NVLink(96GB)}.
\end{enumerate}

\begin{table}[h]
	\centering
	\caption{This table lists several significant parameters of the PINNs.  \label{tabpara}}
	
	\resizebox{\linewidth}{!}{ 
		\begin{tabularx}{1.08\textwidth}{cccc} 
			\toprule
			\textbf{Data Size} & \textbf{Neurons}  & \textbf{Training Optimizer} & \textbf{Activation Function} \\
			\midrule
			$ N_{fs} = N_{fd} = 15876 $ & Input : $ 2 \times [2] \times [70] $  &  Adam for $ 7000 $ epochs & \\		
			$ N_{\Gamma_{s}} = N_{\Gamma_{d}} = 380 $ & Hidden : $ 2 \times [70] \times [70] \times 4 $ & Initial LR : $ 10^{-3} $ &  $ \mathit{tanh(x)} $ or   \\		  		  
			$ N_{\Gamma} = 126 $ & Output : $ 2 \times [70] \times [2] \, $ (No Bias) & L-BFGS  for $ 3000 $ epochs  & $ \mathit{tanh} \circ \mathit{sin}(\frac{2 \pi \mathit{x}}{\mathit{T}})  $  \\
			$ N = 32640 $ &  Total Parameters : $ 40460 $  & Initial LR : $ 10^{-1} $ &  \\
			\bottomrule
		\end{tabularx}
	}
\end{table}

\subsection{Metrics for error}\label{Metrics for error}

\setlength{\parindent}{2em} 
We use the relative Euclidean norm ($ err\mathcal{L}_{2} $) to assess the accuracy of the PINNs. Inspired by the finite volume method, we could replace the continuous equations $ u(x) $ in the tiny neighborhood as the function value at the paired point $ u(x_{i}) $ to estimate the $ err\mathcal{L}_{2} $:
\begin{align}
	err\mathcal{L}_{2}\left( u \right) = \frac{ \left\|u_{NN} - u\right\|_{2}}{\left\| u \right\|_{2} \,} \approx \frac{\sqrt{\sum_{i=1}^{N} \left| u_{NN}(x_{i}) - u(x_{i})\right|^2}}{\sqrt{\sum_{i=1}^{N} \left|u(x_{i})\right|^2}},
	\label{eq:L2}
\end{align}

\setlength{\parindent}{0em} 
where the $ N $ represents the number of points of a specific category in the PINNs training process, the $ u $ represents the analytical solutions, the $ u_{NN} $ represents PINNs numerical solutions of specific physical quantities, and the $ x \in \Omega $ represents a specific point. 

\subsection{Numerical examples}\label{sec:Numerical examples }
We focus on the coupled Stokes-Darcy problem with the discontinuous BJS interface, so we use analytical solutions (\ref{eq:ana_solu}) of \cite{ref-journal5} for different kinds of PINNs. Among them, the non-homogeneous term $ \mathbf{f} $ and the boundary condition $ \mathbf{g}_{\Gamma} $ are naturally determined by the analytical solutions (\ref{eq:ana_solu}).

\begin{align}
	\begin{aligned}
		\mathbf{u}_s &=
		\begin{pmatrix}
			u_s\\
			v_s	
		\end{pmatrix}
		= 
		\begin{pmatrix}
			- \sin^{2}(\pi x) \sin(\pi y) \cos(\pi y)\\
			\sin(\pi x) \cos(\pi x) \sin^{2}(\pi y)	
		\end{pmatrix}
		,
		\mathbf{u}_d  =
		\begin{pmatrix}
			u_d\\
			v_d	
		\end{pmatrix}
		= 
		\begin{pmatrix}
			\frac{1}{2} \sin(2\pi x) \cos(2\pi y)\\
			-\frac{1}{2} \cos(2\pi x) \sin(2\pi y)
		\end{pmatrix}
		,\\ 	    
		p_s  &= p_d  =  \sin(\pi x) \cos(\pi y),
		\label{eq:ana_solu}
	\end{aligned}
\end{align}

\setlength{\parindent}{2em} We set that the Stokes domain is $ \Omega_{s} = [0, 1] \times [0, 1] $ , while the Darcy domain is $ \Omega_{d} = [0, 1] \times [-1, 0] $. And we set \( \alpha = 1 \) and $ C_{p} = 0 $.
So the period of the velocity fields and pressure fields are $ T_{\mathbf{u}} = 1, T_{p} = 2 $, respectively, as well as the interface is \( \Gamma = [0, 1] \times \{0\} \).  Besides, we explore how various combinations of $ \mathbb{K} (\mathbb{K} = \kappa \mathbb{I}) $ and $ \nu $ affect the ability of different kinds of PINNs.

\subsection{Analysis of numerical results}\label{sec:Analysis of numerical results}

\subsubsection{Alleviate the gradient competition}\label{sec:Alleviate the gradient competition} 
Next we will analyse the numerial results.

\setlength{\parindent}{2em}
Firstly, we analyze the result of several algorithms under $ \mathbb{K} = 10^{-4}\mathbb{I} $ and $ \nu = 1 $, because $ \nu / \kappa  \gg 1 $ leads to gradient competition among \( u_d \), \( v_d \), \( p_d \), and \( \nu / \kappa  \gg \nu \) leads to gradient competition between Stokes and Darcy equations in Section~\ref{sec:Algorithm design}. We could draw the following conclusions:

\vspace{-0.5cm}

\begin{figure}[H]
	\centering
	\subfloat[\centering]{\includegraphics[width=7.5cm]{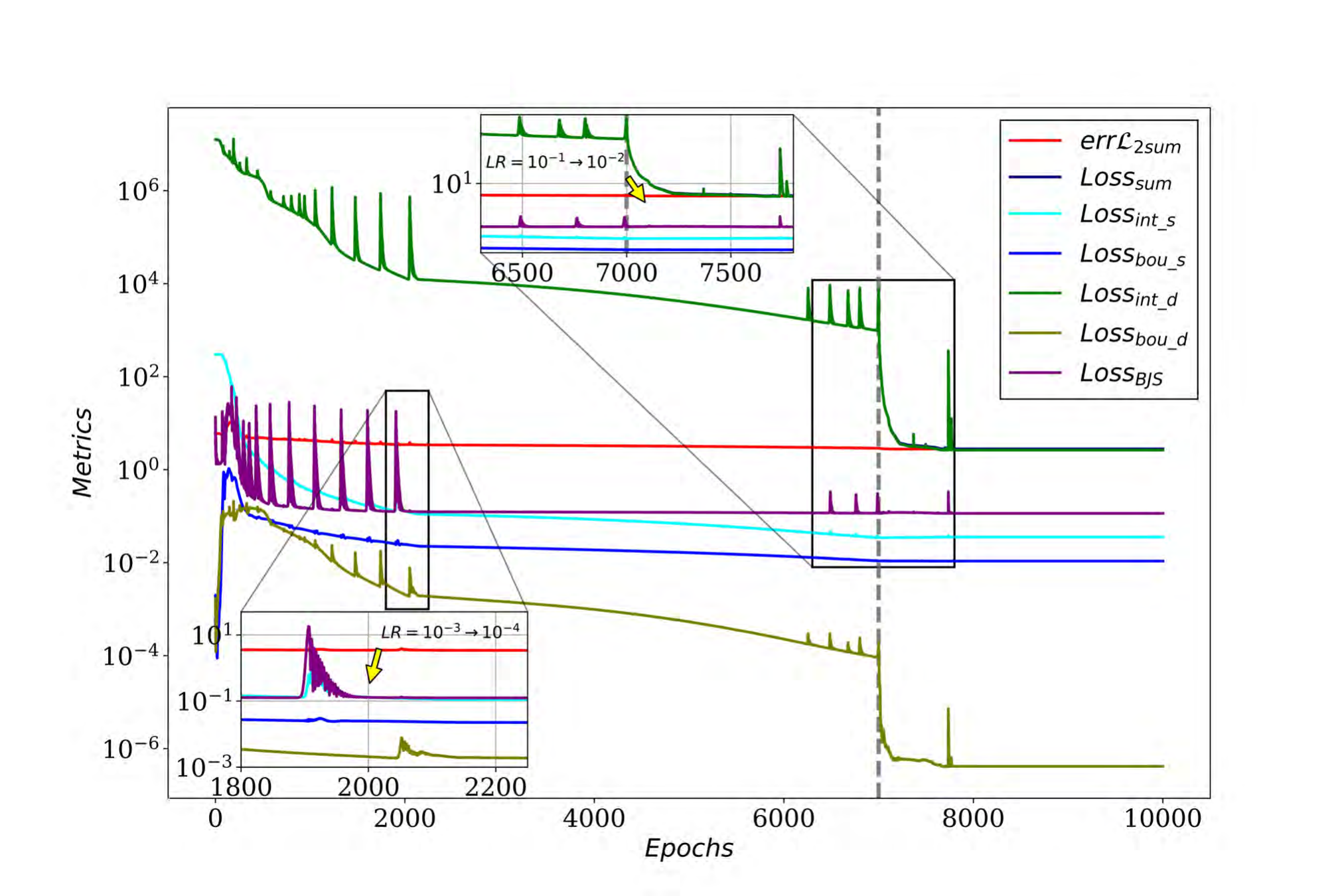}}
	\hfill
	\subfloat[\centering]{\includegraphics[width=7.5cm]{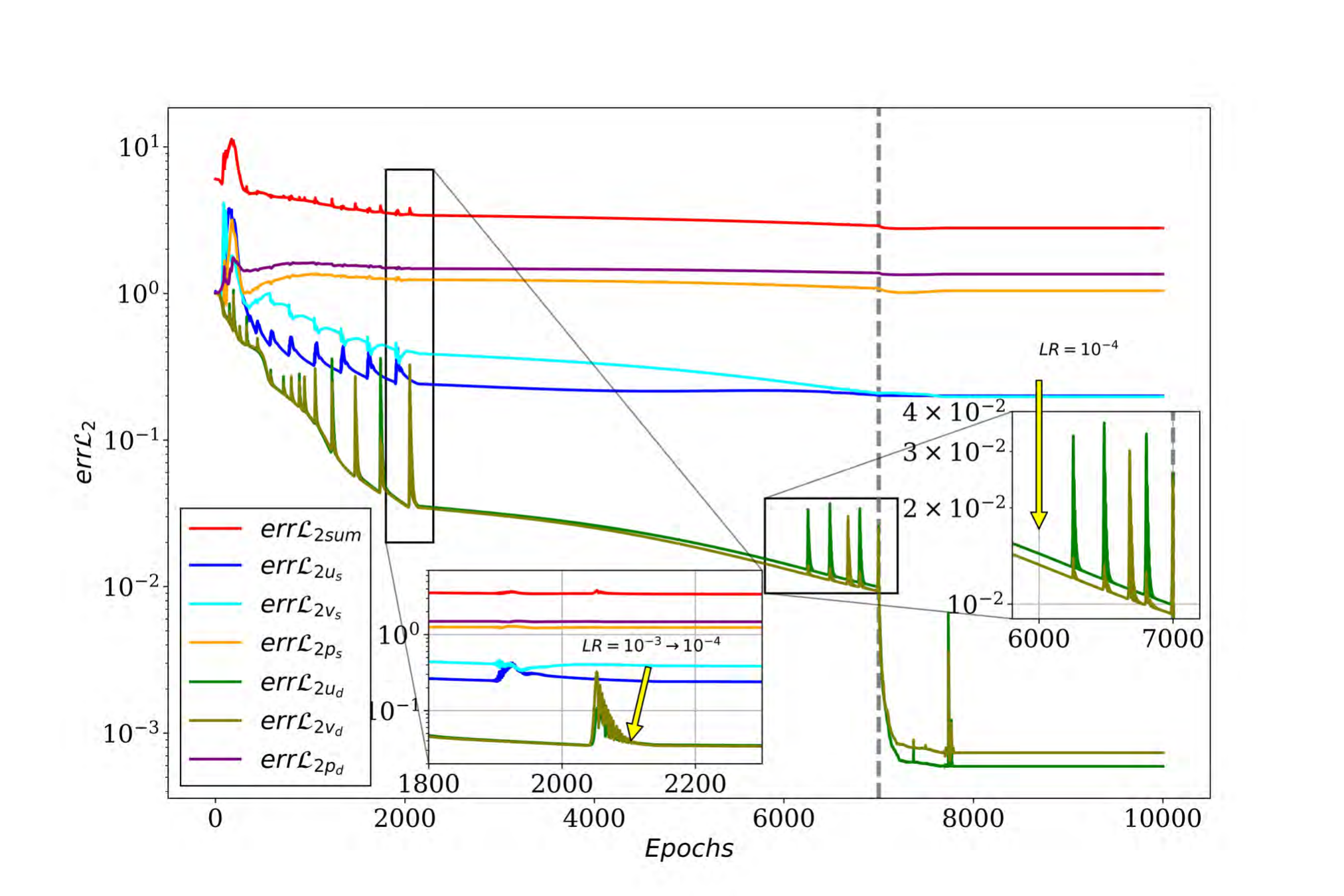}}\
	\\
	\subfloat[\centering]{\includegraphics[width=7.5cm]{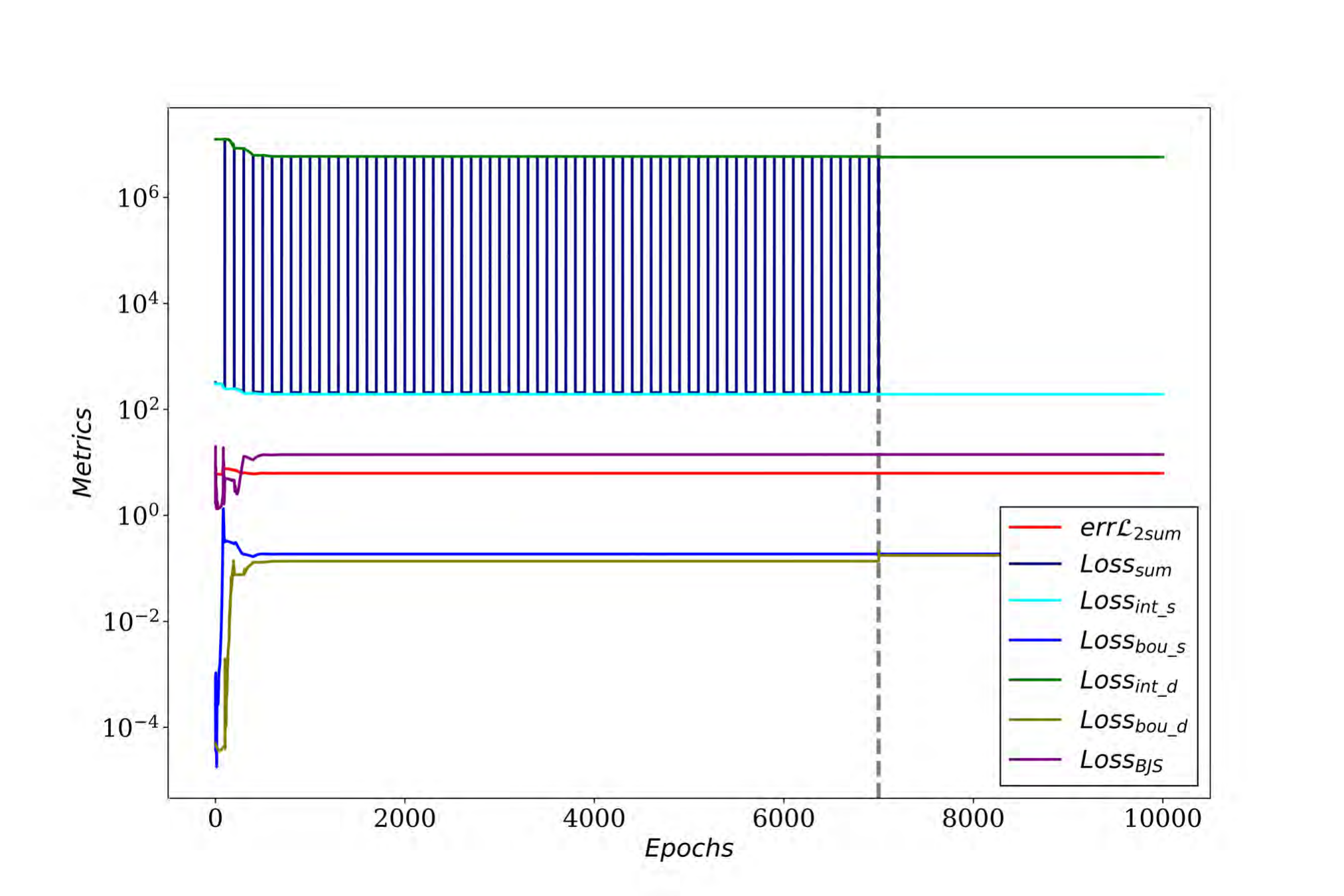}}
	\hfill
	\subfloat[\centering]{\includegraphics[width=7.5cm]{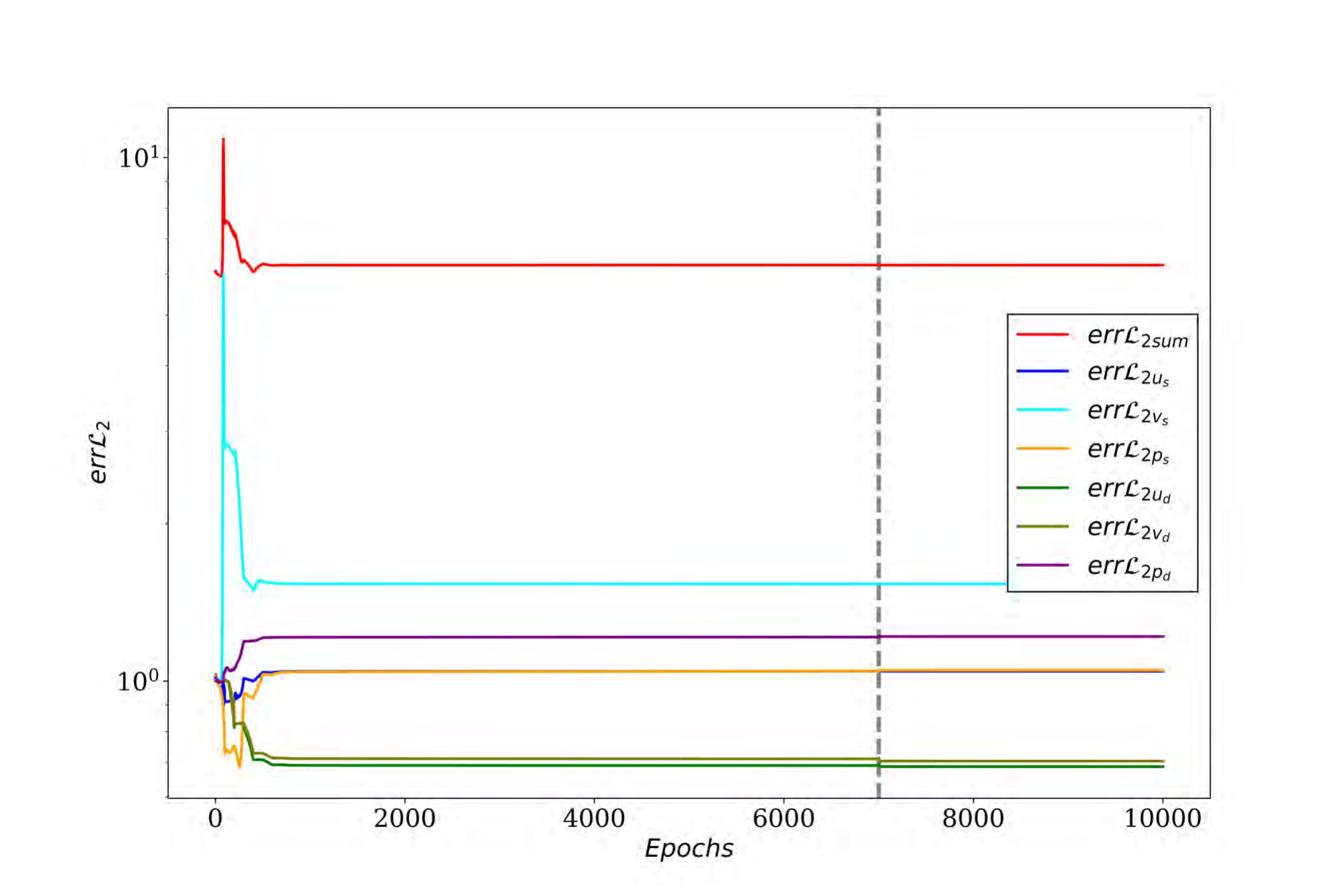}}
	\caption{These images (\textbf{a})-(\textbf{d}) compare the abilities of PINNs and AT-PINNs. The dashed gray line means that we end up using the Adam optimizer and then use the L-BFGS optimizer. \textbf{By row}: PINNs, AT-PINNs;  \textbf{By column}: $ \mathit{Loss} $ , $ err\mathcal{L}_{2} $. \label{Loss_and_L2_12}}
\end{figure} 

\vspace{-0.5cm}

\begin{itemize}
\item In Fig.~\ref{Loss_and_L2_12}, we could observe that the gradient update of \( p_d \) is too tiny, so this fact causes the baseline PINNs to ignore the training of \( p_d \), while the training of \( u_d \) and \( v_d \) is perfect. In other words, the total loss $\mathcal{J}(\mathbf{x},\Theta) $ converges to zero and \( u_{dNN} \) and \( v_{dNN} \) converge to \( u_d \) and \( v_d \), respectively. Results are $ err\mathcal{L}_2(u_d) = 0.05929 \% $ and $ err\mathcal{L}_2(v_d) = 0.07349\% $. But \( p_{sNN} \) and \( p_{dNN} \) do not converge to \( p_s \) and \( p_d \) at all. Results are $
err\mathcal{L}_2(p_s) = 104.1\% $ and $ err\mathcal{L}_2(p_d) = 135.4 \% $. These results verify our inference of Section~\ref{sec:Gradient competition and MF-PINNs}.

\item The AT-PINNs aims to decouple the Stokes and Darcy equations. During the early training stage, the regional decomposed PINNs are trained alternately by using different total loss $\mathcal{J}(\mathbf{x},\Theta) $ in Section~\ref{sec:Algorithm design}. Compared with the baseline PINNs, the AT-PINNs accelerates training by reducing the number of parameters that updates at each epoch, and it saves much time. However, in Fig.~\ref{Loss_and_L2_12}, AT-PINNs may lead to suboptimal outcomes, such as $
err\mathcal{L}_2(u_s) = 104.4 \%, err\mathcal{L}_2(v_s) = 153.3 \%, err\mathcal{L}_2(p_s) = 104.9 \% $ and $  err\mathcal{L}_2(p_d) = 121.6 \% 
$. This is because Adam must rely on historical gradient data for updating and AT-PINNs does not handle the coupling conditions on the interface. 

\end{itemize}

\begin{figure}[H]
	\centering
	\subfloat[\centering]{\includegraphics[width=7.5cm]{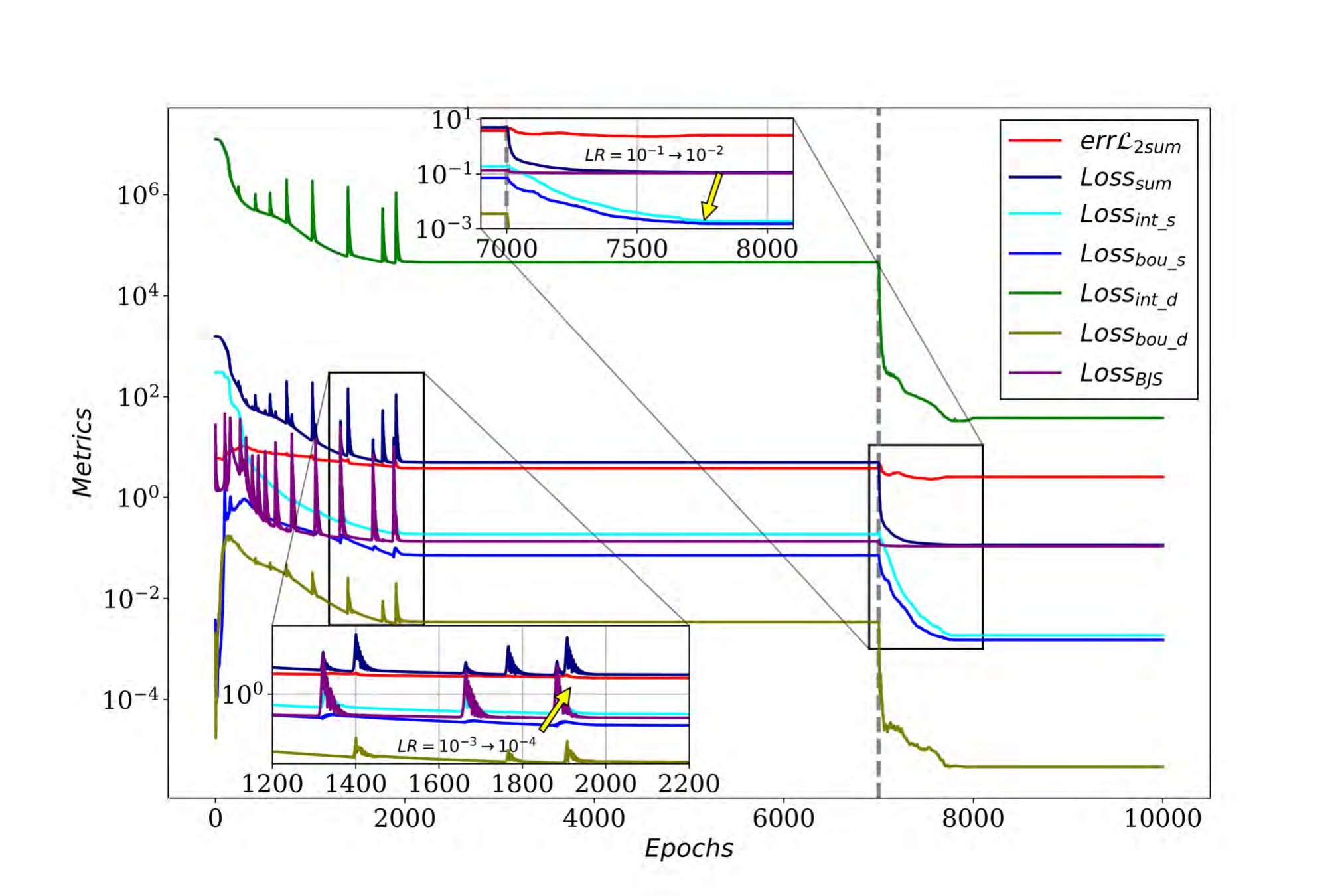}}
	\hfill
	\subfloat[\centering]{\includegraphics[width=7.5cm]{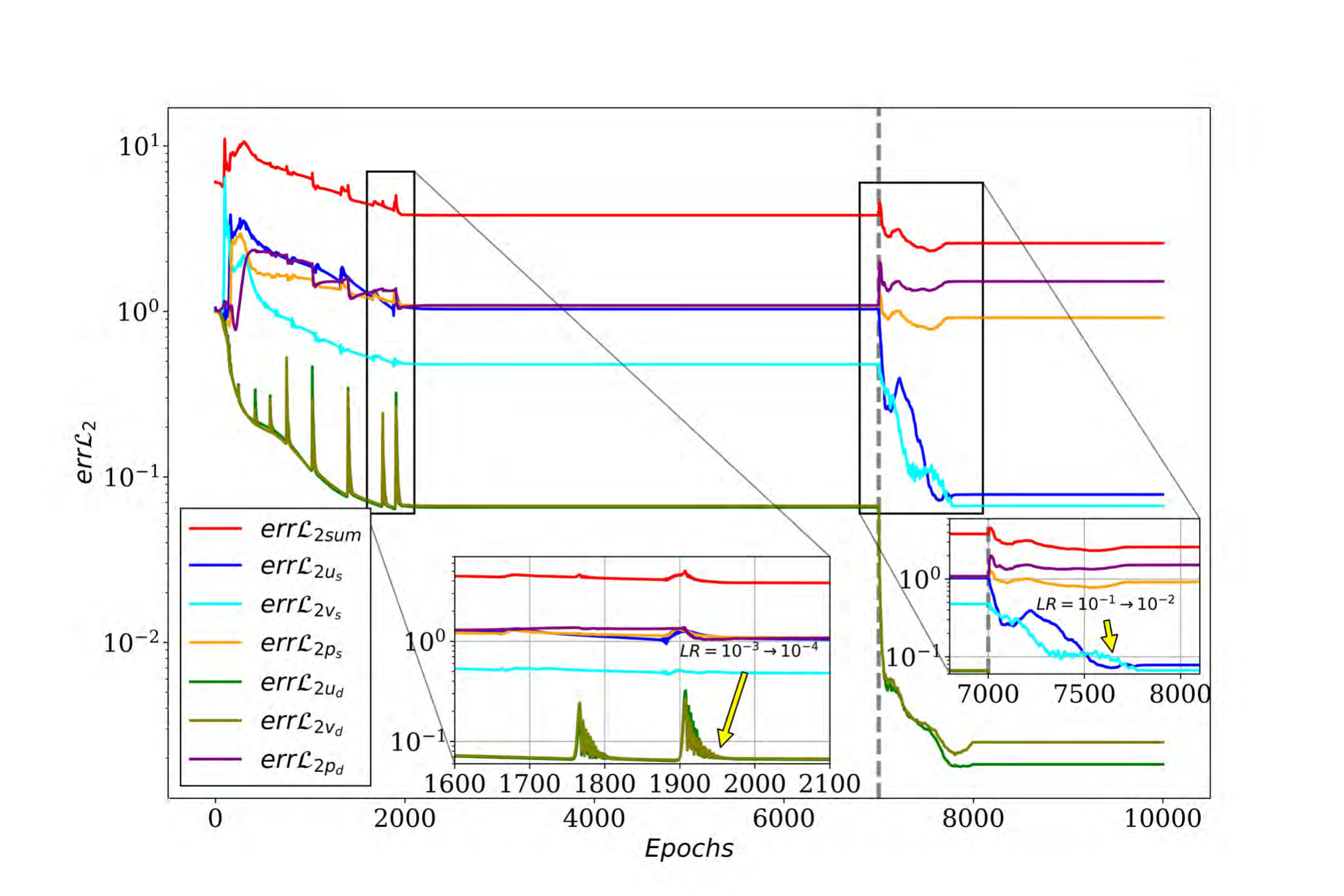}}
	\\	
	\subfloat[\centering]{\includegraphics[width=7.5cm]{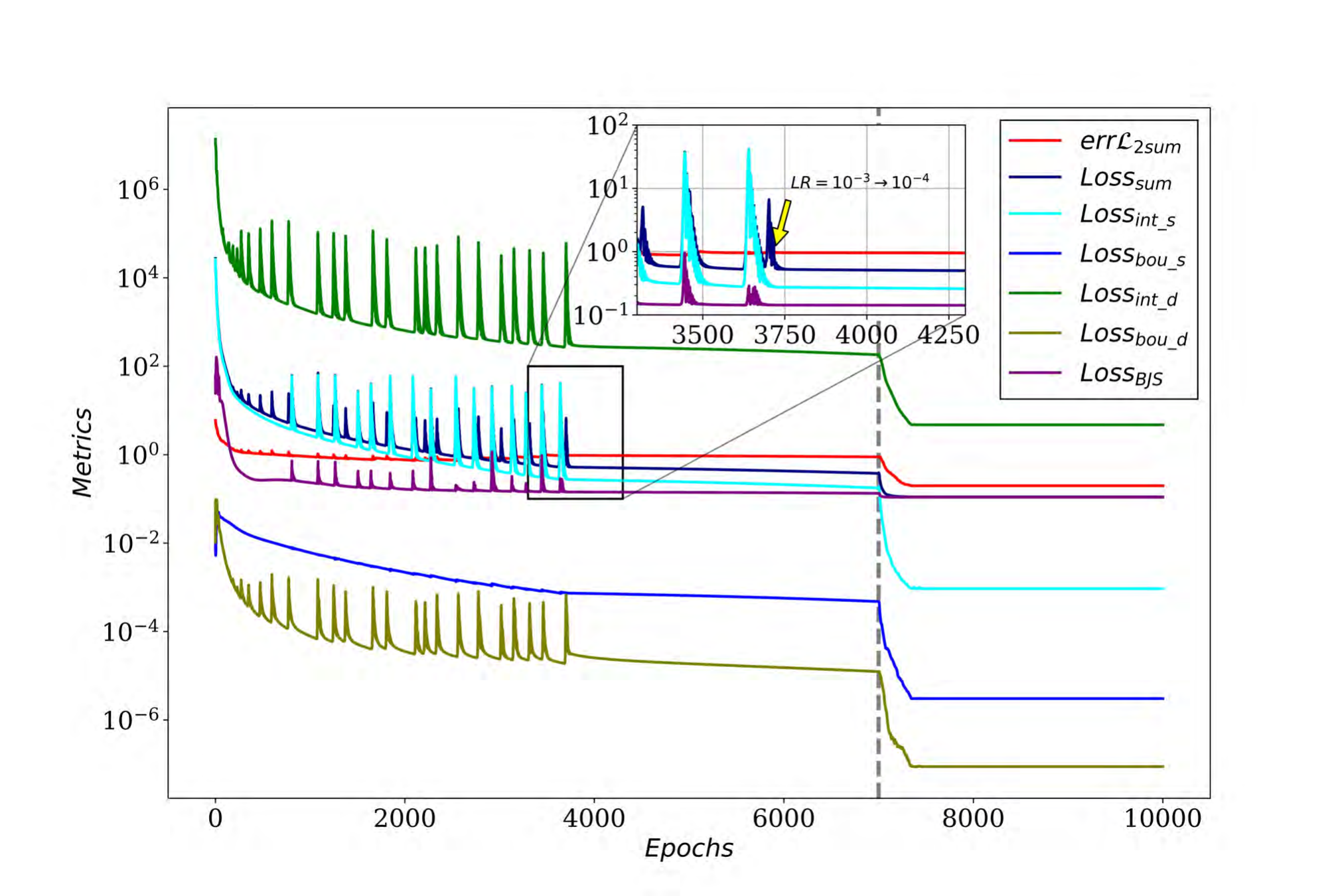}}
	\hfill
	\subfloat[\centering]{\includegraphics[width=7.5cm]{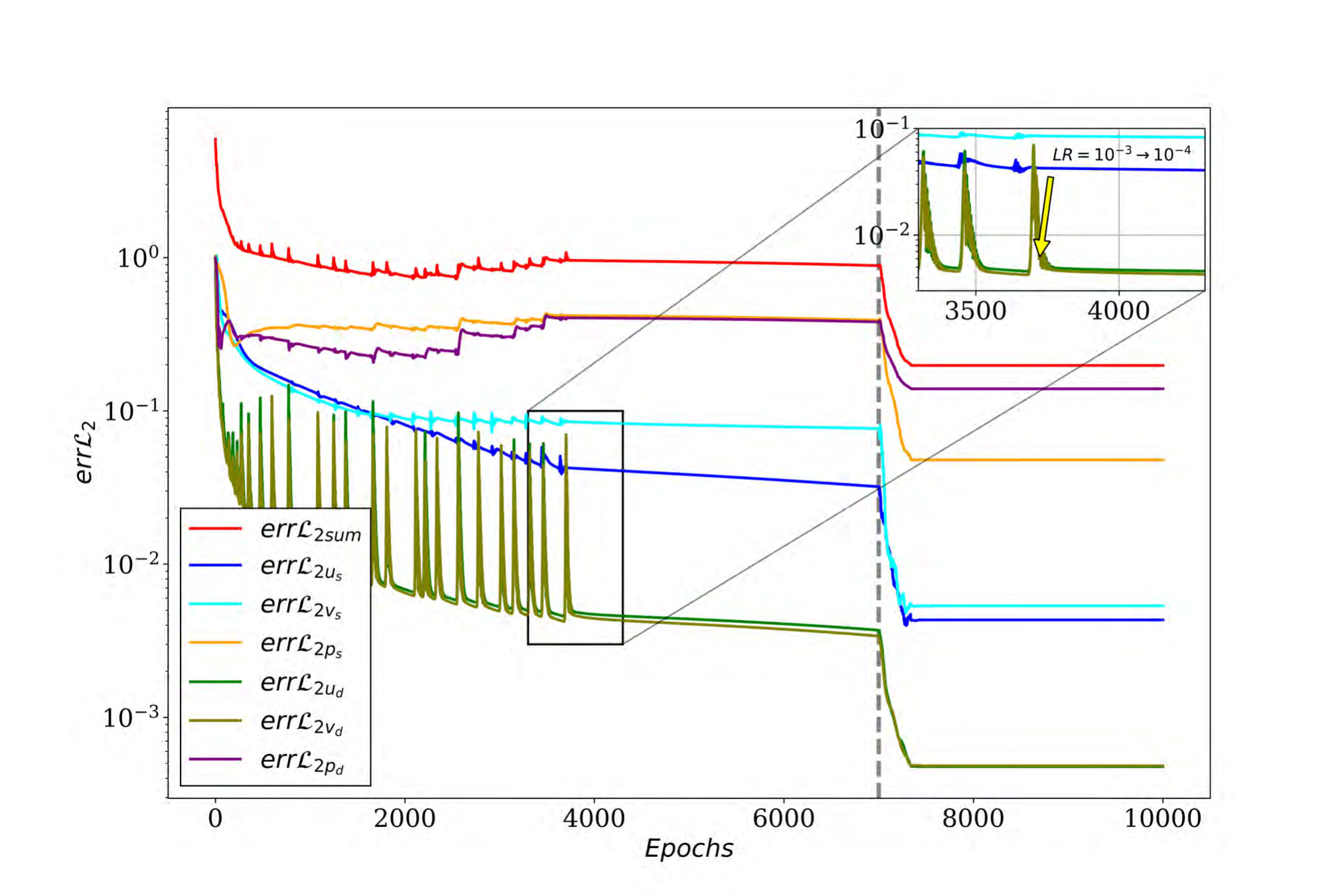}}
	\caption{These images (\textbf{a})-(\textbf{d}) compare the abilities of MW-PINNs and our MF-PINNs. The dashed grey line means that we end up using the Adam optimizer and then use the L-BFGS optimizer. \textbf{By row}: MW-PINNs, MF-PINNs;  \textbf{By column}: $ \mathit{Loss} $ , $ err\mathcal{L}_{2} $. \label{Loss_and_L2_34}}
\end{figure} 

\vspace{-1cm}

\begin{figure}[H]
	\centering
	\subfloat[\centering]{\includegraphics[width=5.0cm]{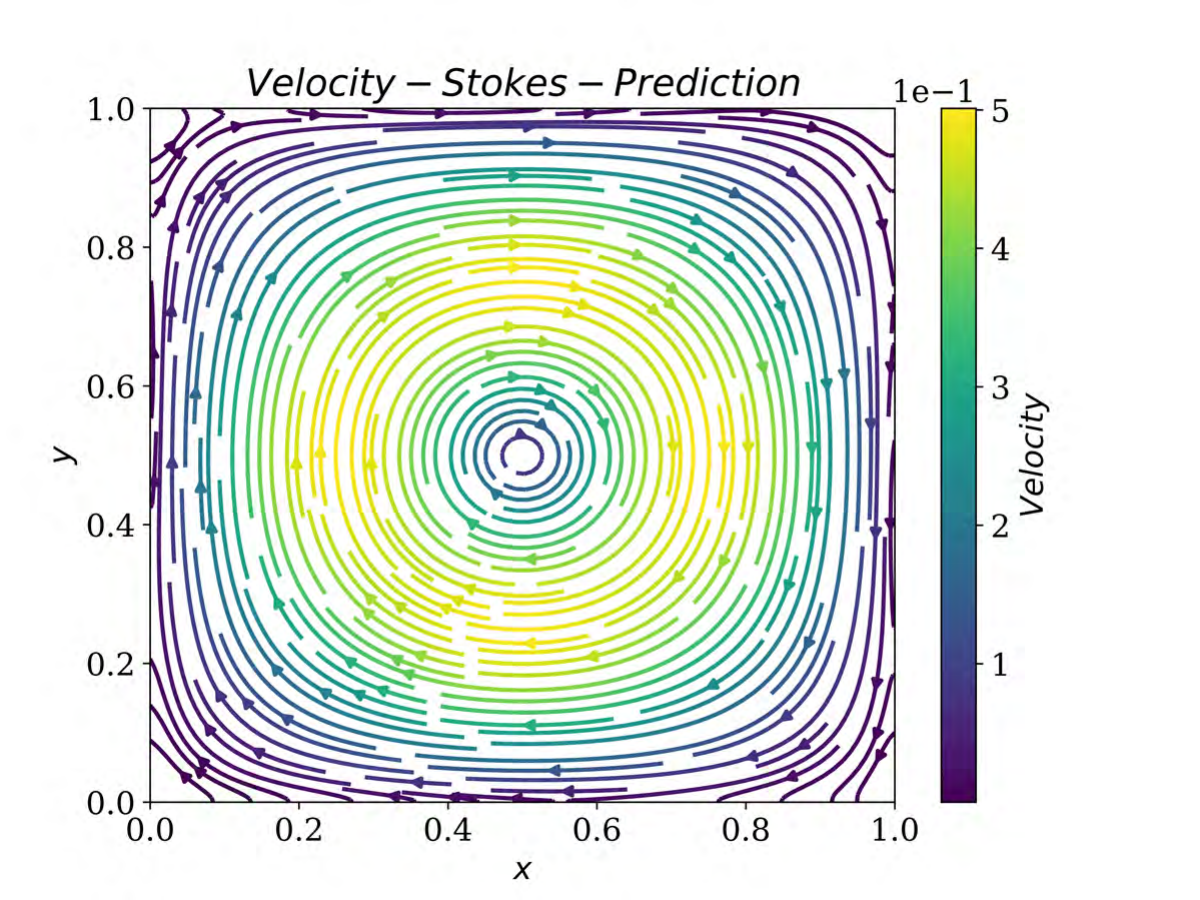}}
	\hfill
	\subfloat[\centering]{\includegraphics[width=5.0cm]{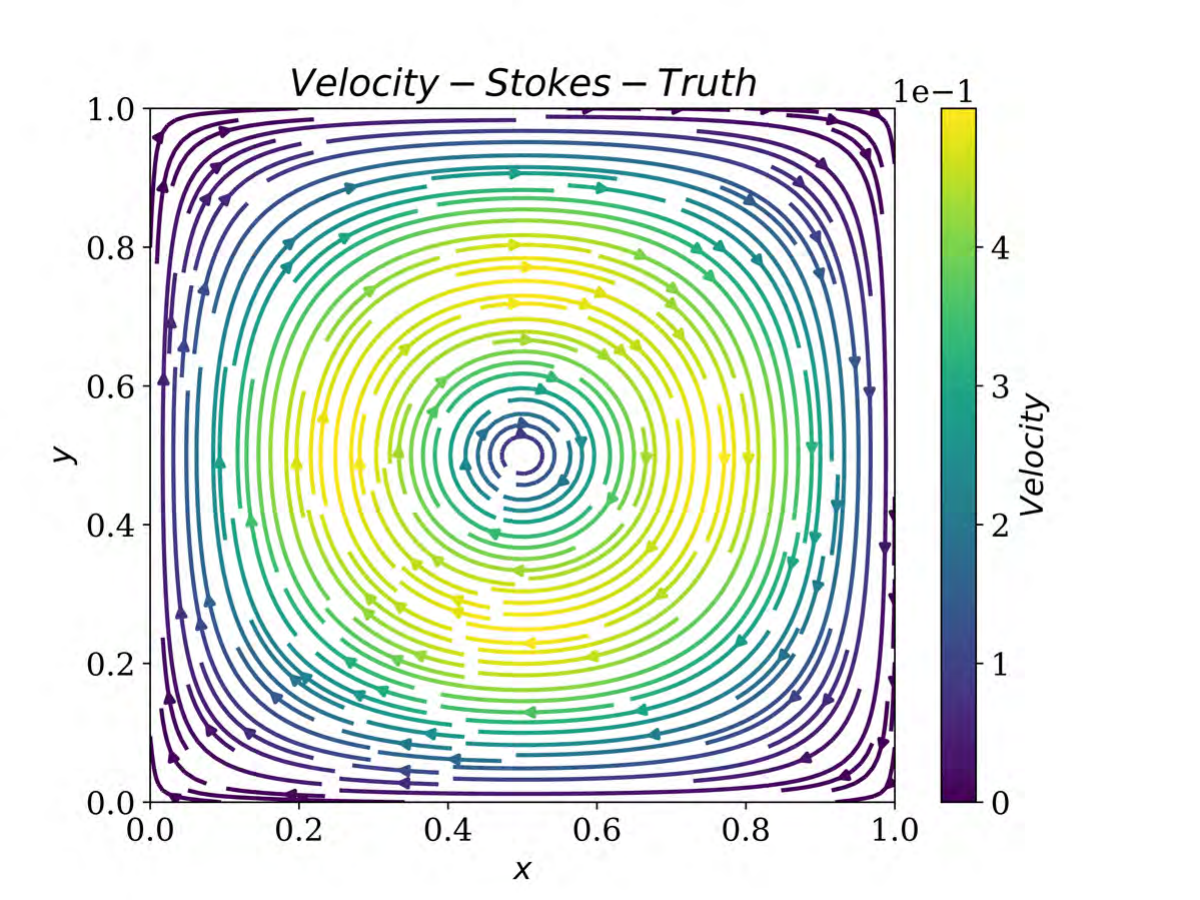}}
	\hfill
	\subfloat[\centering]{\includegraphics[width=5.0cm]{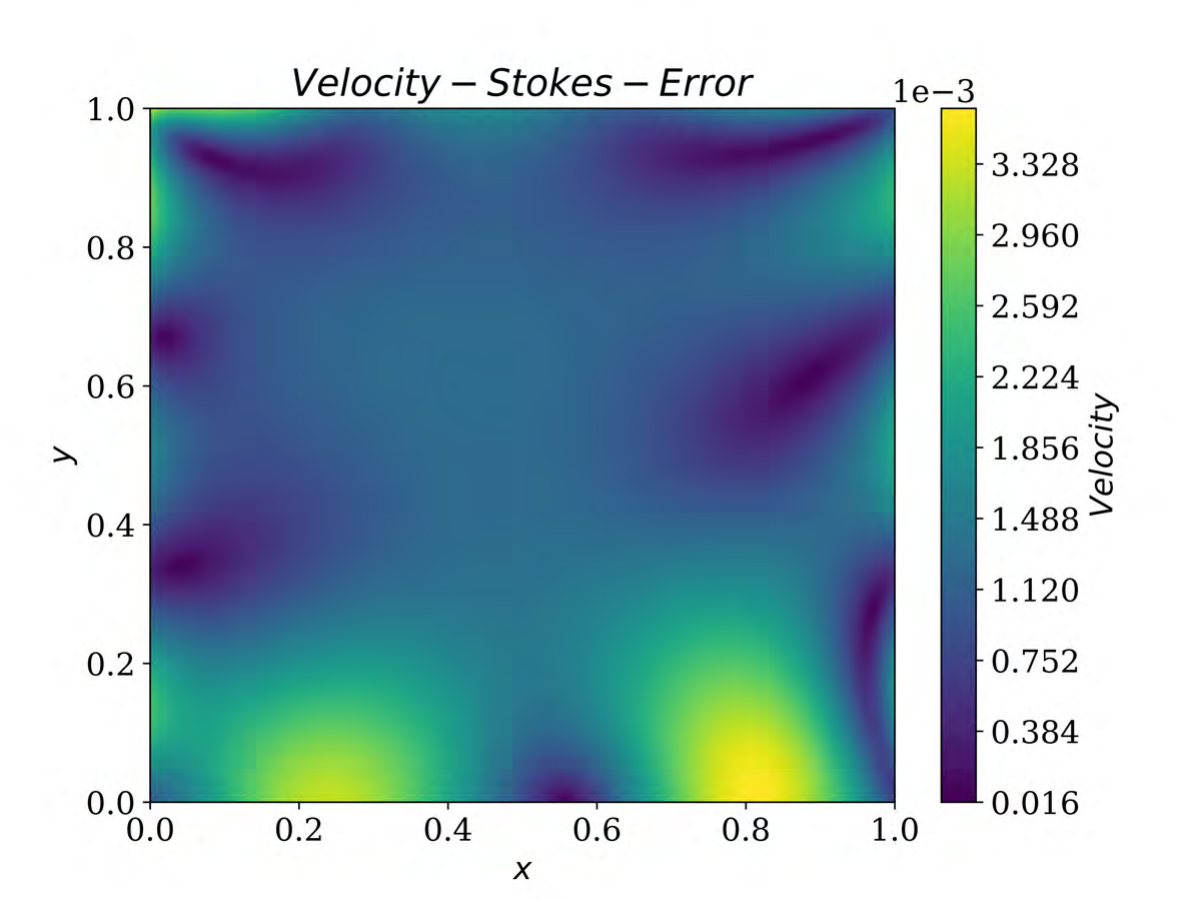}}
	\\	
	\subfloat[\centering]{\includegraphics[width=5.0cm]{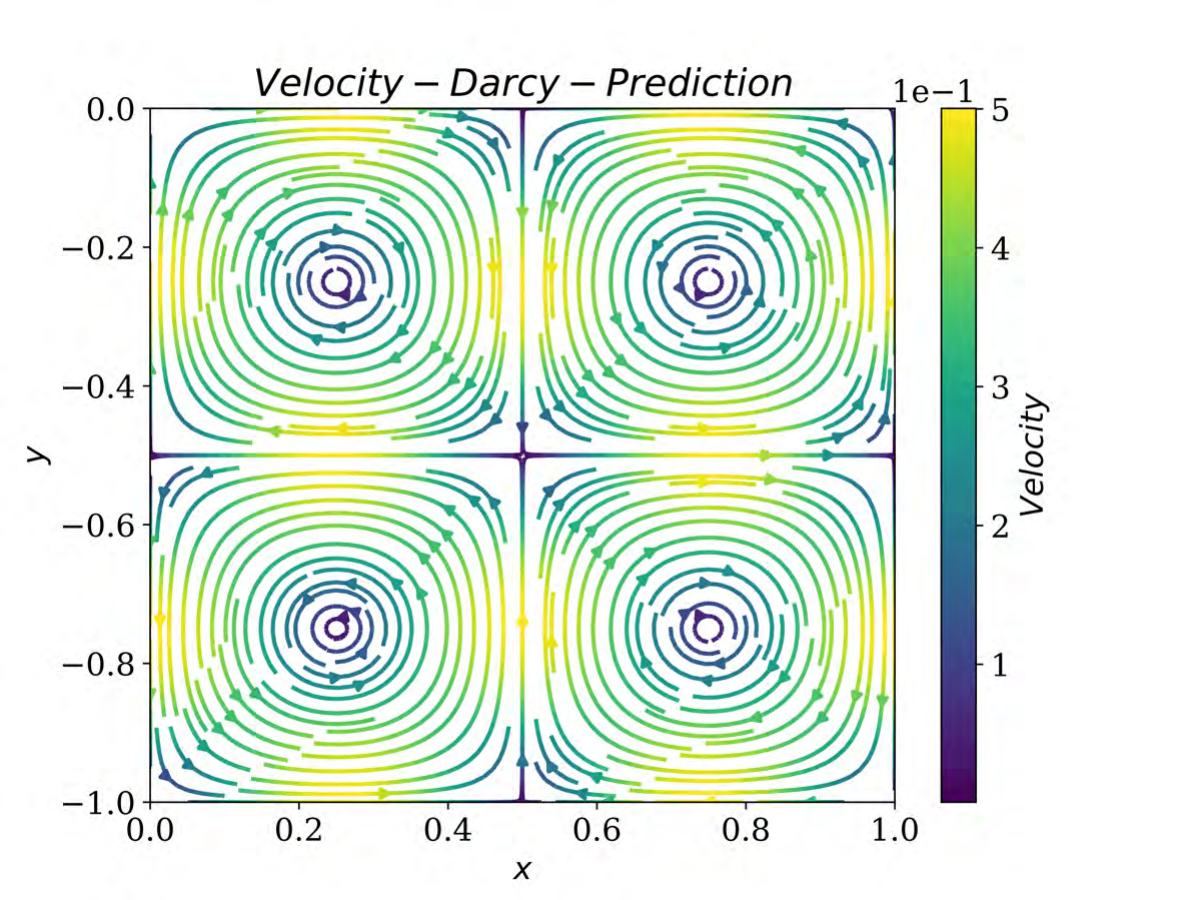}}
	\hfill
	\subfloat[\centering]{\includegraphics[width=5.0cm]{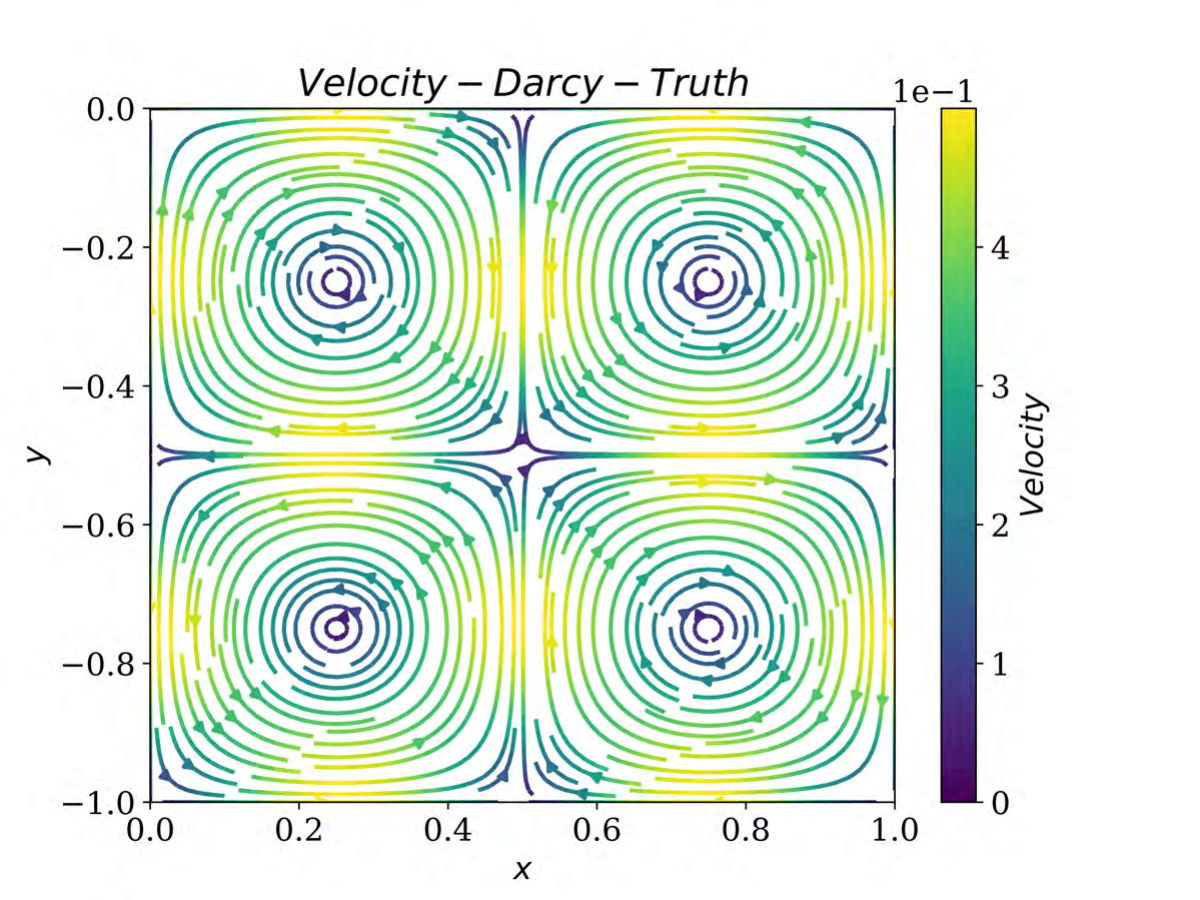}}
	\hfill
	\subfloat[\centering]{\includegraphics[width=5.0cm]{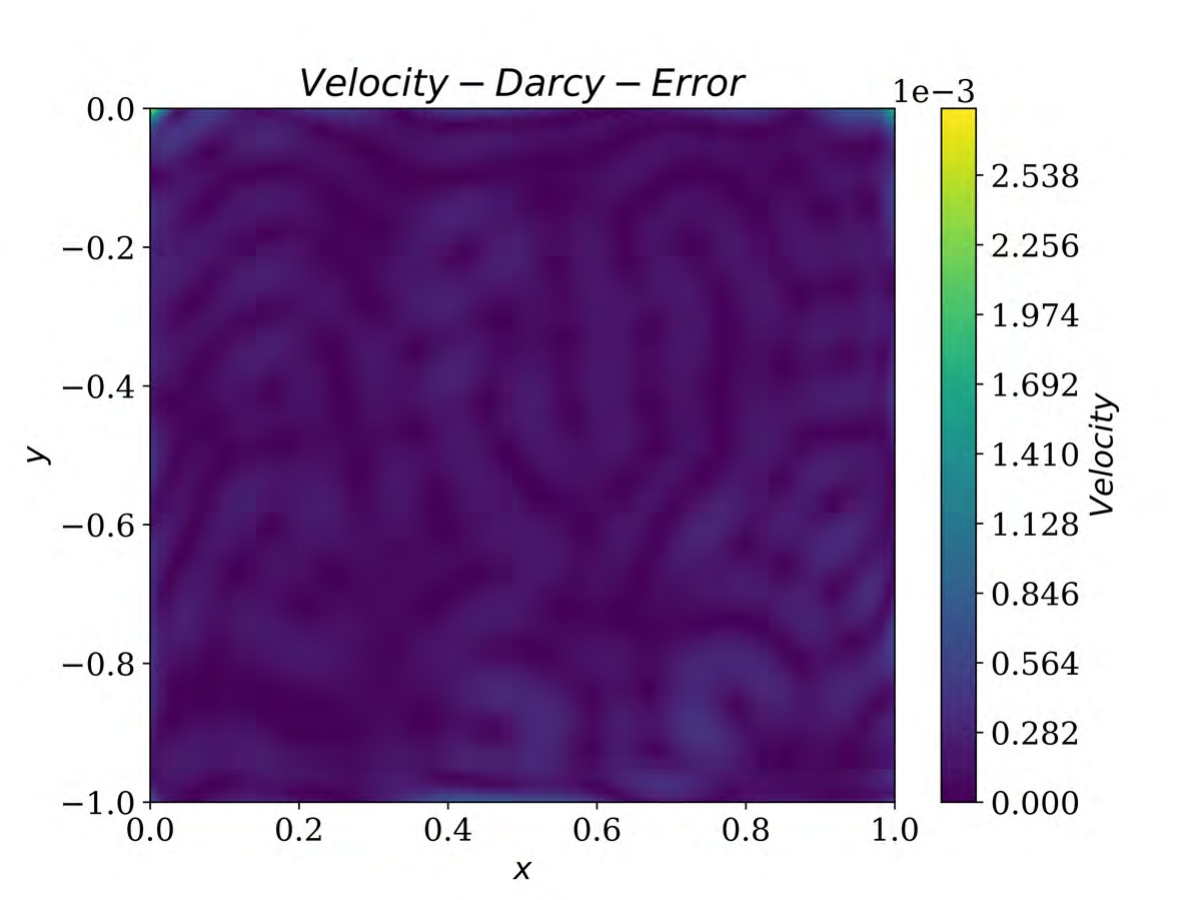}}	
	\caption{These images (\textbf{a})-(\textbf{f}) show the ability of our MF-PINNs to predict the velocity fields of the coupled Stokes-Darcy equations, under $ \mathbb{K} = 10^{-4}\mathbb{I} $ and $ \nu = 1 $. In these images, the colorful lines stand for the streamlines, the arrows stand for the direction of velocity, and the colorbars stand for the value of velocity. \textbf{By row}: Stokes domain, Darcy domain; \textbf{By column}: MF-PINNs numerical solutions, analytical solutions, absolute error.  \label{figvelocity}}
\end{figure}

\begin{figure}[H]
	\centering
	\subfloat[\centering]{\includegraphics[width=5.0cm]{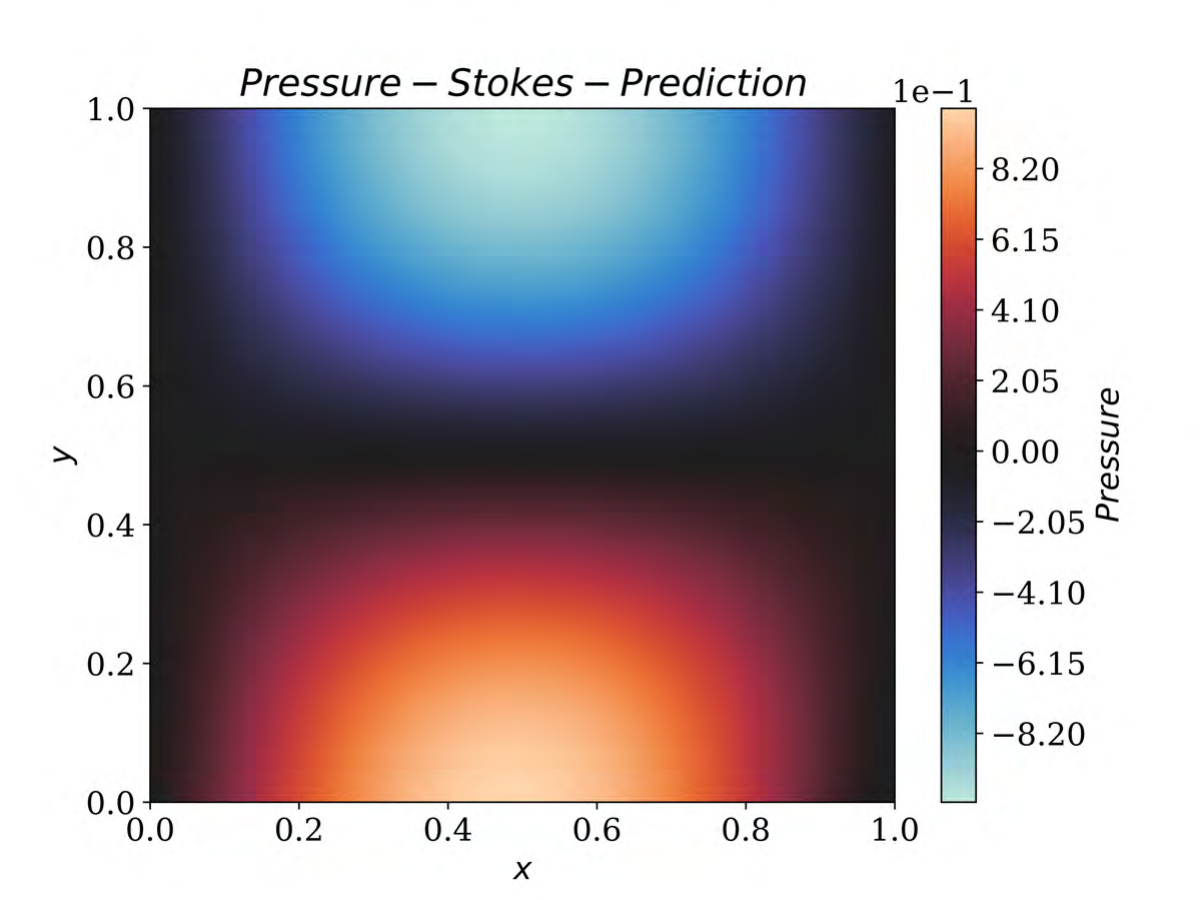}}
	\hfill
	\subfloat[\centering]{\includegraphics[width=5.0cm]{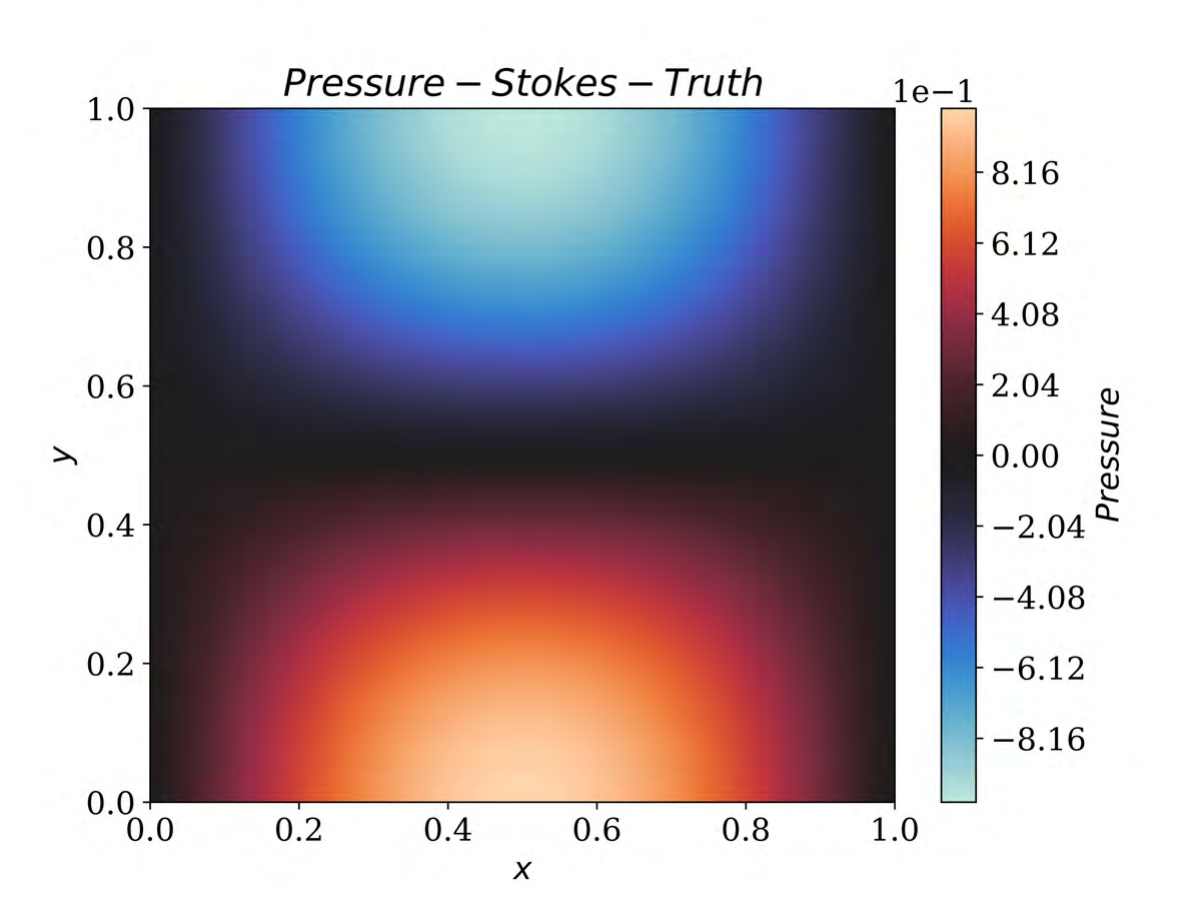}}
	\hfill
	\subfloat[\centering]{\includegraphics[width=5.0cm]{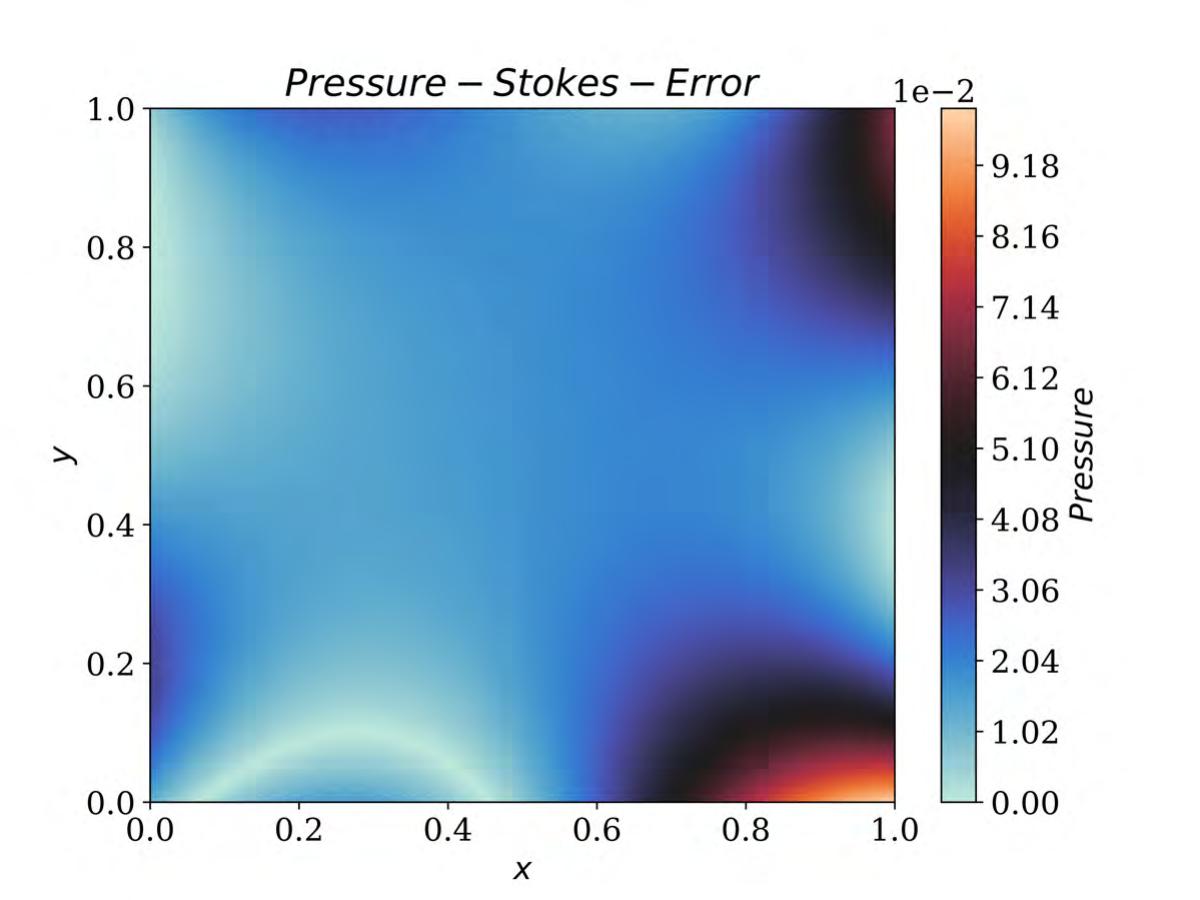}}
	\\
	\subfloat[\centering]{\includegraphics[width=5.0cm]{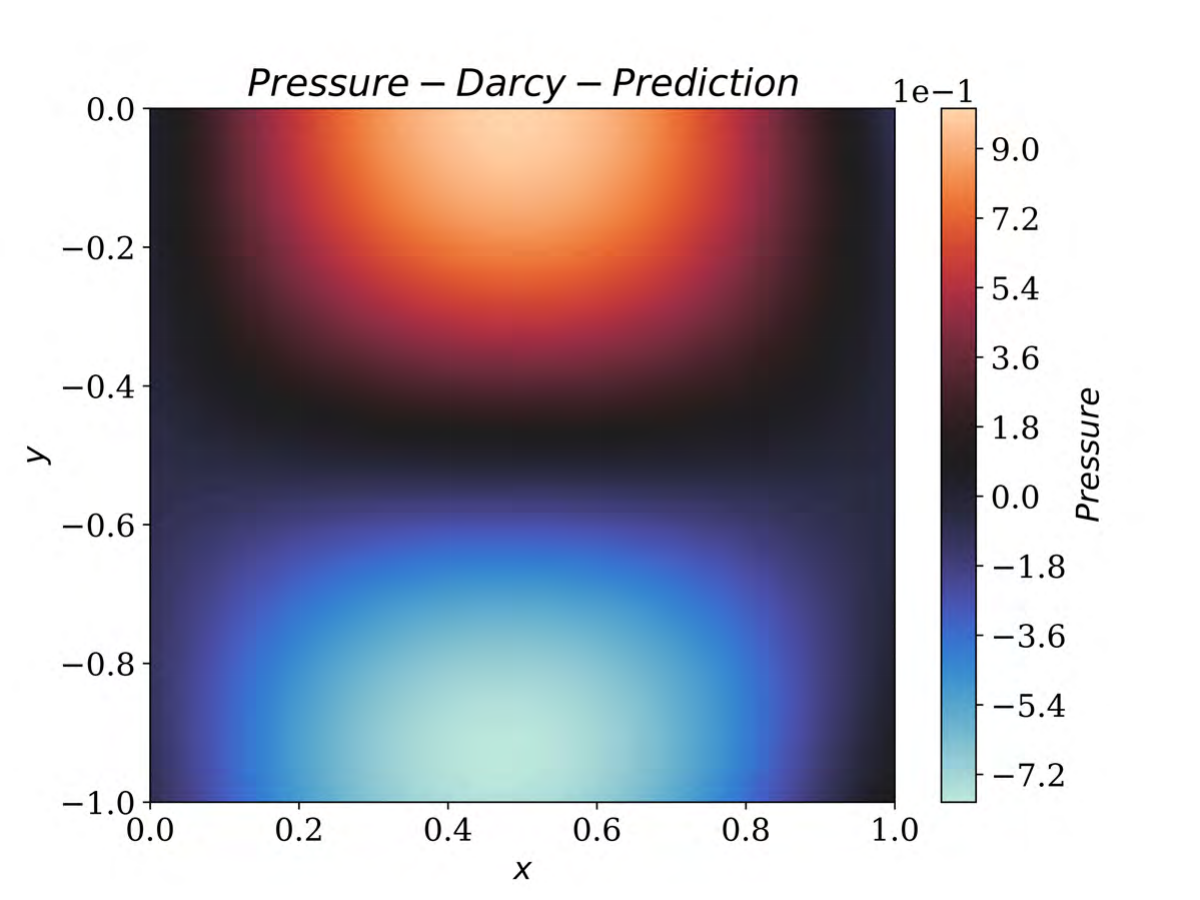}}
	\hfill
	\subfloat[\centering]{\includegraphics[width=5.0cm]{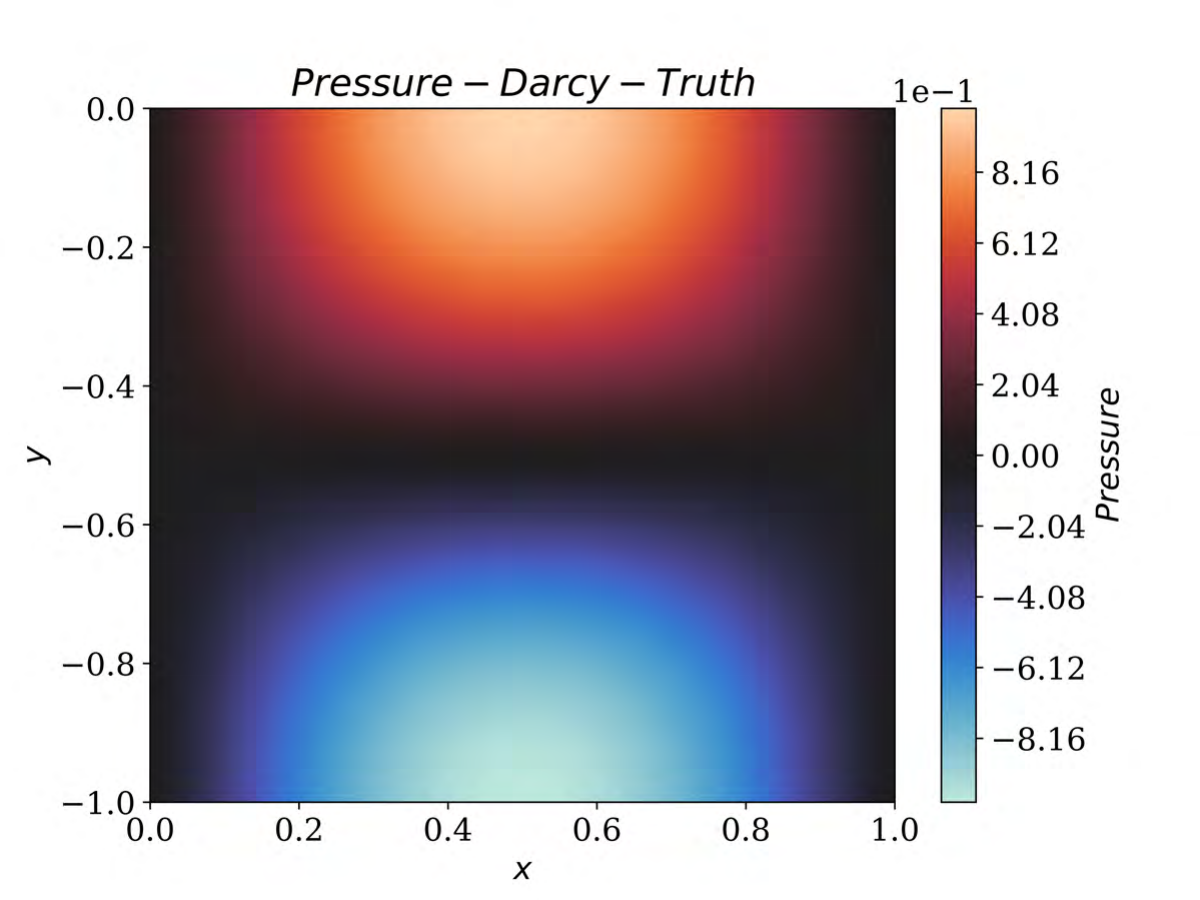}}
	\hfill
	\subfloat[\centering]{\includegraphics[width=5.0cm]{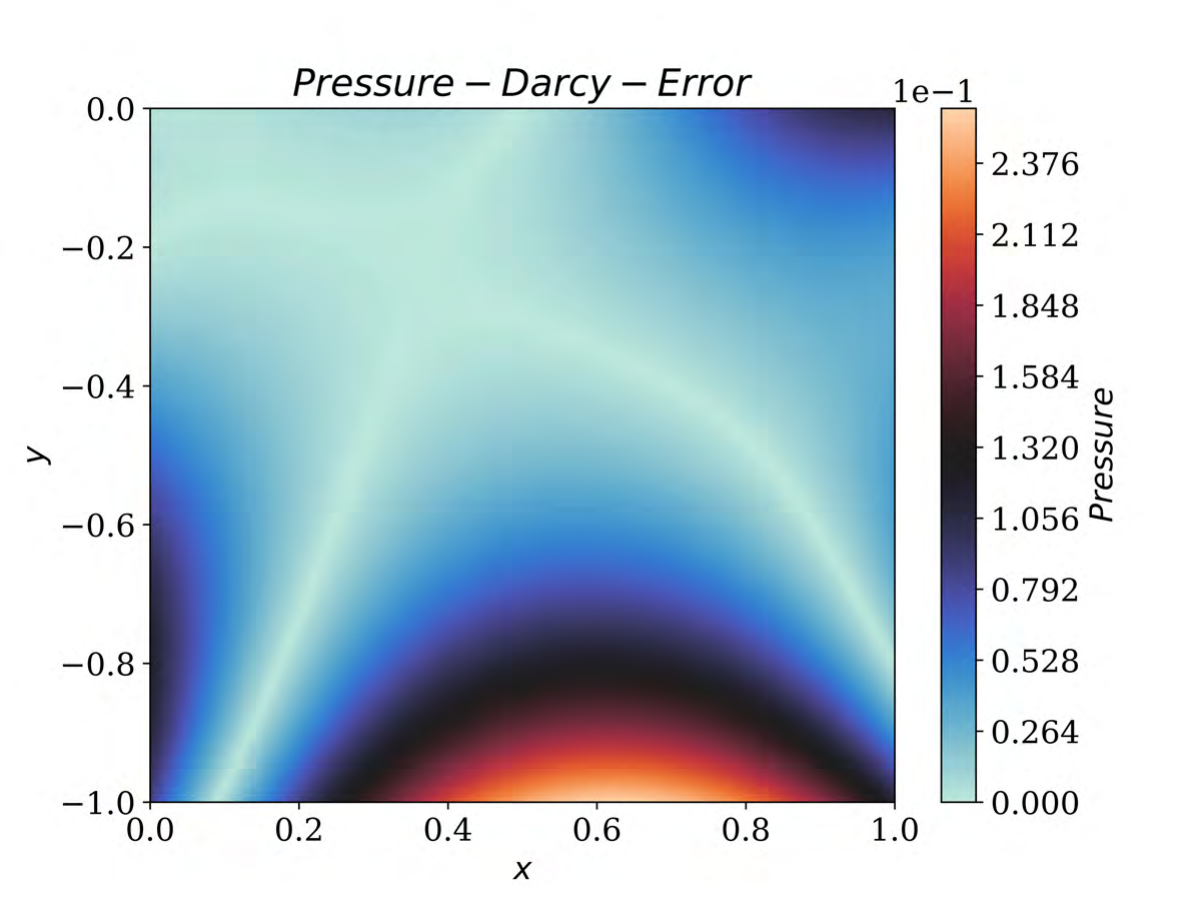}}	
	\caption{These images (\textbf{a})-(\textbf{f}) show the ability of our MF-PINNs to predict the pressure fields of the coupled Stokes-Darcy equations, under $ \mathbb{K} = 10^{-4}\mathbb{I} $ and $ \nu = 1 $. \textbf{By row}: Stokes domain, Darcy domain; \textbf{By column}: MF-PINNs numerical solutions, analytical solutions, absolute error.   \label{figpressure}}
\end{figure}

\begin{figure}[H]
	\centering
	\subfloat[\centering]{\includegraphics[width=5.0cm]{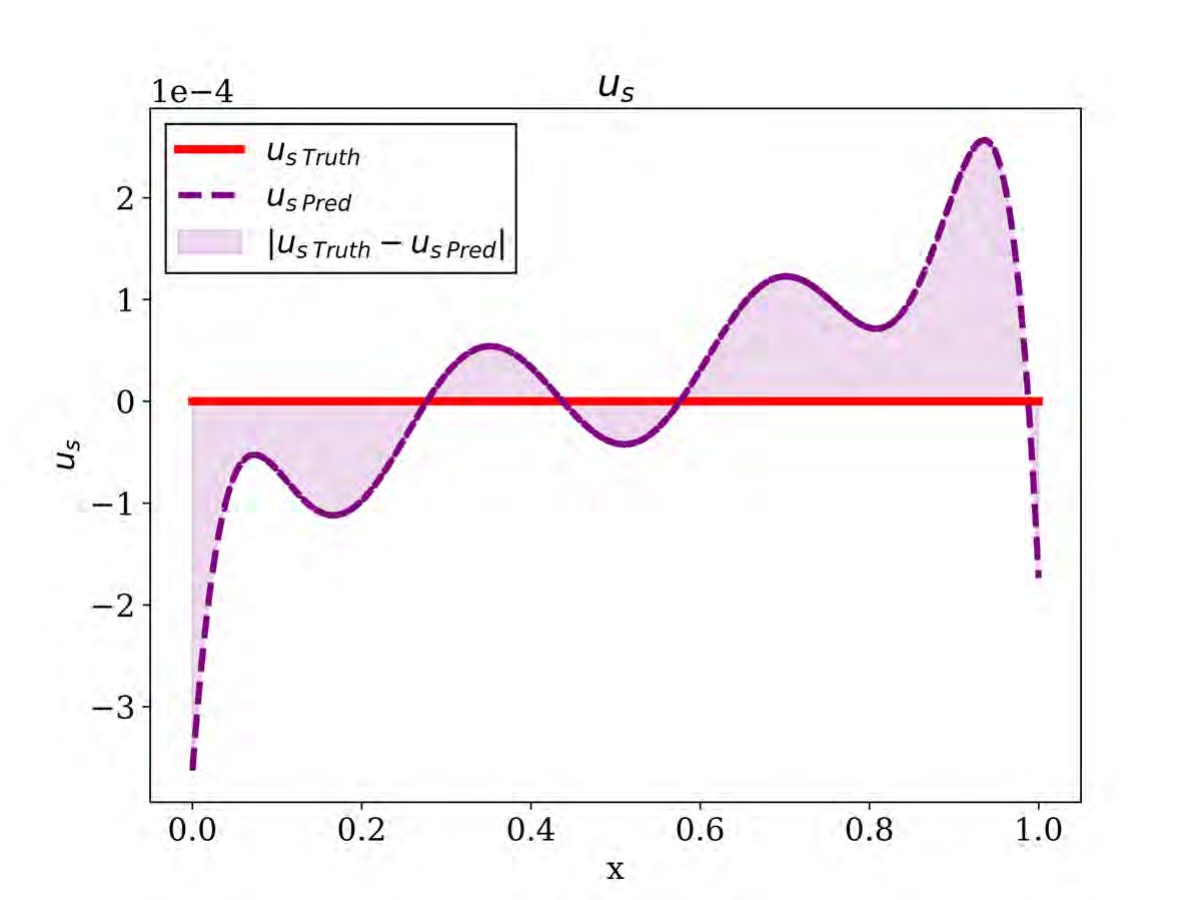}}
	\hfill
	\subfloat[\centering]{\includegraphics[width=5.0cm]{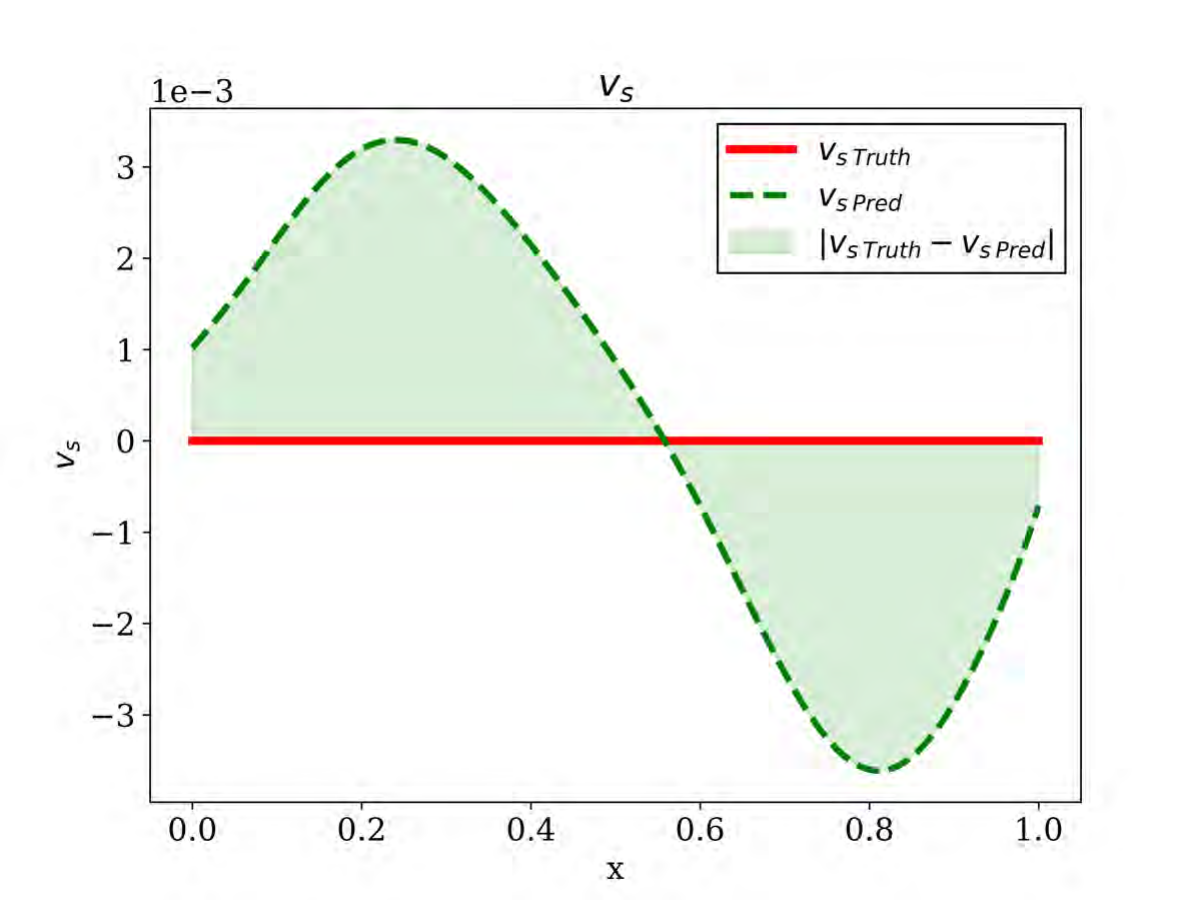}}
	\hfill
	\subfloat[\centering]{\includegraphics[width=5.0cm]{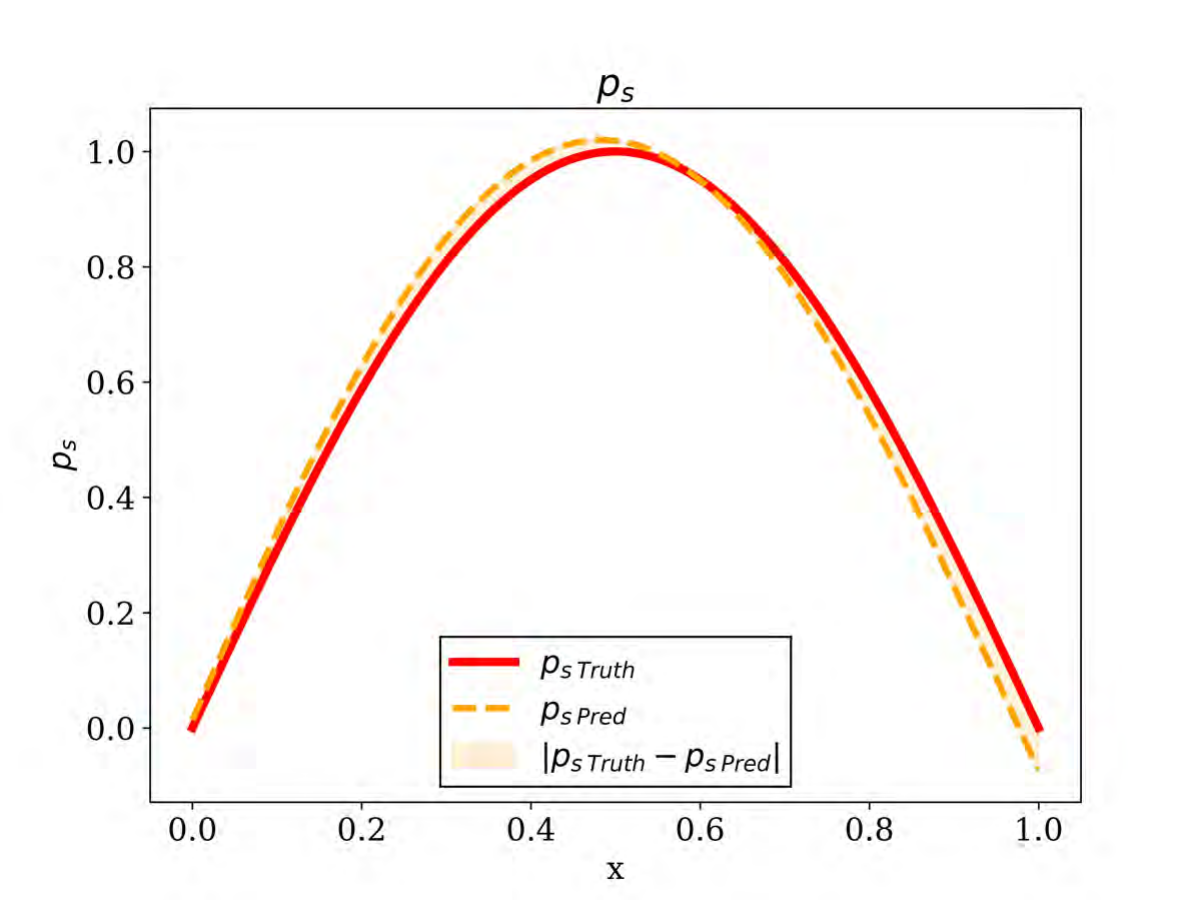}}
	\\
	\subfloat[\centering]{\includegraphics[width=5.0cm]{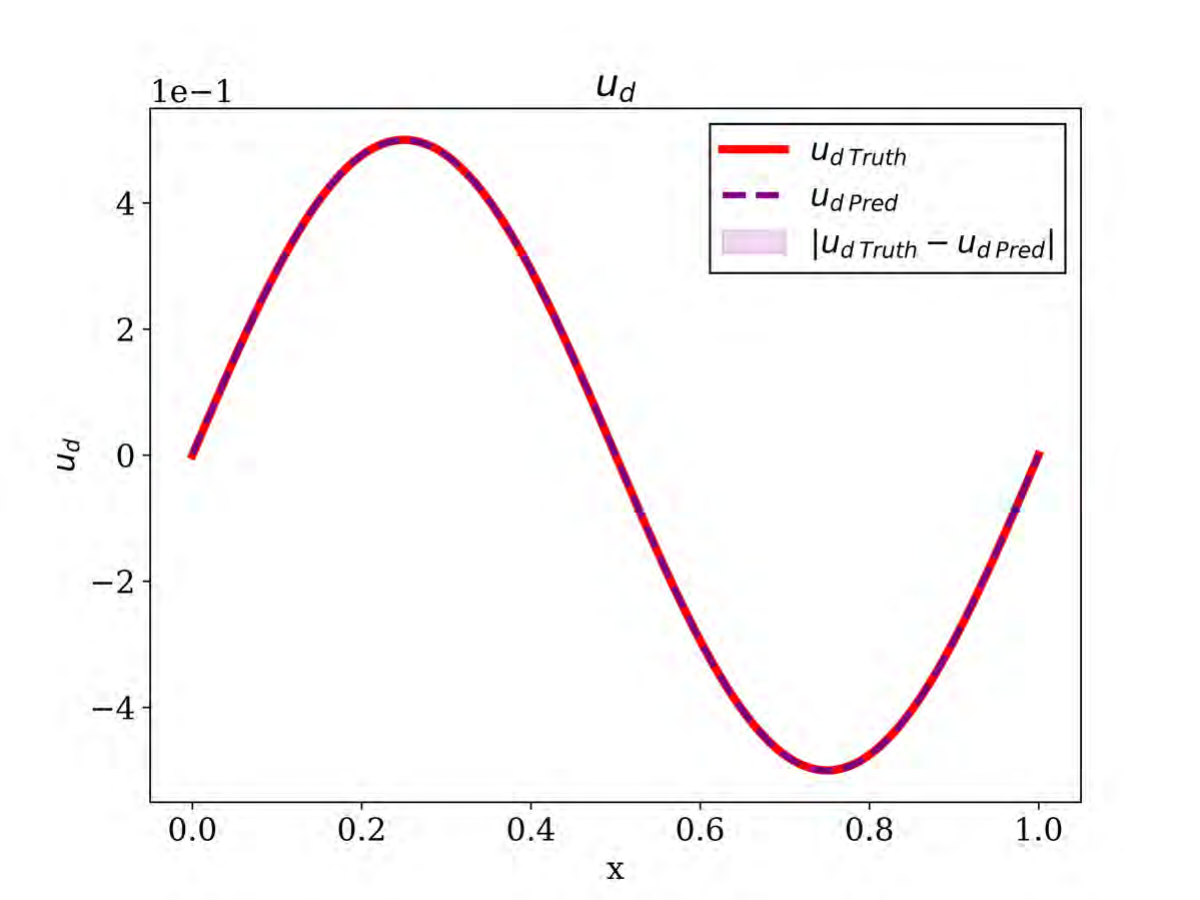}}
	\hfill
	\subfloat[\centering]{\includegraphics[width=5.0cm]{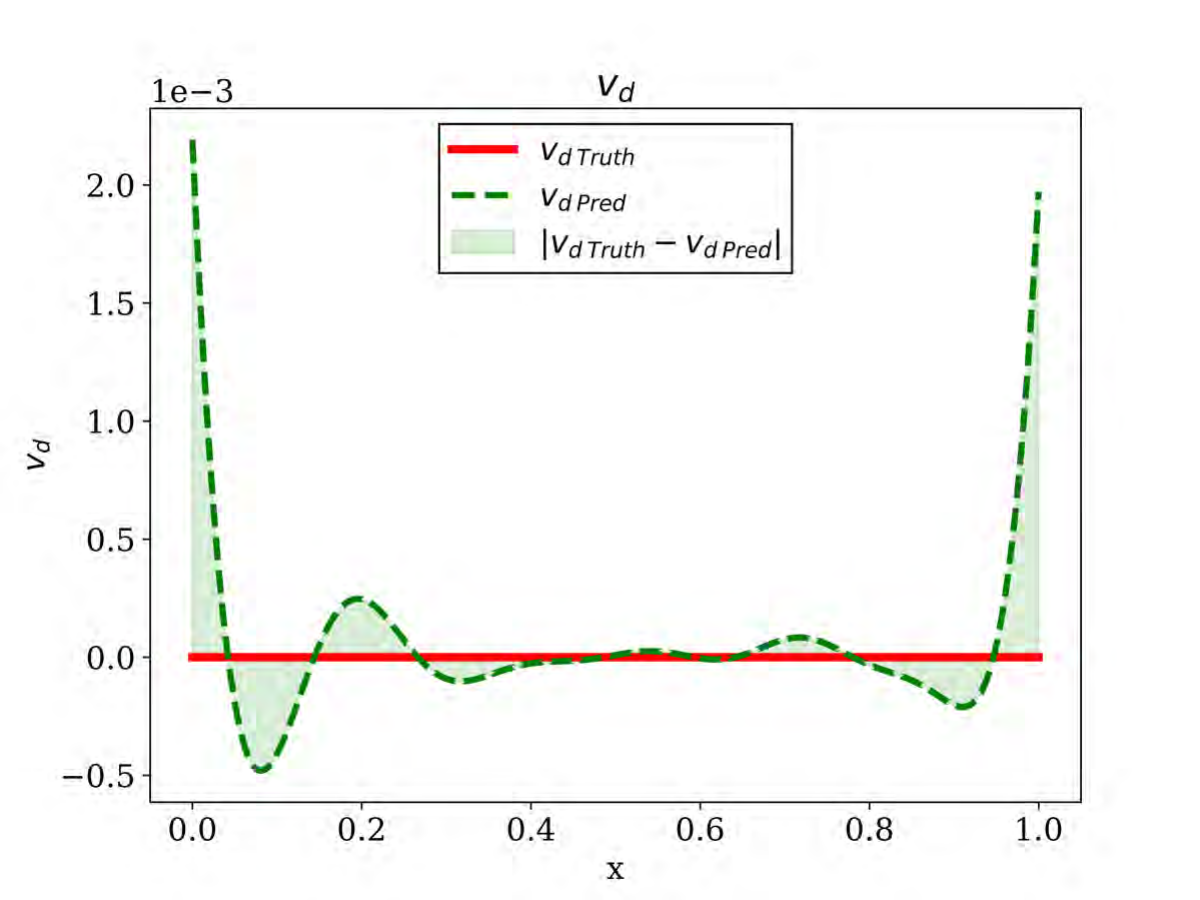}}
	\hfill
	\subfloat[\centering]{\includegraphics[width=5.0cm]{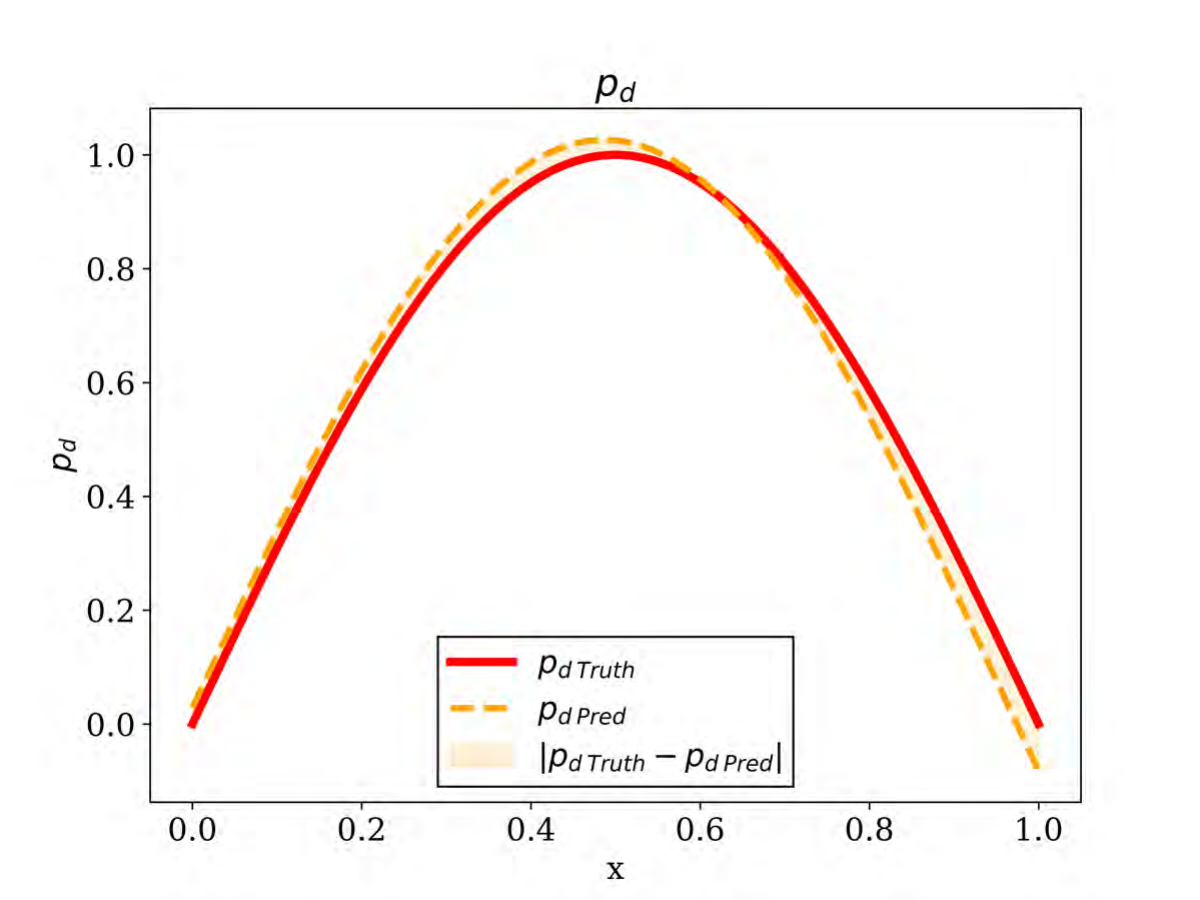}}
	\caption{These images (\textbf{a})-(\textbf{f}) show the absolute error between our MF-PINNs numerical solutions and analytical solutions in the interface, under $ \mathbb{K} = 10^{-4}\mathbb{I} $ and $ \nu = 1 $. We notice that the constant becomes $ \frac{2}{\pi} $ in (\ref{eq:Pressure}).
		\textbf{By row}: Stokes domain, Darcy domain; \textbf{By column}: x direction of velocity field, y direction of velocity field, pressure field.  \label{figInter}}
\end{figure}

\begin{itemize}
\item The MW-PINNS make full use of different weights, $ 1 / \kappa  $ and  $ \nu / \kappa  $, to assemble the total loss $\mathcal{J}(\mathbf{x},\Theta) $. Compared with the baseline PINNs, the MW-PINNs successfully mitigate the gradient competition between the Stokes and Darcy equations caused by \( \nu / \kappa  \gg \nu \). These evidences are $ err\mathcal{L}_2(u_s) = 7.819 \%, err\mathcal{L}_2(v_s) = 6.679 \%, err\mathcal{L}_2(u_d) = 0.1832 \% $  and $ err\mathcal{L}_2(v_d) = 0.2491 \% $. However, the gradient competition among \( u_d \), \( v_d \) and \( p_d \) caused by \( \nu / \kappa  \gg 1 \) could not be mitigated. The evidence is that the $ err\mathcal{L}_{2}(p_d) $ does not decrease in the early training stage of L-BFGS, and $ p_{dNN} $ does not converge to the  $ p_d $ finally in Fig.~\ref{Loss_and_L2_34}. Results are  
$ err\mathcal{L}_2(p_s) = 91.52 \% $ and $ err\mathcal{L}_2(p_d) = 151.6 \%
$.

\item  Compared with the MW-PINNs, our MF-PINNs mitigates the gradient competition among \( u_d \), \( v_d \) and \( p_d \) caused by \( \nu / \kappa \gg 1 \) and between the Stokes and Darcy equations caused by \( \nu / \kappa \gg v \) as is shown in Fig.~\ref{Loss_and_L2_34}. Results are $
err\mathcal{L}_2(u_s) = 0.4324 \%, err\mathcal{L}_2(v_s) = 0.5342 \%, err\mathcal{L}_2(p_s) = 4.789 \%, err\mathcal{L}_2(u_d) = 0.04768 \%, err\mathcal{L}_2(v_d) = 0.04825 \% $ and $ err\mathcal{L}_2(p_d) = 13.91 \%
$. Our Fig.~\ref{figvelocity} and Fig.~\ref{figpressure} show the prediction, truth and error of all the physical fields. These images show the advantages of our MF-PINNs for all the physical fields under extreme $ \kappa = 10^{-4} $ and $ \nu = 1 $. Additionally, Fig.~\ref{figInter} shows the interface of all the physical fields between the Stokes and Darcy domains, and they validate the ability of our MF-PINNs to handle coupled systems.

\end{itemize}

\vspace{-3cm}
\vspace{-2cm}
\setlength{\parindent}{2em}
Secondly, we consider the performance of MF-PINNs in Table~\ref{tabresult_small} under $ \mathbb{K} = \kappa \mathbb{I}, \kappa \leqslant 1 $ and $  \nu \leqslant 1  $. We could therefore make the following deductions:
\begin{enumerate}
	\item Our MF-PINNs could effectively train the velocity fields under extreme cases, like the group $  \kappa=1, \nu=10^{-4} $. Similarly, our MF-PINNs could effectively train the pressure fields under the group $  \kappa=10^{-4}, \nu=1 $. On the opposite side, neither baseline PINNs, MW-PINNs, nor AT-PINNs could not alleviate the problem of gradient competition of velocity fields and pressure fields, even though MW-PINNs could balance the training of physical fields between the Stokes and Darcy domains better than baseline PINNs.
	\item Compared with other PINNs models, our MF-PINNs could also handle the gradient competition of each physical field effectively, in other extreme cases, like the group $  \kappa=10^{-2},\nu=1 $, the group $  \kappa=1,\nu=10^{-2} $, the group $  \kappa=10^{-4}, \nu=10^{-2} $ and the group $ \kappa=10^{-2}, \nu=10^{-4} $.	
	\item However, compared with the MW-PINNs, the performance of our MF-PINNs seems not ideal under the group $ \kappa=10^{-4}, \nu=10^{-4} $. Therefore, we hope to find several more reasonable combinations of coefficients for the new total loss $ \widetilde{\mathcal{J}}(\mathbf{x}, \Theta) $.
	\item  The performances of baseline PINNs and MW-PINNs are very similar under common cases, like the group $ \kappa=1, \nu = 1 $ and the group $ \kappa=10^{-2}, \nu=10^{-2} $. Our MF-PINNs performs better, but it requires much more time.
	\item Besides, AT-PINNs does save a lot of time, but the performance of AT-PINNs is clearly inferior to that of baseline PINNs under most groups of $ \mathbb{K} = \kappa \mathbb{I}, \kappa \leqslant 1 $ and $  \nu \leqslant 1  $.
	\item Among these several kinds of PINNs, the errors of MW-PINNs and our MF-PINNs are closer to that of AS-DNN. This fact reflects that our MF-PINNs is closer to the maximum fitting ability of NN under the size of data and most groups of $ \mathbb{K} = \kappa \mathbb{I}, \kappa \leqslant 1 $ and $  \nu \leqslant 1  $.
\end{enumerate}	
\clearpage

\begin{table}[H] 
	\small 
	\caption{
		This table lists the performance of different kinds of PINNs under different combinations of $ \mathbb{K} $ and $ \nu $ values  ( $ \mathbb{K} = \kappa \mathbb{I}, \kappa \leqslant 1 $ and $  \nu \leqslant 1  $ ). 
		\label{tabresult_small}} 
	
	\resizebox{\linewidth}{!}{ 
		\begin{tabularx}{1.30\textwidth}{c|cccccccc}
			\toprule
			\textbf{Arguments} & \textbf{Algorithm}	& $ \mathbf{err\mathcal{L}_{2}\left(u_s\right)} $ & $  \mathbf{err\mathcal{L}_{2}\left(v_s\right)} $ & $ \mathbf{err\mathcal{L}_{2}\left(p_s\right)} $ & $ \mathbf{err\mathcal{L}_{2}\left(u_d\right)} $ & $ \mathbf{err\mathcal{L}_{2}\left(v_d\right)} $ & $ \mathbf{err\mathcal{L}_{2}\left(p_d\right)} $ & \textbf{Time} \\ 
			\midrule
			
			NULL & AS-DNN & $ 6.179  \times  10^{-3} $  & $ 6.479  \times  10^{-3}  $ &  $ 2.855  \times  10^{-3} $ &  $ 5.251  \times  10^{-3} $ &  $ 5.063  \times  10^{-3} $ &  $ 3.432  \times  10^{-3} $ &  602s \\
			\midrule
			$  $ & PINNs   & $ 1.012  \times  10^{0} $  & $ 9.191  \times  10^{-1}  $ &  $ \underline{3.486  \times  10^{-4}} $ &  $ \underline{1.009  \times  10^{0}} $ &  $ \underline{1.009  \times  10^{0}} $ &  $\underline{ 2.946  \times  10^{-4}} $ &  1177s \\
			
			$  \kappa=1 $  & AT-PINNs  &  $ 1.318  \times  10^{0} $ &  $ 1.066  \times  10^{0} $ & $ 4.974  \times  10^{-4} $ &  $ 1.081  \times  10^{0} $ &  $ 1.045  \times  10^{0} $ &  $ 6.130  \times  10^{-3} $ &   1161s  \\
			
			$  \nu=10^{-4} $  & MW-PINNs   &  $ \underline{1.187  \times  10^{-1}} $ &  $ \underline{1.119  \times  10^{-1}} $ & $ 6.542  \times  10^{-4} $ &  $ 1.073  \times  10^{0} $ &  $ 1.085  \times  10^{0} $ &  $ 6.564  \times  10^{-4} $ &   1948s  \\
			
			$  $  & MF-PINNs $\uparrow$ &  $ \mathbf{1.096  \times  10^{-2}} $ &  $ \mathbf{1.513  \times  10^{-2}} $ & $ \mathbf{1.935  \times  10^{-4}} $ &  $ \mathbf{8.094  \times  10^{-2}} $ &  $\mathbf{ 6.349  \times  10^{-2}} $ &  $ \mathbf{2.101  \times  10^{-4}} $ &  2505s \\
			\midrule
			$  $ & PINNs  & $ 2.001 \times  10^{-1} $  & $ 1.974  \times  10^{-1}  $ &  $ 1.041  \times  10^{0} $ &  $ \underline{5.929  \times  10^{-4}} $ &  $ \underline{7.349  \times  10^{-4}} $ &  $ 1.354  \times  10^{0} $ &  2161s \\
			
			$  \kappa=10^{-4} $  & AT-PINNs  &  $ 1.044  \times  10^{0} $ &  $ 1.533  \times  10^{0} $ & $ 1.049  \times  10^{0} $ &  $ 6.862  \times  10^{-1} $ &  $ 7.032  \times  10^{-1} $ &  $ \underline{1.216  \times  10^{0}} $ &   918s  \\
			
			$ \nu=1 $  & MW-PINNs  &  $ \underline{7.819  \times  10^{-2}}$ &  $ \underline{6.679  \times  10^{-2}} $ & $ \underline{9.152  \times  10^{-1}} $ &  $ 1.832  \times  10^{-3} $ &  $ 2.491  \times  10^{-3} $ &  $ 1.516  \times  10^{0} $ &   2294s  \\
			
			$  $  & MF-PINNs $\uparrow$ &  $ \mathbf{4.324  \times  10^{-3}} $ &  $ \mathbf{5.342  \times  10^{-3}} $ & $ \mathbf{4.789  \times  10^{-2}} $ &  $ \mathbf{4.768  \times  10^{-4}} $ &  $ \mathbf{4.825  \times  10^{-4}} $ &  $ \mathbf{1.393  \times  10^{-1}} $ &  3174s \\
			\midrule
			$  $ & PINNs   & $\underline{ 1.475  \times  10^{-2}} $  & $ \underline{2.036  \times  10^{-2}}  $ &  $\underline{ 7.736  \times  10^{-4}} $ &  $ 9.943  \times  10^{-1} $ &  $ 9.944  \times  10^{-1} $ &  $ 3.217  \times  10^{-4} $ &  1216s \\
			
			$  \kappa=1 $  & AT-PINNs  &  $ 1.543  \times  10^{-1} $ &  $ 1.908  \times  10^{-1} $ & $ 5.487  \times  10^{-3} $ &  $ 1.001  \times  10^{0} $ &  $ 1.008  \times  10^{0} $ &  $ 1.063  \times  10^{-2} $ &   1141s  \\
			
			$  \nu=10^{-2} $  & MW-PINNs   &  $ 2.407  \times  10^{-2} $ &  $ 2.887  \times  10^{-2} $ & $ 1.438  \times  10^{-3} $ &  $ \underline{9.704  \times  10^{-1}} $ &  $\underline{ 9.634  \times  10^{-1}} $ &  $ \mathbf{2.192  \times  10^{-4}} $ &   1322s  \\
			
			$  $  & MF-PINNs $\uparrow$ &  $ \mathbf{1.324  \times  10^{-2}} $ &  $ \mathbf{1.878  \times  10^{-2}} $ & $ \mathbf{6.918  \times  10^{-4}} $ &  $ \mathbf{1.430  \times  10^{-2}} $ &  $ \mathbf{1.418  \times  10^{-2}} $ &  $ \underline{2.343  \times  10^{-4}} $ &  2588s \\
			\midrule
			$  $ & PINNs  & $ 4.832  \times  10^{-2} $  & $ 3.718  \times  10^{-2}  $ &  $ 6.949  \times  10^{-1} $ &  $ 1.587 \times  10^{-2} $ &  $ 5.930  \times  10^{-2} $ &  $ 1.115  \times  10^{0} $ &  2498s \\
			
			$  \kappa=10^{-2} $  & AT-PINNs  &  $ 1.187 \times  10^{-1} $ &  $ 2.764  \times  10^{-1} $ & $ 8.793  \times  10^{-1} $ &  $ 5.022  \times  10^{-1} $ &  $ 4.878  \times  10^{-1} $ &  $ 1.127  \times  10^{0} $ &    1014s\\
			
			$ \nu=1 $  & MW-PINNs  &  $ \underline{2.544  \times  10^{-2}} $ &  $ \underline{2.474  \times  10^{-2}} $ & $ \underline{3.713  \times  10^{-1}} $ &  $ \underline{1.167  \times  10^{-2}} $ &  $ \underline{2.743  \times  10^{-2}} $ &  $ \underline{5.375  \times  10^{-1}} $ &   1347s  \\
			
			$  $ & MF-PINNs $\uparrow$ & $ \mathbf{7.157  \times  10^{-3}} $  & $ \mathbf{7.467  \times  10^{-3}}  $ &  $ \mathbf{1.356  \times  10^{-1}} $ &  $ \mathbf{2.905  \times  10^{-3}} $ &  $ \mathbf{5.332  \times  10^{-3}} $ &  $ \mathbf{1.633  \times  10^{-1}} $ &  3077s \\
			\midrule
			$  $ & PINNs  & $ 1.974  \times  10^{-1} $  & $ 1.984  \times  10^{-1}  $ &  $ 1.636  \times  10^{-1} $ &  $ 1.290 \times  10^{-2} $ &  $ 1.475  \times  10^{-2} $ &  $ 3.103  \times  10^{-1} $ &  2191s \\
			
			$  \kappa=10^{-4}$  & AT-PINNs  &  $ 6.896  \times  10^{-1} $ &  $ 9.729  \times  10^{-1} $ & $ 1.312  \times  10^{-1} $ &  $ 1.601  \times  10^{-1} $ &  $ 1.678  \times  10^{-1} $ &  $ 6.870  \times  10^{-1} $ &   1283s  \\
			
			$  \nu=10^{-2} $  & MW-PINNs  &  $ \underline{2.229 \times  10^{-2}} $ &  $ \underline{2.405  \times  10^{-2}} $ & $ \underline{1.267 \times  10^{-1}} $ &  $ \mathbf{7.705  \times  10^{-3}} $ &  $ \underline{1.197  \times  10^{-2}} $ &  $ \underline{2.178  \times  10^{-1}} $ &    1712s\\
			
			$  $ & MF-PINNs $\uparrow$ & $ \mathbf{7.766  \times  10^{-3}} $  & $ \mathbf{1.219  \times  10^{-2}}  $ &  $ \mathbf{8.323  \times  10^{-2}} $ &  $ \underline{8.133  \times  10^{-3}} $ &  $ \mathbf{8.496  \times  10^{-3}} $ &  $ \mathbf{1.425  \times  10^{-1}} $ &  2373s \\
			
			\midrule
			$  $ & PINNs  & $ 1.140  \times  10^{0} $  & $ 1.064  \times  10^{0}  $ &  $ \underline{3.098  \times  10^{-4}} $ &  $ 1.012 \times  10^{0} $ &  $ 1.016  \times  10^{0} $ &  $ \mathbf{2.902  \times  10^{-4}} $ &  1172s \\
			
			$  \kappa=10^{-2}$  & AT-PINNs  &  $ 1.477  \times  10^{0} $ &  $ 1.319  \times  10^{0} $ & $ 6.653  \times  10^{-4} $ &  $ 1.024  \times  10^{0} $ &  $ 1.022  \times  10^{0} $ &  $ 7.349  \times  10^{-3} $ &   1187s  \\
			
			$  \nu=10^{-4} $  & MW-PINNs  &  $ \underline{6.285 \times  10^{-2}} $ &  $ \underline{9.743  \times  10^{-2}} $ & $ 3.850 \times  10^{-4} $ &  $ \underline{9.711  \times  10^{-1}} $ &  $ \underline{9.658  \times  10^{-1}} $ &  $ \underline{4.748  \times  10^{-4}} $ &    1988s\\
			
			$  $ & MF-PINNs $\uparrow$ & $ \mathbf{5.531  \times  10^{-3}} $  & $ \mathbf{7.903  \times  10^{-3}}  $ &  $ \mathbf{2.608  \times  10^{-4}} $ &  $ \mathbf{1.082  \times  10^{-1}} $ &  $ \mathbf{1.106  \times  10^{-1}} $ &  $ 2.491  \times  10^{-3} $ &  2519s \\
			
			\midrule
			$  $ & PINNs  & $ 1.034  \times  10^{0} $  & $ 1.033  \times  10^{0}  $ &  $ \mathbf{1.170  \times  10^{-3}} $ &  $ \mathbf{1.919 \times  10^{-2}} $ &  $ \mathbf{1.578  \times  10^{-2}} $ &  $ \mathbf{1.063  \times  10^{-3}} $ &  1434s \\
			
			$  \kappa=10^{-4} $  & AT-PINNs  &  $ 1.007  \times  10^{0} $ &  $ 1.084  \times  10^{0} $ & $ 4.226  \times  10^{-3} $ &  $ 2.147  \times  10^{-1} $ &  $ 2.141  \times  10^{-1} $ &  $ 1.189  \times  10^{-2} $ &   1147s  \\
			
			$  \nu=10^{-4} $  & MW-PINNs $\uparrow$ &  $ \underline{8.132 \times  10^{-2}} $ &  $ \underline{1.949  \times  10^{-1}} $ & $ \underline{3.466 \times  10^{-3}} $ &  $ \underline{7.448  \times  10^{-2}} $ &  $ \underline{7.802  \times  10^{-2}} $ &  $ \underline{9.472  \times  10^{-3}} $ &    1885s\\
			
			$  $ & MF-PINNs  & $ \mathbf{1.706  \times  10^{-2}} $  & $ \mathbf{2.491  \times  10^{-2}}  $ &  $ 2.063  \times  10^{-2} $ &  $ 4.968  \times  10^{-1} $ &  $ 5.293  \times  10^{-1} $ &  $ 7.171 \times  10^{-2} $ &  2698s \\
			
			\midrule
			$  $ & PINNs $\uparrow$ & $ \underline{4.067  \times  10^{-3}} $  & $ \underline{4.663  \times  10^{-3}}  $ &  $ \mathbf{5.480  \times  10^{-4}} $ &  $ \underline{5.531 \times  10^{-3}} $ &  $ \underline{6.937  \times  10^{-3}} $ &  $ \mathbf{2.785  \times  10^{-4}} $ &  1206s \\
			
			$  \kappa=10^{-2}$  & AT-PINNs  &  $ 1.521  \times  10^{-1} $ &  $ 2.441  \times  10^{-1} $ & $ 1.436  \times  10^{-2} $ &  $ 8.583  \times  10^{-1} $ &  $ 8.000  \times  10^{-1} $ &  $ 2.668  \times  10^{-2} $ &   1220s  \\
			
			$  \nu=10^{-2} $  & MW-PINNs  &  $ 1.295 \times  10^{-2} $ &  $ 1.169  \times  10^{-2} $ & $ \underline{9.049 \times  10^{-4}} $ &  $ 1.422  \times  10^{-2} $ &  $ 8.422  \times  10^{-3} $ &  $ \underline{4.918  \times  10^{-4}} $ &    1339s\\
			
			$  $ & MF-PINNs  & $ \mathbf{1.720  \times  10^{-3}} $  & $ \mathbf{2.579  \times  10^{-3}}  $ &  $ 1.110  \times  10^{-3} $ &  $ \mathbf{5.286  \times  10^{-3}} $ &  $ \mathbf{5.136  \times  10^{-3}} $ &  $ 3.200 \times  10^{-3} $ &  2589s \\
			
			\midrule
			$  $ & PINNs  & $ 2.283  \times  10^{-2} $  & $\underline{ 2.172  \times  10^{-2}}  $ &  $ 1.073  \times  10^{-1} $ &  $ \underline{7.982 \times  10^{-3}} $ &  $ \underline{8.275  \times  10^{-3}} $ &  $ 2.437  \times  10^{-2} $ &  1320s \\
			
			$  \kappa=1 $  & AT-PINNs  &  $ 3.108  \times  10^{-2} $ &  $ 8.320  \times  10^{-2} $ & $ 1.946  \times  10^{-1} $ &  $ 9.185  \times  10^{-1} $ &  $ 8.696  \times  10^{-1} $ &  $ 8.420  \times  10^{-2} $ &   1199s  \\
			
			$  \nu=1 $  & MW-PINNs  &  $ \underline{2.006 \times  10^{-2}} $ &  $ 2.291  \times  10^{-2} $ & $ \underline{1.061 \times  10^{-1}} $ &  $ 1.074  \times  10^{-2} $ &  $ 9.551  \times  10^{-3} $ &  $ \underline{1.015  \times  10^{-2}} $ &    1326s\\
			
			$  $ & MF-PINNs $\uparrow$  & $\mathbf{ 3.181  \times  10^{-3}} $  & $ \mathbf{2.764  \times  10^{-3}}  $ &  $ \mathbf{1.725  \times  10^{-2}} $ &  $ \mathbf{4.606  \times  10^{-3}} $ &  $ \mathbf{4.361  \times  10^{-3}} $ &  $ \mathbf{3.805 \times  10^{-3}} $ &  3168s \\
			
			\bottomrule
		\end{tabularx}
	}
	The \textbf{bold} marks the lowest error of each physical quantity in each arguments group, while the \underline{underline} marks the second-lowest error of each physical quantity in each arguments group. The upward arrows $\uparrow$ mark the relatively best methods under the same parameters. 	
\end{table}
\setlength{\parindent}{2em}
Thirdly, we consider the performance of MF-PINNs in Table~\ref{tabresult_large} under the groups $ \mathbb{K} = \kappa \mathbb{I}, \kappa \geqslant 1 $ and $  \nu \geqslant 1  $. Consequently, we could draw the following conclusions:
\begin{enumerate}
	\item Our MF-PINNs could effectively train the velocity fields under extreme cases, like the group $  \kappa=1, \nu=10^{4} $. Similarly, our MF-PINNs could effectively train the pressure fields under group $  \kappa=10^{4}, \nu=1 $. On the opposite side, it seems that MW-PINNs has few improvements compared to baseline PINNs. What's worse, neither the baseline PINNs, MW-PINNs nor AT-PINNs could not alleviate the problem of gradient competition.
	\item Compared with other PINNs models, our MF-PINNs could also handle the gradient competition of each physical field effectively, under other complex cases, like the group $  \kappa=10^{2}, \nu=1 $, the group $  \kappa=1, \nu=10^{2} $, the group $  \kappa=10^{4}, \nu=10^{2} $ and the group $ \kappa=10^{2}, \nu=10^{4} $.
	\item Our MF-PINNs performs better than MW-PINNs, while MW-PINNs performs better than baseline PINNs under the group $ \kappa=10^{4}, \nu=10^{4} $ and the group $ \kappa=10^{2}, \nu=10^{2} $.
	\item In the cases of the group $ \kappa=10^{4}, \nu=10^{4} $ and group $ \kappa=10^{2}, \nu=10^{2} $, the performance of our MF-PINNs is better than that of MW-PINNs. And in the cases of the group $\kappa=1, \nu=1 $, the performance of MW-PINNs is like that of baseline PINNs. 	
	\item Besides, AT-PINNs is quick to stop training, but the results of AT-PINNs are not as good as the baseline PINNs obviously under the group $ \mathbb{K} = \kappa \mathbb{I}, \kappa \geqslant 1 $ and the group $  \nu \geqslant 1  $.
	\item Among these several kinds of PINNs, the errors of MW-PINNs and our MF-PINNs are closer to that of AS-DNN. This fact reflects that our MF-PINNs is closer to the maximum fitting ability of NN under the size of data and most groups $ \mathbb{K} = \kappa \mathbb{I}, \kappa \geqslant 1 $ and $  \nu \geqslant 1  $.
\end{enumerate}
\begin{table}[H] 
	\small 
	\caption{
		This table lists the performance of different kinds of PINNs under different combinations of $ \mathbb{K} $ and $ \nu $ values ( $ \mathbb{K} = \kappa \mathbb{I}, \kappa \geqslant 1 $ and $  \nu \geqslant 1  $ ). 
		\label{tabresult_large}}  
	\resizebox{\linewidth}{!}{ 
		\begin{tabularx}{1.30\textwidth}{c|cccccccc}
			\toprule
			\textbf{Arguments} & \textbf{Algorithm}	& $ \mathbf{err\mathcal{L}_{2}\left(u_s\right)} $ & $  \mathbf{err\mathcal{L}_{2}\left(v_s\right)} $ & $ \mathbf{err\mathcal{L}_{2}\left(p_s\right)} $ & $ \mathbf{err\mathcal{L}_{2}\left(u_d\right)} $ & $ \mathbf{err\mathcal{L}_{2}\left(v_d\right)} $ & $ \mathbf{err\mathcal{L}_{2}\left(p_d\right)} $ & \textbf{Time} \\ 
			\midrule
			
			NULL & AS-DNN & $ 5.441  \times  10^{-3} $  & $ 6.615  \times  10^{-3}  $ &  $ 4.039  \times  10^{-3} $ &  $ 4.736  \times  10^{-3} $ &  $ 4.576  \times  10^{-3} $ &  $ 2.921  \times  10^{-3} $ &  611s \\
			\midrule
			$  $ & PINNs   & $ 8.506  \times  10^{-1} $  & $ 8.465 \times  10^{-1}  $ &  $ 4.544  \times  10^{0} $ &  $ \underline{3.615  \times  10^{-2}} $ &  $ \underline{3.317  \times  10^{-2}} $ &  $ 4.385  \times  10^{0} $ &  2671s \\
			
			$  \kappa=1 $  & AT-PINNs  &  $ 1.546  \times  10^{0} $ &  $ 1.275  \times  10^{0} $ & $ 5.273  \times  10^{0} $ &  $ 1.881  \times  10^{-1} $ &  $ 2.014 \times  10^{-1} $ &  $ 5.498  \times  10^{0} $ &   1820s  \\
			
			$  \nu=10^{4} $  & MW-PINNs   &  $ \underline{3.252  \times  10^{-1}} $ &  $ \underline{5.312  \times  10^{-1}} $ & $ \underline{2.262  \times  10^{0}} $ &  $ 1.165  \times  10^{-1} $ &  $ 5.557  \times  10^{-2} $ &  $ \underline{1.803  \times  10^{0}} $ &   3806s  \\
			
			$  $  & MF-PINNs $\uparrow$ &  $\mathbf{ 2.336  \times  10^{-1}} $ &  $ \mathbf{3.984  \times  10^{-1}} $ & $ \mathbf{6.628  \times  10^{-1}} $ &  $ \mathbf{1.686  \times  10^{-2}} $ &  $ \mathbf{1.702  \times  10^{-2}} $ &  $\mathbf{ 2.950  \times  10^{-1}} $ &  4387s \\
			\midrule
			$  $ & PINNs  & $ \underline{2.438  \times  10^{-2}} $  & $ \underline{2.837  \times  10^{-2}}  $ &  $ \underline{1.230  \times  10^{-1}} $ &  $ 1.006  \times  10^{0} $ &  $ 1.008  \times  10^{0} $ &  $ \underline{4.523  \times  10^{-2}} $ &  1422s \\
			
			$  \kappa=10^{4} $  & AT-PINNs  &  $ 5.280  \times  10^{-2} $ &  $ 1.611  \times  10^{-1} $ & $ 2.974  \times  10^{-1} $ &  $ 1.022  \times  10^{0} $ &  $ \underline{1.024  \times  10^{-1}} $ &  $ 2.209  \times  10^{-1} $ &   1263s  \\
			
			$ \nu=1 $  & MW-PINNs  &  $ 4.604  \times  10^{-2} $ &  $ 7.590  \times  10^{-2} $ & $ 2.603  \times  10^{-1} $ &  $ \underline{1.003  \times  10^{0}} $ &  $ 1.012  \times  10^{0} $ &  $ 7.821  \times  10^{-2} $ &   1726s  \\
			
			$  $  & MF-PINNs $\uparrow$ &  $ \mathbf{2.814  \times  10^{-3}} $ &  $\mathbf{ 2.205  \times  10^{-3}} $ & $ \mathbf{1.433  \times  10^{-2}} $ &  $ \mathbf{6.335  \times  10^{-3}} $ &  $ \mathbf{6.114  \times  10^{-3}} $ &  $ \mathbf{5.755  \times  10^{-4}} $ &  3821s \\
			\midrule
			$  $ & PINNs   & $ 3.235  \times  10^{-1} $  & $ 4.478  \times  10^{-1}  $ &  $ 3.944  \times  10^{0} $ &  $ 7.348  \times  10^{-2} $ &  $ \underline{5.276  \times  10^{-2}} $ &  $ 2.985  \times  10^{0} $ &  2695s \\
			
			$  \kappa=1 $  & AT-PINNs  &  $ 3.406  \times  10^{-1} $ &  $ 8.945  \times  10^{-1} $ & $ 2.954  \times  10^{0} $ &  $ 8.897  \times  10^{-1} $ &  $ 6.241  \times  10^{-1} $ &  $ 1.451  \times  10^{1} $ &   1871s  \\
			
			$  \nu=10^{2} $  & MW-PINNs   &  $ \underline{7.680  \times  10^{-2}} $ &  $ \underline{7.207  \times  10^{-2}} $ & $ \underline{1.621  \times  10^{0}} $ &  $ \underline{7.186  \times  10^{-2}} $ &  $ 1.047  \times  10^{-1} $ &  $ \underline{2.315  \times  10^{0}} $ &   2508s  \\
			
			$  $  & MF-PINNs $\uparrow$ &  $ \mathbf{1.500 \times  10^{-2}} $ &  $ \mathbf{1.123  \times  10^{-2} }$ & $ \mathbf{8.136  \times  10^{-1}} $ &  $ \mathbf{6.666  \times  10^{-3}} $ &  $ \mathbf{6.114  \times  10^{-3}} $ &  $\mathbf{ 7.706  \times  10^{-1}} $ &  5637s \\
			\midrule
			$  $ & PINNs  & $ \underline{2.688  \times  10^{-2}} $  & $ \underline{2.561  \times  10^{-2}}  $ &  $ \underline{1.331  \times  10^{-1}} $ &  $ 9.958 \times  10^{-1} $ &  $ 9.982  \times  10^{-1} $ &  $ 3.174  \times  10^{-2} $ &  1368s \\
			
			$  \kappa=10^{2} $  & AT-PINNs  &  $ 5.010 \times  10^{-2} $ &  $ 1.061  \times  10^{-1} $ & $ 2.061  \times  10^{-1} $ &  $ 1.012  \times  10^{0} $ &  $ 1.024  \times  10^{0} $ &  $ 8.993  \times  10^{-2} $ &    1205s\\
			
			$ \nu=1 $  & MW-PINNs  &  $ 3.121  \times  10^{-2} $ &  $ 2.564  \times  10^{-2} $ & $ 1.447  \times  10^{-1} $ &  $ \underline{9.306  \times  10^{-1}} $ &  $ \underline{9.325  \times  10^{-1}} $ &  $ \underline{2.012  \times  10^{-2}} $ &   1440s  \\
			
			$  $ & MF-PINNs $\uparrow$ & $ \mathbf{4.673  \times  10^{-3}} $  & $ \mathbf{6.455  \times  10^{-3}}  $ &  $ \mathbf{2.765  \times  10^{-2}} $ &  $ \mathbf{2.337  \times  10^{-3}} $ &  $ \mathbf{2.512  \times  10^{-3}} $ &  $ \mathbf{1.268  \times  10^{-2}} $ &  3500s \\
			\midrule
			$  $ & PINNs  & $ 4.119  \times  10^{-1} $  & $ 8.610  \times  10^{-1}  $ &  $ 1.924  \times  10^{0} $ &  $ \underline{1.000 \times  10^{0} }$ &  $ 1.000  \times  10^{0} $ &  $ 1.624  \times  10^{0} $ &  2337s \\
			
			$  \kappa=10^{4}$  & AT-PINNs  &  $ 4.804  \times  10^{-1} $ &  $ 6.985 \times  10^{-1} $ & $ 1.334  \times  10^{1} $ &  $ 1.002  \times  10^{0} $ &  $ \underline{9.732  \times  10^{-1}} $ &  $ 1.334  \times  10^{1} $ &   1774s  \\
			
			$  \nu=10^{2} $  & MW-PINNs  &  $ \underline{5.641 \times  10^{-2}} $ &  $ \underline{5.177  \times  10^{-2}} $ & $ \underline{1.126 \times  10^{0}} $ &  $ 1.003  \times  10^{0} $ &  $ 9.970  \times  10^{-1} $ &  $\mathbf{ 6.557 \times  10^{-2}} $ &    2058s\\
			
			$  $ & MF-PINNs $\uparrow$ & $ \mathbf{1.369  \times  10^{-2}} $  & $ \mathbf{1.061  \times  10^{-2}}  $ &  $ \mathbf{9.956  \times  10^{-1}} $ &  $ \mathbf{6.484  \times  10^{-2}} $ &  $ \mathbf{6.555  \times  10^{-2}} $ &  $\underline{ 7.649  \times  10^{-1}} $ &  7396s \\
			
			\midrule
			$  $ & PINNs  & $ 3.098  \times  10^{0} $  & $ 1.106  \times  10^{0}  $ &  $ 8.637  \times  10^{0} $ &  $ 1.000 \times  10^{0} $ &  $ 1.000  \times  10^{0} $ &  $ 8.553  \times  10^{0} $ &  2144s \\
			
			$  \kappa=10^{2}$  & AT-PINNs  &  $ 2.433 \times  10^{0} $ &  $ 1.333  \times  10^{0} $ & $ \underline{2.095 \times  10^{0}} $ &  $ 1.014  \times  10^{0} $ &  $ 1.047  \times  10^{0} $ &  $ 1.883  \times  10^{1} $ &    2770s\\
			
			$  \nu=10^{4} $  & MW-PINNs  &  $ \underline{2.047 \times  10^{0}} $ &  $ \underline{8.288  \times  10^{-1}} $ & $ 6.471 \times  10^{0} $ &  $ \mathbf{1.290  \times  10^{-1}} $ &  $ \mathbf{1.352  \times  10^{-1}} $ &  $ \underline{6.482  \times  10^{0}} $ &    2805s\\
			
			$  $ & MF-PINNs $\uparrow$ & $ \mathbf{2.906  \times  10^{-1}} $  & $ \mathbf{1.882  \times  10^{-1}}  $ &  $ \mathbf{3.701  \times  10^{-1}} $ &  $ \underline{1.464  \times  10^{-1}} $ &  $ \underline{1.539  \times  10^{-1}} $ &  $ \mathbf{1.477  \times  10^{-1}} $ &  4014s \\
			
			\midrule
			$  $ & PINNs  & $ 1.408  \times  10^{0} $  & $ 1.565  \times  10^{0}  $ &  $ 4.623  \times  10^{0} $ &  $ 1.000 \times  10^{0} $ &  $ 1.000  \times  10^{0} $ &  $ 3.257  \times  10^{0} $ &  3072s \\
			
			$  \kappa=10^{4}$  & AT-PINNs  &  $ 1.562  \times  10^{0} $ &  $ 1.455  \times  10^{0} $ & $ 2.027  \times  10^{0} $ &  $ 2.262  \times  10^{0} $ &  $ 3.733  \times  10^{0} $ &  $ 1.929  \times  10^{1} $ &   2547s  \\
			
			$  \nu=10^{4} $  & MW-PINNs  &  $ \underline{3.484 \times  10^{-1}} $ &  $ \mathbf{5.449 \times  10^{-1}} $ & $ \underline{1.286 \times  10^{0}} $ &  $ \underline{7.459  \times  10^{-1}} $ &  $ \underline{9.900  \times  10^{-1}} $ &  $ \underline{4.087  \times  10^{-1}} $ &    2837s\\
			
			$  $ & MF-PINNs $\uparrow$ & $ \mathbf{2.000  \times  10^{-1}} $  & $ \underline{6.306  \times  10^{-1}}  $ &  $ \mathbf{4.114  \times  10^{-1}} $ &  $ \mathbf{1.689  \times  10^{-1}} $ &  $ \mathbf{1.697  \times  10^{-1}} $ &  $ \mathbf{2.032 \times  10^{-1}} $ &  3995s \\
			
			\midrule
			$  $ & PINNs  & $ 1.900  \times  10^{-1} $  & $ 1.147  \times  10^{0}  $ &  $ 3.038  \times  10^{0} $ &  $ 1.012 \times  10^{0} $ &  $ 2.336  \times  10^{0} $ &  $ 2.990  \times  10^{0} $ &  1904s \\
			
			$  \kappa=10^{2}$  & AT-PINNs  &  $ 8.916  \times  10^{-1} $ &  $ 8.958  \times  10^{-1} $ & $ 2.961  \times  10^{0} $ &  $ 1.904  \times  10^{0} $ &  $ 1.227  \times  10^{0} $ &  $ 2.152  \times  10^{0} $ &   2272s  \\
			
			$  \nu=10^{2} $  & MW-PINNs  &  $ \underline{6.667 \times  10^{-2}} $ &  $\underline{ 4.141  \times  10^{-2}} $ & $ \underline{7.902 \times  10^{-1}} $ &  $ \underline{8.544  \times  10^{-2}} $ &  $ \underline{6.701  \times  10^{-2}} $ &  $ \underline{4.041  \times  10^{-1}} $ &    2285s\\
			
			$  $ & MF-PINNs $\uparrow$ & $ \mathbf{2.014  \times  10^{-2}} $  & $ \mathbf{1.450  \times  10^{-2}}  $ &  $ \mathbf{3.157  \times  10^{-1}} $ &  $ \mathbf{5.781  \times  10^{-2}} $ &  $ \mathbf{5.812  \times  10^{-2}} $ &  $ \mathbf{5.913 \times  10^{-2}} $ &  6679s \\
			
			\midrule
			$  $ & PINNs  & $ 2.867  \times  10^{-2} $  & $ 2.729  \times  10^{-2}  $ &  $ 1.478  \times  10^{-1} $ &  $ 1.335 \times  10^{-2} $ &  $ 1.652  \times  10^{-2} $ &  $ 4.096  \times  10^{-2} $ &  1324s \\
			
			$  \kappa=1 $  & AT-PINNs  &  $ 4.077  \times  10^{-2} $ &  $ 8.280  \times  10^{-2} $ & $ 2.055  \times  10^{-1} $ &  $ 8.397  \times  10^{-1} $ &  $ 8.074  \times  10^{-1} $ &  $ 8.716  \times  10^{-2} $ &   1255s  \\
			
			$  \nu=1 $  & MW-PINNs  &  $ \underline{1.725 \times  10^{-2}} $ &  $ \underline{2.180  \times  10^{-2}} $ & $ \underline{7.160 \times  10^{-2}} $ &  $ \underline{1.207  \times  10^{-2}} $ &  $ \underline{1.160  \times  10^{-2}} $ &  $\underline{ 2.102  \times  10^{-2}} $ &    1420s\\
			
			$  $ & MF-PINNs $\uparrow$ & $ \mathbf{2.996  \times  10^{-3}} $  & $ \mathbf{2.524  \times  10^{-3}}  $ &  $ \mathbf{1.936  \times  10^{-2}} $ &  $ \mathbf{2.919  \times  10^{-3}} $ &  $ \mathbf{3.026  \times  10^{-3}} $ &  $ \mathbf{1.608 \times  10^{-3}} $ &  3329s \\	
			\bottomrule
		\end{tabularx}
	}
	The \textbf{bold} marks the lowest error of each physical quantity in each arguments group, while the \underline{underline} marks the second-lowest error of each physical quantity in each arguments group. The upward arrows $\uparrow$ mark the relatively best methods under the same parameters. 	
\end{table}
\clearpage
\subsubsection{Ablation experiments for MF-PINNs}\label{sec:Ablation experiments for MF-PINNs} To verify the effect of other improvements for MF-PINNs, we conduct the following ablation experiments.

\setlength{\parindent}{2em} Firstly, we conduct several experiments on different combinations of activation functions. Table~\ref{tabAF} lists the results of the different combinations of activation functions. All the experiments adopt MF-PINNs under the group $ \mathbb{K} = 10^{-4}\mathbb{I} , \nu = 1 $. Thus, we could draw the following conclusions:
\begin{enumerate}
	\item The smoothness of $ ReLU(\theta \textbf{x}) $ is too limited, and it could not be used in Stokes-Darcy problems, unless it is replaced by $ Softplus(\theta \textbf{x}) $, $ Swish(\theta \textbf{x}) $, etc.
	\item The adaptive $ tanh(\theta \textbf{x}) $ converges faster than $ sigmoid(\theta \textbf{x}) $, while it seems that the effects of adaptive $ sigmoid(\theta \textbf{x}) $ are much better than $ tanh(\theta \textbf{x}) $ in our MF-PINNs.
	\item  The adaptive $ sin( 2 \pi \theta \textbf{x} / T ) $ is quite effective for high-frequency problems, but it may not be the best choice for low-frequency problems. In addition, if the parameter $ 2 \pi \theta / T $ is too large, like $ sin(2 \pi \theta \textbf{x}) $ in this case, it may be very risky to cause gradient explosion during backwarding.
	\item The pre-positioned Fourier feature layers are one of the effective ways for our MF-PINNs. Furthermore, accurate periodic characteristics are quite crucial for training PINNs. For example, the performance of $ tanh(\theta) \circ sin(\mathbf{x})$ would not be as suitable as that of $ tanh(\theta) \circ  sin(\pi \mathbf{x})$ in this case.
	\item In this case, it is obvious to see the period of the velocity fields and pressure fields, $ T_{\mathbf{u}} = 1, T_{p} = 2 $. Hence, the least common multiple of their periods is $ T = 2 $, and the periods of Fourier feature operators had better be several integer multiples of $ T = 2 $. In Table~\ref{tabAF}, the performance of $ tanh(\theta) \circ  sin(\pi \mathbf{x}) $ is not as effective as $ tanh(\theta) \circ sin(2 \pi \mathbf{x}) $ for our MF-PINNs. And this difference is especially reflected in the error of pressure field, $ err\mathcal{L}_{2}\left(p\right)$. 
\end{enumerate}

\begin{table}[H]	
	\caption{This table shows that different combinations of activation functions (AF) lead to changing accuracy of our MF-PINNs  under group $ \mathbb{K} = 10^{-4}\mathbb{I}$, $ \nu = 1 $. \label{tabAF}}
	
	\resizebox{\linewidth}{!}{ 
		\begin{tabularx}{1.350\textwidth}{cc|ccccccc}
			\toprule
			\textbf{First AF} & \textbf{Other AF} & $ \mathbf{err\mathcal{L}_{2}\left(u_s\right)} $ & $  \mathbf{err\mathcal{L}_{2}\left(v_s\right)} $ & $ \mathbf{err\mathcal{L}_{2}\left(p_s\right)} $ & $ \mathbf{err\mathcal{L}_{2}\left(u_d\right)} $ & $ \mathbf{err\mathcal{L}_{2}\left(v_d\right)} $ & $ \mathbf{err\mathcal{L}_{2}\left(p_d\right)} $ & 
			\textbf{Time} \\ 
			\midrule
			$ ReLU(\theta \mathbf{x}) $  & $ ReLU(\theta \mathbf{x}) $ &  $ Inf $ &  $ Inf $ & $ Inf $ &  $ Inf $ &  $ Inf $ &  $ Inf $ &  4582s  \\	
			$ Softplus(\theta \mathbf{x}) $  & $ Softplus(\theta \mathbf{x}) $  &  $ 1.731  \times  10^{-2} $ &  $ 1.954 \times 10^{-2} $ & $ 7.732  \times  10^{-1} $ &  $ 2.629 \times  10^{-3} $ &  $ 2.708  \times  10^{-3} $ &  $ 2.372  \times  10^{0} $ &  7477s  \\
			$ sigmoid(\theta \mathbf{x}) $  & $ sigmoid(\theta \mathbf{x}) $  &  $ 2.811  \times  10^{-2} $ &  $ 2.745 \times 10^{-2} $ & $ 1.630  \times  10^{-1} $ &  $ 6.859  \times  10^{-4} $ &  $ 6.695  \times  10^{-4} $ &  $ \underline{2.299  \times  10^{-1}} $ &  5652s  \\	
			$ tanh(\theta \mathbf{x})$ & $ tanh(\theta \mathbf{x}) $  &  $ \mathbf{ 4.287  \times  10^{-3}} $ &  $ \mathbf{4.367 \times 10^{-3}} $ & $ 8.676  \times  10^{-1} $ &  $ 9.146  \times  10^{-4} $ &  $ 1.072  \times  10^{-3} $ &  $ 1.371  \times  10^{0} $ &  3756s  \\
			\midrule	
			$ sin (\theta \mathbf{x}) $  & $ sin(\theta \mathbf{x}) $  &  $ 6.273  \times  10^{-3} $ &  $ \underline{6.381 \times 10^{-3}} $ & $ 8.854  \times  10^{-1} $ &  $ 1.817  \times  10^{-3} $ &  $ 1.869  \times  10^{-3} $ &  $ 1.384  \times  10^{0} $ &  3616s  \\
			$ sin( \pi \theta \mathbf{x}) $  & $ sin( \pi \theta \mathbf{x}) $  &  $ 6.999  \times  10^{-3} $ &  $ 7.427 \times 10^{-3} $ & $ \underline{1.137  \times  10^{-1}} $ &  $ \mathbf{4.568 \times  10^{-4}} $ &  $ \mathbf{4.304  \times  10^{-4}} $ &  $ 2.303  \times  10^{-1} $ &  3652s  \\
			$ sin(2 \pi \theta \mathbf{x}) $  & $ sin(2 \pi \theta\mathbf{x}) $  &  $ 1.000  \times  10^{0} $ &  $ 1.000 \times 10^{0} $ & $ 1.179  \times  10^{1} $ &  $ 2.911 \times  10^{-1} $ &  $ 2.885  \times  10^{-1} $ &  $ 1.179  \times  10^{1} $ &  2359s  \\
			\midrule
			\makecell[cc]{$ tanh(\theta) \circ $ \\ $ sin(\mathbf{x}) $}  & $  tanh(\theta \mathbf{x}) $  &  $ 5.218  \times  10^{-3} $ &  $ 6.560 \times 10^{-3} $ & $ 1.231  \times  10^{0} $ &  $ 6.244  \times  10^{-4} $ &  $ 1.081  \times  10^{-3} $ &  $ 1.907  \times  10^{0} $ &  4424s  \\		
			\makecell[cc]{$ tanh(\theta) \circ $ \\ $ sin(\pi \mathbf{x}) $}  & $  tanh(\theta \mathbf{x}) $  &  $ 8.621  \times  10^{-3} $ &  $ 9.590 \times 10^{-3} $ & $ 2.326  \times  10^{-1} $ &  $ 7.893  \times  10^{-4} $ &  $ 7.703  \times  10^{-4} $ &  $ 3.653  \times  10^{-1} $ &  3163s  \\
			\makecell[cc]{$ tanh(\theta) \circ $ \\ $ sin(2 \pi \mathbf{x}) $ } & $  tanh(\theta \mathbf{x}) $  &  $\underline{ 4.964  \times  10^{-3}} $ &  $ 6.485 \times 10^{-3} $ & $ \mathbf{6.737  \times  10^{-2}} $ &  $ \underline{4.964  \times  10^{-4}} $ &  $ \underline{4.927  \times  10^{-4}} $ &  $ \mathbf{1.068  \times  10^{-2}} $ &  3662s  \\
			\bottomrule
		\end{tabularx}
	}
	\noindent{\footnotesize{The \textbf{bold} marks the lowest error of each physical quantity, while the \underline{underline} marks the second-lowest error of each physical quantity. The term \textbf{Time} refers to the total time record taken for $ 10000 $ epochs. }}	
\end{table}

\begin{figure}[H]
	\centering
	\subfloat[\centering]{\includegraphics[width=7.2cm]{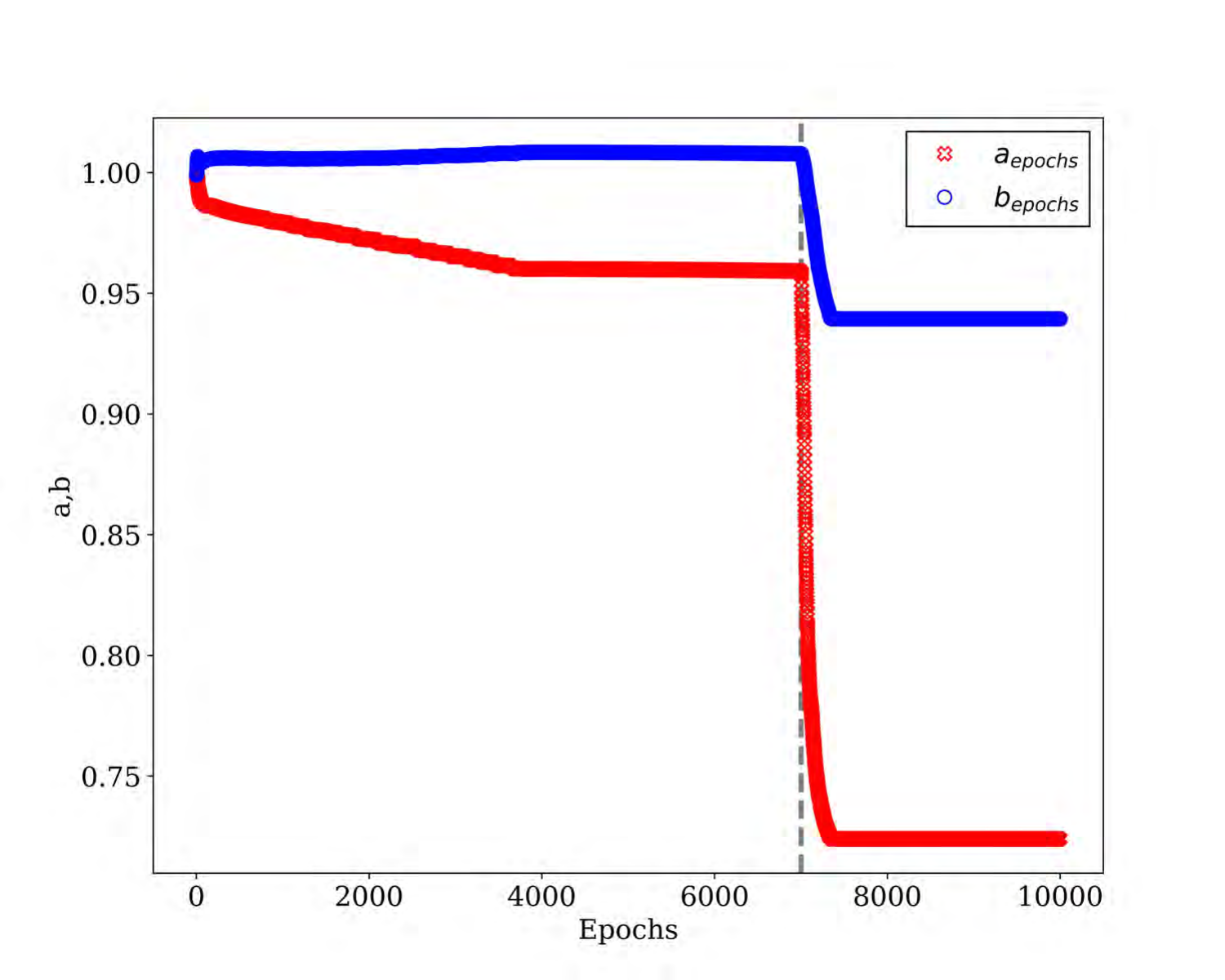}}
	\hfill 
	\subfloat[\centering]{\includegraphics[width=7.2cm]{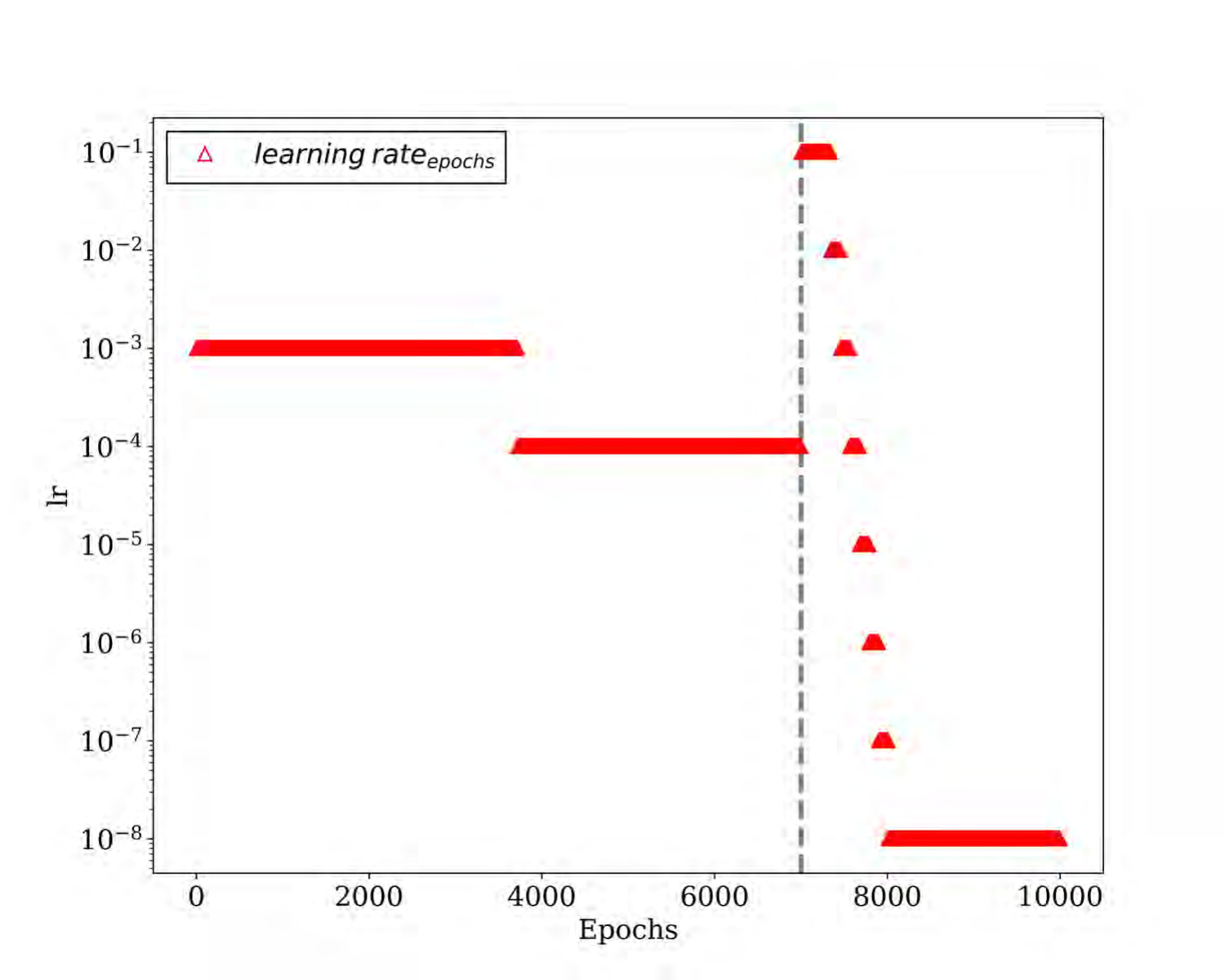}}
	\caption{These two pictures show the training process of our MF-PINNs under  $ \mathbb{K} = 10^{-4}\mathbb{I}$ and $ \nu = 1 $. The dashed grey line means we end up using the Adam optimizer and then use the L-BFGS optimizer. (\textbf{a}) the dynamic change of adaptive parameters $ a $ and $ b $ of the adaptive activation functions. (\textbf{b}) the dynamic change of learning rate.  \label{ab_lr}}
\end{figure}

\setlength{\parindent}{2em} Secondly, we use the adaptive activation function strategy for our MF-PINNs to accelerate converging according to Section~\ref{sec:Activation functions with high-frequency features}. Fig.~\ref{ab_lr} shows the dynamic change of the adaptive parameters $ a,b $ during the training process. The activation functions $ \mathcal{F}_s = tanh(0.7240x) $ and $ \mathcal{F}_d = tanh(0.9394x) $ are suitable for this particular example under the group $ \mathbb{K} = 10^{-4}\mathbb{I}  $,  $ \nu = 1 $.

\setlength{\parindent}{2em}
Lastly, we verify the effect of the adaptive strategy for learning rate decay via \texttt{ReduceLROnPlateau} we adopted in Section~\ref{sec:Optimizer and learning rate decay} under group $ \mathbb{K} = 10^{-4}\mathbb{I}, \nu = 1 $. Furthermore, we could draw the following conclusions:
\begin{enumerate}
	\item In Fig.~\ref{ab_lr}, at the beginning of training during Adam and L-BFGS stages, we set high initial learning rates in Table~\ref{tabpara} to accelerate  $ u_{NN} $ to converge effectively. Consequently, the $ err\mathcal{L}_2 $ of group  Adam:$ \, 10^{-3} $ and  L-BFGS: $\, 10^{-1} $ is generally lower than that of group  Adam $:\, 10^{-3} $ and  L-BFGS: $\, 10^{-3} $.	
	\item Midway through training, we make the learning rate adaptively decay. This work avoids $ u_{NN} $ oscillating around the optimal point $ u $ because of a relatively high learning rate. 
	The evidence is obvious as follows: 
	\begin{enumerate}
		\item In Table~\ref{tabLR}, the group with the adaptive learning rate decay strategy for the Adam optimizer has a lower error. This is because the strategy for the Adam has a higher training efficiency, and it brings the parameter group closer to the optimal point when the optimizer is changed to L-BFGS.
		\item In Table~\ref{tabLR}, although the pressure fields are easy to be ignored, the groups with the adaptive learning rate decay strategy for the L-BFGS could perform more outstandingly. 
		\item In Fig.~\ref{Loss_and_L2_34}, the total loss $\mathcal{J}(\mathbf{x},\Theta) $  oscillates violently and the phenomenon, loss spikes, keep appearing. So it causes the error of MF-PINNs to increases rather than decreases. After adjusting the learning rate of the Adam optimizer from $10^{-3} $ to $ 10^{-4} $  at the $3718^{th} $ epoch in Fig.~\ref{ab_lr}, the total loss oscillations  are mitigated, and the $ err\mathcal{L}_2(u_d) $ and $ err\mathcal{L}_2(v_d) $ begin to decrease steadily again in Fig.~\ref{Loss_and_L2_34}. 
	\end{enumerate}	
	
	\item In Table~\ref{tabLR}, the groups without any strategy for decaying the learning rate of L-BFGS require more epochs to stop the oscillation. This trouble means more time and computing resources are consumed. At the end of training of L-BFGS, the learning rate decreases adaptively, interval by interval. When the learning rate decreases to $ 10^{-8} $, $\mathcal{J}(\mathbf{x},\Theta) $  and $ err\mathcal{L}_2 $ barely change in value. Hence, we could infer that $ u_{NN}(\mathbf{x},\Theta) $ has nearly enough reached the optimal point $ u(\mathbf{x},\overline{\Theta}) $.
	
\end{enumerate}

\setlength{\parindent}{2em}
These facts prove that the strategies we have defined in Section~\ref{sec:Optimizer and learning rate decay} are very necessary and highly efficient for training PINNs. 

\begin{table}[H]
	\caption{This table shows how different initial learning rates (I-LR) and learning rate decay strategies (LRD) affect the accuracy of our MF-PINNs under the group $ \mathbb{K} = 10^{-4}\mathbb{I}$, $ \nu = 1 $. If all the $err\mathcal{L}_2$ no longer oscillate later, we approximately consider that the PINNs solutions converge at the $ n^{th} $ epoch (CE) during the L-BFGS stage.\label{tabLR}}
	
	\resizebox{\linewidth}{!}{ 
		\begin{tabularx}{1.50\textwidth}{c|cccccccccc}
			\toprule
			\textbf{I-LR} & \textbf{LRD-A} & \textbf{LRD-L} & $ \mathbf{err\mathcal{L}_{2}\left(u_s\right)} $ & $ \mathbf{err\mathcal{L}_{2}\left(v_s\right)} $ & $ \mathbf{err\mathcal{L}_{2}\left(p_s\right)} $ & $ \mathbf{err\mathcal{L}_{2}\left(u_d\right)} $ & $ \mathbf{err\mathcal{L}_{2}\left(v_d\right)} $ & $ \mathbf{err\mathcal{L}_{2}\left(p_d\right)} $ & 
			\textbf{CE} &
			\textbf{Time}\\ 
			
			\midrule
			
			\textit{Adam:}  &  \XSolidBrush  & \XSolidBrush &  $ \underline{1.195  \times  10^{-2}} $ &  $\underline{ 2.209 \times 10^{-2}} $ & $ \underline{1.365  \times  10^{-1}} $ &  $\underline{1.565 \times  10^{-3}}  $ &  $ 1.805  \times  10^{-3} $ &  $ 2.375  \times  10^{-1} $ & $ \geqslant 3000^{th} $ &  10027s  \\
			
			$ 10^{-3} $ &  \CheckmarkBold  & \XSolidBrush &  $ 3.866  \times  10^{-2} $ &  $ 5.955 \times 10^{-2} $ & $ 2.540  \times  10^{-1} $ &  $ 4.406 \times  10^{-3} $ &  $ 4.718  \times  10^{-3} $ &  $ 2.473  \times  10^{-1} $ & $ \geqslant 2989^{th} $ &  3827s  \\
			
			\textit{L-BFGS:}  &  \XSolidBrush  & \CheckmarkBold  &  $1.459  \times  10^{-2} $ &  $ 4.341 \times 10^{-2} $ & $ 3.221  \times  10^{-1} $ &  $ 1.764 \times  10^{-3} $ &  $ \underline{1.757  \times  10^{-3}} $ &  $ 4.602  \times  10^{-1} $ & $ 1611^{th} $ &  6161s  \\
			
			$ 10^{-3} $ &  \CheckmarkBold  & \CheckmarkBold  &  $ 4.154  \times  10^{-2} $ &  $ 7.835 \times 10^{-2} $ & $ 2.726  \times  10^{-1} $ &  $ 3.950 \times  10^{-3} $ &  $ 4.466  \times  10^{-3} $ &  $ \underline{2.116  \times  10^{-1}} $ & $ 1^{st} $ &  2017s  \\
			
			\midrule
			
			\textit{Adam:}  &  \XSolidBrush  & \XSolidBrush &  $ 5.596  \times  10^{-3} $ &  $ 7.626 \times 10^{-3} $ & $ 1.972  \times  10^{-1} $ &  $ \mathbf{2.516 \times  10^{-4}} $ &  $ \mathbf{2.596  \times  10^{-4}} $ &  $ 2.858  \times  10^{-1} $ & $ \geqslant 3000^{th} $ &  6131s  \\
			
			$ 10^{-3} $ &  \CheckmarkBold  & \XSolidBrush &  $ 4.200  \times  10^{-3} $ &  $ 4.270 \times 10^{-3} $ & $ 1.537  \times  10^{-1} $ &  $ 2.603 \times  10^{-4} $ &  $ 2.741  \times  10^{-4} $ &  $ 2.275  \times  10^{-1} $ & $ \geqslant 3000^{th} $ &  5198s  \\
			
			\textit{L-BFGS:}  &  \XSolidBrush  & \CheckmarkBold  &  $5.690  \times  10^{-3} $ &  $ 6.345 \times 10^{-3} $ & $ 3.532  \times  10^{-1} $ &  $ 4.785 \times  10^{-4} $ &  $ 5.433  \times  10^{-4} $ &  $ 5.927  \times  10^{-1} $ & $ 597^{th} $ &  3086s  \\
			
			$ 10^{-1} $ &  \CheckmarkBold  & \CheckmarkBold  &  $ \mathbf{2.715  \times  10^{-3}} $ &  $ \mathbf{3.103 \times 10^{-3}} $ & $ \mathbf{5.404  \times  10^{-2}} $ &  $ 4.959 \times  10^{-4} $ &  $ 6.254  \times  10^{-4} $ &  $\mathbf{ 6.951  \times  10^{-2}} $ & $ 569^{th} $ &  3365s  \\
			\bottomrule
		\end{tabularx}
	}
	\noindent{\footnotesize{The \underline{underline} marks the minimum error of the former group, while the \textbf{bold} marks the minimum error of the latter group. }}
\end{table}

\section{Conclusions and prospects}\label{sec:Conclusions and prospects}  In this paper, we conclude that extreme physical constants always produce ill-conditional numerical formulations in conventional methods. To improve PINNs, we conclude with the following suggestions.

\begin{enumerate}
	\item \textbf{From the perspective of physical laws:}	
	\begin{enumerate}
		\item 	The multiple physics fields are usually coupled through physical constants, such as Reynolds number, permeability tensor, etc. When they are either extremely high or low, they may lead to gradient competition between the multiple physics fields and failed training for conventional PINNs.
		\item	For the problems above, our MF-PINNs decouples the velocity field and the pressure field by combining the VP form and the SV form. This improvement could effectively alleviate the gradient competition among multiple physics fields.
		\item At present, the idea of decoupling must rely on the linear differential operators in the equations. However, it may be uncertain to generalize it to other more complex systems or models, such as Euler's equations, compressible flows, and shock waves. 
	\end{enumerate} 
	\item	\textbf{From the perspective of the activation functions and training parallel PINNs:}
	\begin{enumerate}
		\item   It is necessary to select activation functions with sufficient smoothness, because they directly determine whether the PINNs numerical solutions are well-defined or not.
		\item We could obtain the physical periodicity from the boundary conditions and non-homogeneous terms. The period of the activation functions had better be integer multiples of the original problem. Otherwise, the opposite operation may waste many computing resources.
		\item We conclude that different activation functions are effective for different problems, and combining different types of activation functions may improve the abilities of PINNs. For example, the $ tanh $ is suitable for discontinuity problems, while the $ sin $ is appropriate for high-frequency problems.		
		\item We find that increasing the initial learning rate of L-BFGS appropriately and using adaptive strategies for learning rate decay are important to our MF-PINNs.
	\end{enumerate}
\end{enumerate}

\setlength{\parindent}{2em}
Though our MOD-PINNS overcomes some shortcomings in this paper, we have to admit that it has not been studied and applied further. How to select the optimal weights for different equation forms in loss functions? Could our MF-PINNs have the potential to solve complex turbulence hidden in Navier-Stokes systems, when the Reynolds numbers are extremely high or low? These topics are worth further exploring and studying.  

\clearpage

\section*{CRediT authorship contribution statement}\label{sec:CRediT authorship contribution statement} \textbf{Li Shan:} Conceptualization, Methodology, Investigation, Writing – review \& editing, Supervision, Project administration, Funding acquisition. \textbf{Xi Shen:} Methodology, Visualization, Coding, Validation, Writing - original draft preparation.

\section*{Declaration of competing interest}\label{sec:Declaration of competing interest}
The authors declare that they have no known competing financial interests or personal relationships that could have appeared to influence the work reported in this paper.

\section*{Data availability}\label{sec:Data availability}
Data will be made available on request. The code and data associated with this paper are available at \href{https://github.com/shxshx48716/MF-PINNs.git}{https://github.com/shxshx48716/MF-PINNs.git}. 

\section*{Acknowledgements}\label{sec:Acknowledgements} 
The authors of this paper gratefully acknowledge Prof. J. Zhao of Capital Normal University for his expert guidance on methodology and reviewing of the entire manuscript.

\end{document}